\documentclass[12pt]{article}
\usepackage{amssymb}
\def\be{\begin{equation}}
\def\ee{\end{equation}}
\def\bea{\begin{eqnarray}}
\def\eea{\end{eqnarray}}
\renewcommand{\theequation}{\arabic{section}.\arabic{equation}}

\newcommand\eq[1]{Eq.~(\ref{#1})}

\usepackage[dvips]{graphicx}
\begin{document}
\pagestyle{empty}
\title{Cosmological Perturbations \\
in an\\ 
Inflationary Universe}

\vspace{5.0cm}

\author{{\sc Karim Ali Malik} \\ \\ \\ \\ \\ \\ \\ \\ \\
        Submitted for the Degree of Doctor of Philosophy \\ \\
        Department of Computer Science and Mathematics \\ \\
        University of Portsmouth, UK.}

\date{January 2001}
\maketitle
\thispagestyle{empty}
%

\newpage
\begin{center}
\vspace*{12.0cm}
To my parents.

\end{center}
\newpage

\newpage
\begin{center}
\vspace*{12.0cm}
Parturient montes, nascetur ridiculus mus \cite{Horaz}.

\end{center}
\newpage

\newpage

\vspace*{6.0cm}

\begin{center}
\section*{Acknowledgments}
\end{center}

\vspace*{2.0cm}

I would like to thank my supervisor David Wands for his kind guidance
and his patience and express my gratitude to David Matravers and Roy
Maartens. \\

I must also like to thank the Relativity and Cosmology Group for their
constant inspiration and the School of Computer Science and
Mathematics and the University of Portsmouth for funding this work.

%
\newpage
\pagestyle{plain}
\pagenumbering{roman}

\begin{abstract}
%
%
After introducing the perturbed metric tensor in a
Friedmann-\-Robertson--Walker (FRW) background we review how the
metric perturbations can be split into scalar, vector and tensor
perturbations according to their transformation properties on spatial
hypersurfaces and how these perturbations transform under small
coordinate transformations.  This allows us to introduce the notion of
gauge-invariant perturbations and to provide their governing
equations. We then proceed to give the energy- and
momentum-conservation equations and Einstein's field equations for a
perturbed FRW metric for a single fluid and extend this set of
equations to the multi-fluid case including energy transfer.  In
particular we are able to give an improved set of governing equations
in terms of newly defined variables in the multi-fluid case.  We also
give the Klein-Gordon equations for multiple scalar fields. After
introducing the notion of adiabatic and entropic perturbations  we
describe under what conditions curvature perturbations are conserved
on large scales. We derive the ``conservation law'' for the curvature
perturbation for the first time using only the energy conservation
equation \cite{separate}.

We then investigate the dynamics of 
assisted inflation. In this model an arbitrary number of scalar fields
with exponential potentials evolve towards an inflationary scaling
solution, even if each of the individual potentials is too steep to
support inflation on its own.  By choosing an appropriate rotation in
field space we can write down for the first time explicitly 
the potential for the
weighted mean field along the scaling solution and for fields
orthogonal to it.  
This allows us to present analytic solutions
describing homogeneous and inhomogeneous perturbations about the
attractor solution without resorting to slow-roll approximations. We
discuss the curvature and isocurvature perturbation spectra produced
from vacuum fluctuations during assisted inflation \cite{MW}.

Finally we investigate the recent claim that preheating after
inflation may affect the amplitude of curvature perturbations on large
scales, undermining the usual inflationary prediction. We analyze the
simplest model of preheating analytically, and show that in linear
perturbation theory the effect is negligible. The dominant effect is
second-order in the field perturbation and we are able to show that
this too is negligible, and hence conclude that
preheating has no significant influence on large-scale perturbations
in this model. We briefly discuss the likelihood of an effect in other
models \cite{LLMW}.

We end this work with some concluding remarks, possible extensions
and an outlook to future work.

\end{abstract}
%
%
\newpage

\tableofcontents
\newpage

\newpage
\pagenumbering{arabic}
%
%
%
%
\newpage
\section{Introduction}

It has almost become impossible to write about cosmology without
having at least mentioned once that we are now experiencing and living
through ``the golden age'' of cosmology.  This thesis is no exception.\\

The standard inflationary paradigm is an extremely successful model in
explaining observed structures in the Universe (see
Refs.~\cite{LL93,LLbook,David+Tony} for reviews).  In this
cosmological model the universe is dominated at early times by a
scalar field that gives rise to a period of accelerated expansion,
which has become known as inflation. Inhomogeneities originate from
the vacuum fluctuations of the inflaton field, which on being
stretched to large scales become classical perturbations. The field
inhomogeneities generate a perturbation in the curvature of uniform
density hypersurfaces, and later on these inhomogeneities are
inherited by matter and radiation when the inflaton field decays.
Large scale structure then forms in the eras of radiation and matter
domination through gravitational attraction of the seed fluctuations.

Defect models of cosmic structure formation have for a long period
been the only viable competitor for the inflation based models.  In
these models a network of defects, usually cosmic strings, seeds the
growth of cosmic structure continuously through the history of the
Universe. Unfortunately, at least for the researchers involved in the
investigation of these models, recent cosmic microwave background
(CMB) observations have nearly ruled these models out \cite{albrecht}.

The amount of data available in cosmology is increasing rapidly. New
galaxy surveys, like the 2-degree-Field and the
Sloan-Digital-Sky-Survey are beginning to determine the matter power
spectrum, the distribution of galaxies and clusters of galaxies, on
large scales. New balloon based experiments, like Boomerang and
Maxima, have already measured the fluctuations in the CMB with
unprecedented accuracy. Forthcoming satellite missions, like MAP and
Planck, are going to improve these results even more. 

Although it might seem that extending the parameter space of the
cosmological models that are investigated will lead to a decrease in
accuracy, the quality and sheer quantity of the new data will
nevertheless allow us to constrain the parameter space enough to rule
out large classes of models and to give hints in which direction we
should direct our theoretical and experimental attention in future 
\cite{bucher}. 
The combination of data from different CMB experiments with large
scale structure and super-nova observations will be particularly
useful in this respect.

But to take full advantage from these new experiments and the new data
they will provide one has to make accurate theoretical predictions.
In this work we extend the theory of cosmological perturbations.  We
apply the methods we have developed to investigate some models of the early
universe and calculate their observable consequences.

The evolution and the dynamics of the universe 
are governed by  Einstein's theory of relativity, whose
field equations are \cite{MTW}
\be 
\label{Einstein}
G_{\mu\nu}=8\pi G~T_{\mu\nu} \, ,
\ee
where $G_{\mu\nu}$ is the Einstein tensor, $G$ Newton's constant 
and $T_{\mu\nu}$ the energy-momentum tensor.
From the 
contracted Bianchi identity $G_{\mu\nu;}^{~~~\nu} =0$, we find
that the energy-momentum tensor is covariantly conserved,
\be
\label{stressenergy}
T_{\mu\nu;}^{~~~\nu} =0\, .
\ee
In an unperturbed Friedmann-Robertson-Walker (FRW) universe, described
by the line element
\be
ds^2=-dt^2+a^2(t)\left[\frac{dr^2}{1-\kappa r^2}+r^2\left(d\theta^2
+\sin^2\theta d\phi^2\right)\right] \,,
\ee 
the homogeneous spatial hypersurfaces pick out a natural cosmic time
coordinate, $t$, and hence a (3+1) decomposition of spacetime.  Here
$a(t)$ is the scale factor and $r$, $\theta$ and $\phi$ are polar
coordinates. But in the presence of inhomogeneities this choice of
coordinates is no longer unambiguous.

The need to clarify these ambiguities lead to the fluid-flow approach
which uses the velocity field of the matter to define perturbed
quantities orthogonal to the fluid-flow~\cite{Hawking,Lyth_a,
Lyth_b,Lyth_c,EB_a,EB_b}. An alternative school has sought to define
gauge-invariant perturbations in any coordinate system by constructing
quantities that are explicitly invariant under first-order coordinate
transformations~\cite{Bardeen,KS,MFB,ruth,bertschinger2}. Results
obtained in the two formalisms can be difficult to compare. One
stresses the virtue of invariance of metric perturbations under gauge
transformations, while the other claims to be covariant and therefore
manifestly gauge-invariant due to its physically transparent
definition.

We put special emphasis on the fact that perturbations defined on an
unambiguous physical choice of hypersurface can always be written in a
gauge-invariant manner. In the coordinate based formalism
\cite{Bardeen} the choice of hypersurface implies a particular choice
of temporal gauge. By including the spatial gauge transformation
from an arbitrary initial coordinate system all the metric perturbations
can be given in an explicitly gauge-invariant form.
In this language, linear perturbations in the fluid-flow or
``covariant'' approach appear as a particular gauge choice (the
comoving orthogonal gauge~\cite{Bardeen,KS}) whose metric
perturbations can be given in an explicitly gauge-invariant form. But
there are many other possible choices of hypersurface, including the
zero-shear (or longitudinal or conformal Newtonian)
gauge~\cite{Bardeen,KS,MFB,Bertschinger} in which gauge-invariant
quantities may be defined. \\

Whereas there are still more things we do not know than we do know,
and even more things we don't even know we don't know, it is safe to
say that the age of ``golden-ish'' cosmology has begun.

%
%
%
%
%
%
\newpage
\section{Cosmological Perturbation Theory}
\label{cosmopertsect}
%
%
In this section we present the fundamental building blocks of
perturbation theory in a cosmological context. After introducing the
concept of gauge and gauge transformation we give the governing
equations in gauge dependent and gauge invariant form for scalar,
vector and tensor quantities in a perturbed Friedmann-Robertson-Walker
(FRW) spacetime in the single fluid case.  After a short section on
scalar fields we then give the governing equations for multiple
fluids. We conclude this section with an application of the theory to
quantities that are conserved on large scales.

\subsection{Decomposing the metric tensor}
\label{metrictensorsection}

We consider first-order perturbations about a 
FRW background, so that the metric tensor can be split up as
\be
g_{\mu\nu}=g^{(0)}_{\mu\nu} + \delta g_{\mu\nu} \, ,
\ee
where the background metric $g^{(0)}_{\mu\nu}$ is given by
\be
g^{(0)}_{\mu\nu}=a^2(\eta) \left( 
\begin{array} {cc} -1 & 0 \\ 
			 0    & \gamma_{ij}
\end{array} \right) \,,
\ee
and $\gamma_{ij}$ is the metric on the 3-dimensional space with
constant curvature $\kappa$. We will denote covariant spatial
derivatives on this space as $X_{abc\ldots|i}$
\footnote{See appendix for definitions and notation.}. 

The metric tensor has 10 independent components in 4 dimensions.  For
linear perturbations it turns out to be very useful to split the
metric perturbation into different parts labelled scalar, vector or
tensor according to their transformation properties on spatial
hypersurfaces \cite{Bardeen,Stewart}.  The reason for splitting the
metric perturbation into these three types is that they are decoupled
in the linear perturbation equations. \\
{\em Scalar} perturbations can always be constructed from a scalar
quantity, or its derivatives, and any background quantities such
as the 3-metric $\gamma_{ij}$.  We can construct any first-order
scalar metric perturbation in terms of four scalars $\phi$, $B$,
$\psi$ and $E$, where
\begin{eqnarray}
\delta g_{00} &=& - a^2 2\phi \, ,\\
\delta g_{0i} &=& \delta g_{i0} =  a^2 B_{|i} \, ,\\
\delta g_{ij} &=& -2 a^2 (\psi\;\gamma_{ij} - E_{|ij}) \, .
\end{eqnarray}

Any 3-vector, such as $B_{|i}$, constructed from a scalar is
necessarily curl-free, i.e., $B_{|[ij]}=0$. Thus one can distinguish
an intrinsically {\em vector} part of the metric perturbation $\delta
g_{0i}$, which we denote by $-S_i$, which gives a non-vanishing
$\delta g_{0[i|j]}$. Similarly we can define a vector contribution to
$\delta g_{ij}$ constructed from the (symmetric) derivative of a
vector $F_{(i|j)}$. The scalar perturbations are distinguished from
the vector contribution by requiring that the vector part is
divergence-free, i.e., $\gamma^{ij}S_{i|j}=0$.  This means that there
can be no contribution to $\delta g_{00}$ from vector perturbations to
first-order. The decomposition of a
vector field into curl- and divergence-free parts in Euclidean space
is known as Helmholtz's theorem. The curl-free and divergence-free
parts are also called longitudinal and solenoidal, respectively.
%

Finally there is a {\em tensor} contribution to $\delta g_{ij}=a^2
h_{ij}$ which is transverse ($\gamma^{jk}h_{ij|k}=0$), i.e.,
divergence-free, and trace-free ($\gamma^{ij}h_{ij}=0$) which cannot
be constructed from scalar or vector perturbations.

We have introduced four scalar functions, two (spatial) vector valued 
functions with three components each, and a symmetric spatial tensor
with six components. But these functions are subject to several
constraints: $h_{ij}$ is transverse and traceless, which contributes 
four constraints, $F_i$ and $S_i$ are divergence-free,
one constraint each. We are therefore finally left 10 new degrees
of freedom, the same number as the independent components of the
perturbed metric.  

We can now write the most general metric perturbation to first-order as
\be
\delta g_{\mu\nu}=a^2(\eta) \left( 
\begin{array}{cc} 
-2\phi & B_{|i} - S_i\nonumber\\ 
B_{|j} - S_j& -2\psi\gamma_{ij} + 2E_{|ij} + F_{i|j} + F_{j|i} + h_{ij}
 \nonumber
\end{array} 
\right) \,.
\ee
The contravariant metric tensor, including the perturbed part, follows
by requiring (to first order),
$g_{\mu\nu} ~  g^{\nu\lambda}=\delta_{\mu}^{~\lambda}$, 
which gives
\be
\label{gmunu}
g^{\mu\nu}=a^{-2}(\eta) \left( 
\begin{array}{cc} 
-(1-2\phi) & B_{|}^{~i}-S^i \nonumber\\ 
B_{|}^{~j}-S^j &
 (1+2\psi)\gamma^{ij} - 2E_{|}^{~ij} -F^{i~j}_{~|} - F^{j~i}_{~|} - h^{ij}
 \nonumber
\end{array} 
\right)\,.
\ee
Thus we have the general linearly perturbed line element:
\bea
ds^2 = &a^2(\eta)& \left\{ -(1+2\phi) d\eta^2+2(B_{|i}-S_i)d\eta dx^i
\right. \nonumber \\ 
&\qquad& \left. +
\left[(1-2\psi)\gamma_{ij}+2E_{|ij} +2F_{i|j}+h_{ij} \right] dx^idx^j
 \right\} \, .
\label{ds}
\eea
%

\subsection{Coordinate Transformation}

We use Eq.~(\ref{ds}) as a starting point and perform a small
coordinate transformation, where the new coordinate system is denoted
by a tilde.

The homogeneity of a FRW spacetime gives a natural choice of
coordinates in the absence of perturbations. But in the presence of
first-order perturbations we are free to make a first-order change in
the coordinates, i.e., a gauge transformation,
\begin{equation}
\tilde\eta = \eta+\xi^0 \,, \qquad
\tilde x^i = x^i+\xi_{|}^{~i} +\bar\xi^i \,,
\label{gauge}
\end{equation}
where $\xi^0=\xi^0(\eta,x^i)$ and $\xi=\xi(\eta,x^i)$ are arbitrary 
scalar functions, and $\bar\xi^i=\bar\xi^i(\eta,x^i)$ is a 
divergence-free 3-vector. 
The function $\xi^0$ determines the choice of constant-$\eta$
hypersurfaces, that is  the time-slicing, while $\xi_|^{~i}$ and
$\bar\xi^i$ then select the
spatial coordinates within these hypersurfaces.  The choice of
coordinates is arbitrary to first-order and the definitions
of the first-order metric and matter perturbations are thus
gauge-dependent.

The result of the transformation Eq.~(\ref{gauge}) acting on any
quantity is that of taking the Lie derivative of the background value
of that quantity, i.e.~for every tensor $T$ one has \cite{Wald}
\be
\tilde{\delta T}=\delta T + {\cal{L}}_{\xi} T_0 \,,
\ee
where $T_0$ is the background value of that quantity.

We use a slightly different approach
by directly perturbing the line element in Eq.~(\ref{ds}).
Starting with the total differentials of $\xi^0$ and $\xi$ and
$d\bar\xi^i$ 
\bea
d\xi^0 &=& {\xi^0}' d\tilde\eta + \xi^0_{~|i} d\tilde x^i  \, ,  \\
d\xi &=&\xi' d\tilde\eta + \xi_{~|j} d\tilde x^j  \, , \\
d\bar\xi^i &=& \bar\xi^{i\prime} d\tilde\eta + \bar\xi^i_{~|j} 
d\tilde x^j \, ,
\eea
we get using Eq. (\ref{gauge})
\bea
d\eta &=& d\tilde\eta - {\xi^0}' d\tilde\eta - \xi^0_{~|i} d\tilde x^i  
\, , \nonumber \\
dx^i &=& d\tilde x^i - (\xi'^{~i}_{|} + \bar\xi^{i\prime}) d\tilde\eta
 - (\xi^{~i}_{|~j} + \bar\xi^i_{~|j}) d\tilde x^j \, ,  
\label{diffs}
\eea
where we have used the fact that  
$\xi^0(\eta,x^i)=\xi^0(\tilde\eta,\tilde x^i)$ and 
$\xi(\eta,x^i)=\xi(\tilde\eta,\tilde x^i)$, to first order in the
coordinate transformations. \\
Substituting the differentials Eq.~(\ref{diffs}) into Eq.~(\ref{ds})
and using
\be
a(\eta)=a(\tilde\eta)-\xi^0 a'(\tilde\eta) \, ,
\ee
we then get for the line element in the ``new'' coordinate system,
working only to first order in both the metric perturbations and the
coordinate transformations:
\bea
ds^2 &=& a^2\left(\tilde\eta\right) 
\left\{ -\left( 1+2\left(\phi-h \xi^0-\xi^{0\prime} \right) \right) 
d{\tilde\eta}^2 + 2 \left( B+\xi^0 -\xi' \right)_{|i} d\tilde\eta d\tilde x^i 
\right.  \nonumber \\
&\qquad& 
\left.
-2 \left( S_i + \bar\xi'_i \right) d\tilde\eta d\tilde x^i  
+ \left[ \left( 1- 2 \left( \psi+h\xi^0 \right) \right) \gamma_{ij} + 
2 \left( E- \xi \right)_{|ij} \right. \right. \nonumber \\
&\qquad& 
\left. \left. + 2 \left( F_{i|j} - \bar\xi_{i|j} \right) 
+ h_{ij} \right] d\tilde x^i d \tilde x^j \right\}  \, .
\label{dstilde}
\eea
The general form of the line element presented in Eq.~(\ref{ds})
must be invariant under infinitesimal coordinate 
transformations, since the line element $ds^2$ is a scalar 
invariant. 
We can read off the transformation equations for the 
metric perturbations by writing down the line element in the ``new''
coordinate system
\bea
ds^2 = &a^2(\tilde\eta)& \left\{ -(1+2\tilde\phi) d\tilde\eta^2+
2(\tilde B_{|i}-\tilde S_i)d\tilde\eta d\tilde x^i \right.
\nonumber \\
&\qquad& \left.
+ \left[(1-2\tilde\psi)\gamma_{ij}+2\tilde E_{|ij} +
2\tilde F_{i|j}+\tilde h_{ij} \right] d\tilde x^i d\tilde x^j
 \right\} \, .
\eea
The coordinate transformation of Eq.~(\ref{gauge}) induces a change in
the scalar functions $\phi$, $\psi$, $B$ and $E$ defined by Eq.~(\ref{ds})
\begin{eqnarray}
\label{transphi}
\tilde\phi&=&\phi-h \xi^0-\xi^{0\prime} \, ,\\
\label{transpsi}
\tilde\psi&=&\psi+h\xi^0 \, , \\
\label{transB}
\tilde B&=&B+\xi^0 -\xi' \, ,\\
\label{transE}
\tilde E&=&E-\xi \, , 
\label{scaltran}
\end{eqnarray}
where $h=a'/a$ and a dash indicates differentiation with
respect to conformal time $\eta$.
The vector valued functions $S_i$ and $F_i$ transform as
\bea
\tilde F_i &=& F_i - \bar \xi_i , \\ 
\tilde S_i &=& S_i + \bar \xi_i'.
\label{vectran}
\eea
The tensor part of the perturbation, $h_{ij}$, is unaffected 
by the gauge transformations.

\subsection{Time slicing and spatial hypersurfaces}

We now relate the metric perturbations to physical, that is 
observable, quantities~\cite{KS}. 
The unit time-like vector field orthogonal to constant-$\eta$ 
hypersurfaces is
\begin{equation}
N^\mu={1\over a}(1-\phi,-B_{|}^{~i}+S^i) \,,
\label{Nmu}
\end{equation}
and the covariant vector field is
\be
\label{covtimelike}
N_{\mu}=-a(1+\phi, 0) \, .
\ee
The covariant derivative of the time like unit vector field
$N_{\mu}$ can be decomposed uniquely as follows~\cite{Wald}
\footnote{
Note that the covariant time-like vector field defined in 
Eq.~(\ref{covtimelike}) can not support vorticity.}
:
\be
N_{\mu;\nu}=
\sigma_{\mu\nu}
+\frac{1}{3}\theta P_{\mu\nu}-a_{\mu}N_{\nu} \, ,
\ee
where the projection tensor $P_{\mu\nu}$ is given by
\be
P_{\mu\nu}=g_{\mu\nu}+N_{\mu}N_{\nu}.
\ee
%
%
%
The  expansion rate is given as
\be
\label{theta}
\theta=N^{\mu}_{~;\mu} \,,
\ee
the shear is
\be
\sigma_{\mu\nu}=
\frac{1}{2} P^{~\alpha}_{\mu}P^{~\beta}_{\nu}
\left( N_{\alpha; \beta} + N_{\beta; \alpha} \right)
-\frac{1}{3} \theta P_{\mu\nu} \,,
\ee
and the acceleration is
\be
a_{\mu}=N_{\mu;\nu} N^{\nu} \,.
\ee 
Note that on spatial hypersurfaces the vorticity, the shear, the
acceleration and the expansion coincide with their Newtonian
counterparts in fluid dynamics \cite{mase,HawkingEllis}.

The extrinsic curvature of hypersurfaces defined by $N_{\mu}$
is defined as
\be
K_{\mu\nu}=-P_{\nu}^{~\lambda} N_{\mu;\nu} \, .
\ee
The Raychauduri equation~\cite{HawkingEllis} is given solely in 
terms of geometric quantities and gives the evolution of 
the expansion $\theta$ with respect to proper time as
\be
\frac{d \theta}{d \tau}=a^{\mu}_{~;\mu}
-\sigma_{\mu\nu}\sigma^{\mu\nu}
-\frac{1}{3}\theta^2- R_{\mu\nu} N^{\mu} N^{\nu} \, .
\ee
%
%
%
We can write the expansion, acceleration and shear of the vector
field for a perturbed FRW spacetime, considering scalar quantities
first, as
\begin{eqnarray}
\label{expansion}
\theta &=& 3{a'\over a^2} \left(1- \phi \right) - {3\over a} \psi' -
{1\over a} \nabla^2 \left(B-E'\right) \, ,\\
\label{acceleration}
a_i &=& \phi_{|i}\,, \\
\sigma_{ij} &=& a \left( \sigma_{|ij}
  - {1\over3} \gamma_{ij} \nabla^2\sigma \right) \,,
\end{eqnarray}
where the scalar describing the shear is 
\begin{equation}
\label{shearscalar}
\sigma = - B + E' \,,
\end{equation}
and where the time components are zero, i.e.
$a_0=0$, $\sigma_{0\mu}=0$. 
Note that for an unperturbed background the expansion $\theta$ 
coincides with the expansion rate of the spatial volume per unit 
proper time.


The intrinsic spatial curvature on hypersurfaces of
constant conformal time $\eta$ is given by \cite{Bardeen,KS}
\begin{equation}
^{(3)}R = {6\kappa\over a^2}
 + \frac{12\kappa}{a^2} \psi + \frac{4}{a^2} \nabla^2\psi \, .
\end{equation}
For a perturbation with comoving wavenumber $k$,
such that $\nabla^2\psi=-k^2\psi$, we therefore have
\begin{equation}
\delta^{(3)}R = \frac{4}{a^2} \left(3\kappa-k^2\right)~\psi \, ,
\end{equation}
and $\psi$ is often simply referred to as the curvature perturbation. 

For vector perturbations we find $\bar a_{\mu}=0$, 
$\bar\sigma_{00}=\bar\sigma_{0j}=0$, 
and $\bar\theta=0$, 
and the only non-zero first order quantity is the 
shear given by
\be
\tau_{ij}\equiv \bar\sigma_{ij} 
= \frac{1}{2}a \left\{ \left( S_{i|j}+S_{j|i} \right)
+\left( F_{i|j} + F_{j|i} \right)' \right\} \, ,
\label{vecshear}
\ee
where we distinguished  the vector quantities from the scalar ones
by a ``bar''.


There is no tensor contribution to the expansion, the acceleration
and the vorticity, but there is a non-zero contribution of the
tensor pertubations to the shear,
\be
{}^{\rm{(tensor)}}\sigma_{ij}=\frac{1}{2}a\,h'_{ij} \,.
\ee

\subsubsection{Evolution of the curvature perturbation}
\label{curvpertsect}

We can now give a simple expression for the evolution of the 
curvature perturbation $\psi$.
Multiplying Eq.~(\ref{expansion}) through by $(1+\phi)$ 
in order to give the expansion with respect to coordinate 
time, $t\equiv\int a d\eta$, we get 
\begin{equation}
\tilde\theta = (1+\phi)\theta
 = 3H - 3\dot\psi + \nabla^2\tilde\sigma \, ,
\end{equation}
where $\tilde\sigma\equiv\dot E -B/a$. 
We can write this as an equation for the time evolution of $\psi$ in
terms of the perturbed expansion,
$\delta\tilde\theta\equiv\tilde\theta-3H$, and the shear:
\begin{equation}
\label{dotpsi}
\dot\psi = -{1\over3}\delta\tilde\theta + {1\over3} \nabla^2
\tilde\sigma \, .
\end{equation}
Note that this is independent of the field equations and follows
simply from the geometry. It shows that on large scales
($\nabla^2\tilde\sigma \to 0$) the change in the curvature
perturbation, $\dot\psi$, is proportional to the change in the
expansion $\delta\tilde\theta$.

\subsection{The stress-energy tensor for a fluid,\\ 
including anisotropic stress}
\label{perffluid}

Thus far we have concerned ourselves solely with the metric and its
representation under different choices of coordinates. However, in any
non-vacuum spacetime we will also have matter fields to consider. Like
the metric, the coordinate representation of these fields will also be
gauge-dependent.

The stress-energy tensor of a fluid with density $\rho$, isotropic
pressure $p$ and 4-velocity $u^{\mu}$ is given by~\cite{KS}
\begin{equation}
T^{\mu}_{~\nu} = \left( \rho +p\right) u^{\mu}u_{\nu}+p
\delta^{\mu}_{~\nu} + \pi^{\mu}_{~\nu} \, ,
\end{equation}
where we have included the anisotropic stress tensor which decomposes
into a trace-free scalar part, $\Pi$, a vector part, $\pi^i$, and a
tensor part, ${}^{\rm{(tensor)}}\pi^i_{~j}$, according to
\be
\pi^i_{~j}=\Pi_{|~j}^{~i}- \frac{1}{3}
\nabla^2\Pi \delta^{i}_{~j} + \frac{1}{2}\left(\pi^i_{~|j}
+\pi_{j|}^{~i}\right) +{}^{\rm{(tensor)}}\pi^i_{~j} \, .
\ee
The anisotropic stress tensor has only spatial components, $\pi_{ij}$,
and is gauge-invariant. The gauge-invariance can either be shown by
direct calculation as in \cite{KS}, or by observing that $\pi_{ij}$
must be manifestly gauge-invariant due to its being zero in the
background \cite{Stewart}, since the background is FRW and so 
isotropic by definition.
The linearly perturbed velocity can be written as
\bea
u^{\mu} &=& \frac{1}{a} \left[ \left(1-\phi\right), 
~v_{|}^{~i}+v^i\right] \,, \\
u_{\nu} &=& a \left[ -(1+\phi),~v_{|i}+B_{|i}+v_i-S_i \right] \, ,
\eea
where we enforce the constraint
%
$u_{\mu}u^{\mu}=-1$. 
%
We can introduce the velocity potential $v$ since the flow
is irrotational for scalar perturbations.
We then get for the components of the stress energy tensor
\begin{eqnarray}
T^0_{~0} &=& -(\rho_0+\delta\rho) \, , \\
\label{perfT_0i}
T^0_{~i} &=&(\rho_0+p_0)~ \left(B_{|i} +v_{|i} + v_i 
-S_i \right) \, , \\
T^i_{~0} &=& -(\rho_0+p_0)~\left(v_|^{~i}+v^i\right) \, , \\
\label{perfT_ij}
T^i_{~j} &=& (p_0+\delta p) ~\delta^i_{~j}+ \pi^i_{~j} \, .
\end{eqnarray}
Note, that the above definition differs slightly from that 
presented in Ref.~\cite{MFB}.

Coordinate transformations affect the split between spatial and
temporal components of the matter fields and so quantities like the
density, pressure and 3-velocity are gauge-dependent, as described in
Section \ref{scalvecttrans}. Density and pressure are scalar
quantities which transform as given in Eq.~(\ref{transrho}) in the
following section, but the velocity potential transforms as $\tilde v
= v + \xi'$, as given in Eq.~(\ref{transv}).

\subsubsection{Transformations of scalar and vector matter 
quantities}
\label{scalvecttrans}
%
%
Any scalar $\rho$ (including the fluid density or
pressure) which is homogeneous in the background FRW model can be
written as $\rho(\eta, x^i) = \rho_0(\eta) + \delta\rho(\eta,
x^i)$.  The perturbation in the scalar quantity then transforms as
\begin{equation}
\delta \tilde\rho = \delta \rho - \xi^0 \rho'_0 \, .
\label{transrho}
\end{equation}
Physical scalars on the hypersurfaces, such as spatial curvature,
acceleration, shear or the density perturbation $\delta\rho$, only
depend on the choice of temporal gauge, $\xi^0$, but are independent
of the coordinates within the 3-dimensional hypersurfaces determined
by $\xi$. The spatial gauge, determined by $\xi$, can only affect the
components of 3-vectors or 3-tensors on the hypersurfaces but not
3-scalars. \\
Vector quantities that are derived from a potential, such as the
velocity potential $v$, only depend on the shift $\xi$ within the
hypersurface and are independent of $\xi^0$. We therefore find that
the velocity potential transforms as
\begin{equation}
\tilde v = v + \xi^{\prime} \,.
\label{transv}
\end{equation}
The function $\bar\xi^i$ only affects the components of
divergence-free 3-vectors and 3-tensors within the 3-dimensional
hypersurfaces, such as the velocity perturbation $v^i$, which then
transforms as
\be
\tilde v^i= v^i +\bar\xi^{i \prime} \, .
\ee
%

\subsection{Gauge-invariant combinations}
\label{gicomb}

\paragraph{Scalar perturbations}

The gauge-dependence of the metric perturbations lead Bardeen to
propose that only quantities that are explicitly gauge-invariant under
gauge transformations should be considered. The two scalar gauge
functions allow two of the metric perturbations to be eliminated,
implying that one should seek two remaining gauge-invariant
combinations. By studying the transformation
Eqs.~(\ref{transphi}--\ref{transE}), Bardeen constructed two such
quantities~\cite{Bardeen,MFB} \footnote {In Bardeen's notation these
gauge-invariant perturbations are given as $\Phi\equiv\Phi_A Q^{(0)}$
and $\Psi\equiv-\Phi_H Q^{(0)}$.},
\begin{eqnarray}
\label{Phi}
\Phi &\equiv& \phi + h(B-E') + (B-E')'  \,,\\
\label{Psi}
\Psi &\equiv& \psi - h \left( B-E' \right) \,.
\end{eqnarray}
These turn out to coincide with the metric perturbations in a
particular gauge, called variously the orthogonal
zero-shear~\cite{Bardeen,KS}, conformal Newtonian~\cite{Bertschinger}
or longitudinal gauge~\cite{MFB}.
It may therefore appear that this gauge is somehow preferred over
other choices. However any unambiguous choice of time-slicing can be
used to define explicitly gauge-invariant perturbations. The
longitudinal gauge of Ref.~\cite{MFB} provides but one example,
as we shall show in Section \ref{difftimeslice}.

\paragraph{Vector perturbations}

There is one vector valued gauge transformation $\bar\xi^i$ and we have 
introduced two vector functions $S_i$ and $F_i$ into the metric,
Eq.~(\ref{ds}). From these we can construct only one variable 
independent of the vector valued functions $\xi^i$, which is given by
\be 
\tilde S_i=\tilde F_i = S_i + F^{\prime}_i \,. 
\ee

\subsection{Different time slicings}
\label{difftimeslice}

\subsubsection{Longitudinal gauge}

If we choose to work on spatial hypersurfaces with vanishing shear, we
find from Eqs.~(\ref{transB}),(\ref{transE}) and (\ref{shearscalar})
that the shear scalar transforms as $\tilde\sigma=\sigma-\xi^0$ and
this implies that starting from arbitrary coordinates we should
perform a gauge-transformation
\begin{equation}
\xi^0_l = -B+E' \,.
\end{equation}
This is sufficient to determine the scalar metric perturbations
$\phi$, $\psi$, $\sigma$ or any other scalar quantity on these 
hypersurfaces. In addition, the longitudinal gauge is completely 
determined by the spatial gauge choice
\begin{equation}
\xi_l = E \,,
\end{equation}
and hence $\tilde{E}=\tilde{B}=0$. 
The remaining scalar metric perturbations $\phi$, $\psi$ and 
the density perturbation $\delta\rho$ become
\begin{eqnarray}
\tilde \phi_l &=& \phi + h(B-E') + (B-E')' \,, \\
\label{psil}
\tilde \psi_l &=& \psi - h \left( B-E' \right) \,, \\
%
%
\delta \tilde \rho_l &=& \delta \rho + \rho_0'
\left(B-E'\right) \,.
\end{eqnarray}
Note, that $\tilde\phi_l$ and $\tilde\psi_l$ are then identical to
$\Phi$ and $\Psi$ defined in Eqs.~(\ref{Phi}) and~(\ref{Psi}).
These gauge-invariant quantities are simply a coordinate independent
definition of the perturbations in the longitudinal gauge.
{\em Other specific gauge choices may equally be used
to construct quantities that are manifestly gauge-invariant}.

\subsubsection{Uniform curvature gauge}

An interesting alternative gauge choice, defined purely by local
metric quantities is the uniform curvature gauge~\cite{KS,Hwang1_a,
Hwang1_b,Hwang1_c,LS}, also called the off-diagonal
gauge~\cite{pbb_a,pbb_b}. In this gauge one selects spatial
hypersurfaces on which the induced 3-metric is left unperturbed, which
requires $\tilde\psi=\tilde{E}=0$. This corresponds to a gauge
transformation
\begin{equation}
\xi^0_\kappa = -{\psi\over h} \,, \qquad
\xi_\kappa = E \,.
\end{equation}
The gauge-invariant definitions of the remaining metric degrees of
freedom are then from Eqs.~(\ref{transphi}) and~(\ref{transB})
\begin{eqnarray}
\tilde\phi_\kappa &=& \phi + \psi + \left( {\psi\over h}
\right)^{\prime} \,, \\
\tilde{B}_\kappa &=& B-E'-{\psi\over h} \, .
\end{eqnarray}
These gauge-invariant combinations were denoted ${\cal A}$ and ${\cal
B}$ by Kodama and Sasaki~\cite{KS}.
Perturbations of scalar quantities in this gauge become
\begin{equation}
\delta\tilde\rho_\kappa = \delta\rho + \rho_0' {\psi \over h} \,.
\label{Mphi}
\end{equation}

The shear perturbation in the uniform curvature gauge is just given by
$\tilde\sigma_\kappa=-\tilde{B}_\kappa$. This is closely related to
the curvature perturbation in the zero-shear (longitudinal) gauge,
$\tilde\psi_l$, given in equation~(\ref{psil}), 
\be 
\tilde{B}_\kappa = -h\tilde\psi_l = \xi^0_\kappa - \xi^0_l \,.  
\ee 
Gauge-invariant quantities, such as $\tilde{B}_\kappa$ or
$\tilde\psi_l$ are proportional to the displacement between two
different choices of spatial hypersurface, which would vanish for a
homogeneous model.

In some circumstances it is actually more convenient to use the
uniform-curvature gauge-invariant variables instead of $\Phi$ and
$\Psi$. For instance, when calculating the evolution of perturbations
during a collapsing ``pre Big Bang'' era the perturbations
$\tilde\phi_\kappa$ and $\tilde{B}_\kappa$ may remain small even when
$\Phi$ and $\Psi$ become large~\cite{pbb_a,pbb_b}.

Note that the scalar field perturbation on uniform curvature
hypersurfaces,
\be
\delta\varphi_\kappa\equiv\delta\varphi+\varphi_0'\frac{\psi}{h} \,,
\ee
is the gauge-invariant scalar field perturbation used by
Mukhanov~\cite{Mukhanov}.

\subsubsection{Synchronous gauge}

For comparison note that the synchronous gauge, defined by
$\tilde\phi=\tilde{B}=0$, does not determine the time-slicing
unambiguously. There is a residual gauge freedom $\hat\xi^0=X/a$,
where $X(x^i)$ is an arbitrary function of the spatial coordinates,
and it is not possible to define gauge-invariant quantities in general
using this gauge condition~\cite{MS}.
This gauge was originally used by Lifshitz in his pioneering work on 
perturbations in a FRW spacetime \cite{lifshitz}.
He dealt with the residual gauge freedom by eliminating the unphysical
gauge modes through symmetry arguments.

\subsubsection{Comoving orthogonal gauge}

The comoving gauge is defined by choosing spatial coordinates such
that the 3-velocity of the fluid vanishes,
$\tilde{v}=0$. Orthogonality of the constant-$\eta$ hypersurfaces to
the 4-velocity, $u^\mu$, then requires $\tilde{v}+\tilde{B}=0$,
which shows that the momentum vanishes as well. {}From
Eqs.~(\ref{transB}) and~(\ref{transv}) this implies
\begin{eqnarray}
\xi^0_{\rm{m}} &=& -(v+B) \, , \nonumber \\
\xi_{\rm{m}}&=& -\int v d\eta + \hat\xi(x^i) \, ,
\label{xi0m}
\end{eqnarray}
where $\hat\xi(x^i)$ represents a residual gauge freedom,
corresponding to a constant shift of the spatial coordinates.
All the 3-scalars like curvature, expansion, acceleration
and shear are independent of $\hat\xi$.
Applying the above transformation from arbitrary coordinates, the scalar
perturbations in the comoving orthogonal gauge can be written as
\begin{eqnarray}
\tilde \phi_{\rm{m}} &=& \phi+\frac{1}{a} 
\left[ \left( v+B \right) a \right]' \, , \\
%
\tilde \psi_{\rm{m}} &=& \psi - h \left( v+B \right) \, , \\
\label{psim}
\tilde E_{\rm{m}} &=& E + \int v d \eta - \hat\xi \, .
\end{eqnarray}
Defined in this way, these combinations are gauge-invariant under
transformations of their component parts in exactly the same way as,
for instance, $\Phi$ and $\Psi$ defined in Eqs.~(\ref{Phi})
and~(\ref{Psi}), apart from the residual dependence of $\tilde{E}_m$
upon the choice of $\hat\xi$.

Note that the curvature perturbation in the comoving gauge given
above, Eq.~(\ref{psim}) has been used first (with a constant pre-factor)
by Lukash in 1980, \cite{Lukash}. It was later employed by
Lyth and denoted $\cal{R}$ in his seminal paper, \cite{Lyth85},
and in many subsequent works, e.g.~\cite{LL93} and \cite{jimetal}.

The density perturbation on the comoving orthogonal hypersurfaces is
given by Eqs.~(\ref{transrho}) and~(\ref{xi0m}) in gauge-invariant form as
\begin{equation}
\delta\tilde\rho_{\rm{m}} = \delta\rho + \rho_0' \left( v+B \right) \,,
\end{equation}
and corresponds to the gauge-invariant density perturbation
$\epsilon_mE_0Q^{(0)}$ in the notation of Bardeen~\cite{Bardeen}.  The
gauge-invariant scalar density perturbation $\Delta$ introduced in
Refs.~\cite{EB_b} and \cite{EB_a} corresponds to
$\delta\tilde\rho_{{\rm{m}}|i}^{~~~i}/\rho_0$.

If we wish to write these quantities in terms of the metric
perturbations rather than the velocity potential then we
can use the Einstein equations, presented in  Section (\ref{uniden}),
to obtain
\begin{equation}
v+B = \frac{h \phi + \psi' - \kappa \left(B-E'\right)}{h' - h^2 -
\kappa}
\,.
\end{equation}
In particular we note that we can write the comoving curvature
perturbation, given in Eq.~(\ref{psim}), in terms of the longitudinal
gauge-invariant quantities as
\begin{equation}
\tilde \psi_{\rm{m}}
 = \Psi - \frac{h(h\Phi+\Psi')}{h'-h^2-\kappa} \, ,
\end{equation}
which coincides (for $\kappa=0$) with the quantity denoted $\zeta$
by Mukhanov, Feldman and Brandenberger in \cite{MFB}.

\subsubsection{Comoving total matter gauge}
\label{totmat}

The comoving total matter gauge extends the comoving orthogonal gauge
from the single to the multi-fluid case.  Whereas in the comoving
gauge the fluid 3-velocity and the momentum of the single fluid
vanished, in the total matter gauge the
total momentum vanishes,
\be
(\rho+p)\left(\tilde v+\tilde B\right) \equiv \sum_{\alpha}
(\rho_\alpha + p_\alpha) \left(\tilde v_\alpha+\tilde B\right) =0 \, ,
\ee
where $v_{\alpha}$, $\rho_{\alpha}$ and $p_{\alpha}$ are
the velocity, the density and the pressure of the 
fluid species $\alpha$, respectively.
Orthogonality of the constant-$\eta$ hypersurfaces to the total
4-momentum, $u^\mu$, then again requires that $\tilde B=0$,
independently. 
Note that the gauge-invariant velocity $V\equiv v+E'$ 
(actually the velocity in the longitudinal gauge) coincides
with the shear $\sigma$ of the constant-$\eta$ hypersurfaces
in the total matter gauge~\cite{KS}.

\subsubsection{Uniform density gauge}

Alternatively, we can use the matter content to pick out uniform
density hypersurfaces on which to define perturbed quantities. Using
Eq.~(\ref{transrho}) we see that this implies a gauge transformation
\begin{equation}
\xi_{\delta\rho}^0 = {\delta\rho \over \rho_0'} \,.
\end{equation}
On these hypersurfaces the gauge-invariant curvature perturbation
is~\cite{Deru,MS}
\begin{equation}
\label{defzeta}
-\zeta \equiv \tilde\psi_{\delta\rho} 
= \psi + h {\delta\rho \over \rho_0'} \, .
\end{equation}
The sign is chosen to coincide with 
$\zeta$ defined in Refs.~\cite{BST,Bardeen88}.
%
There is another degree of freedom inside the spatial hypersurfaces
and we can choose either $\tilde B$, $\tilde E$ or $\tilde v$ 
to be zero.

\subsection{Gauge-dependent equations of motion}
\label{gd_equ_section}

We now give the governing equations in the homogeneous background 
and the gauge-dependent equations of motions for the scalar, 
vector and tensor perturbations. 
%
%
%
%
%
%
In the following we use the stress energy tensor of a   fluid,
including anisotropic stresses, as given in Section~\ref{perffluid}.

\subsubsection{Conservation of the energy-momentum tensor}

In the homogeneous background the energy conservation equation 
is given by
\be 
\rho^{\prime}=-3h (\rho+p) \,.
\ee
Note, there is no zeroth order momentum conservation equation
as momentum is zero by assumption of isotropy.

The conservation of energy-momentum yields one evolution
equation for the perturbation in the energy density
\be
\label{rhoevol}
\delta\rho' +3h(\delta\rho+\delta p) = (\rho+p)
\left[3\psi' -\nabla^2(v+E')\right] \, ,
\ee
plus an evolution equation for the momentum
\be
\label{peqm}
\left[(\rho+p)(v+B)\right]' +\delta p
+ {2\over3}(\nabla^2+3{\kappa}) \Pi 
= - (\rho+p)\left[\phi +4h(v+B)\right] \, . 
\ee
There is no equivalent to the energy conservation equation in 
the case of vector perturbations since energy is a scalar quantity,
but Eq.~(\ref{stressenergy}) gives a momentum conservation equation 
for the vector perturbations,
\bea
\label{energvec}
\left[(\rho+p)(v_i-S_i)\right]^{\prime} +4h(\rho+p)(v_i-S_i)
&=& -\nabla_k \left(\pi^k_{~|i}+\pi_{i|}^{~k}\right) \nonumber \\
&=& -\left(\nabla^2+2\kappa\right) \pi_i \, ,
\eea
since momentum is a vector.

There is neither an energy nor a momentum conservation equation 
for the tensor matter variables, since, as already mentioned, 
energy is a scalar and momentum a vector quantity, and hence 
they are decoupled from the tensor quantities.

\subsubsection{Einstein's field equations}

The equations of motion for the homogeneous background
with scale factor $a(\eta)$ and Hubble rate $h/a\equiv
a'/a^2$ are
\bea
\label{fried0}
h^2 &=& {8\pi G \over 3} \rho a^2 - \kappa \, , \\
h'&=& - {4\pi G \over 3} (\rho+3p) a^2 \, .
\eea
They are derived from the $0-0$ and the $i-j$ components of the 
unperturbed Einstein equations, respectively.
Equation~(\ref{fried0}) is often referred to as the ``Friedmann 
equation''.

We now give the perturbed Einstein field equations beginning 
with the \emph{scalar perturbations} in the single fluid case.
The first-order perturbed Einstein equations yield two evolution 
equations from the $i-j$ component and its trace, respectively,
\bea
\label{psievol}
\psi'' + 2h\psi' - \kappa\psi + h\phi' + (2h'+h^2)\phi
&=& 4\pi Ga^2 \left(\delta p +{2\over3}\nabla^2 \Pi \right) \, ,\\
\label{Eevol}
\sigma' + 2h\sigma - \phi + \psi &=& 8\pi Ga^2 \Pi \, , 
\eea
plus the energy and momentum constraints
\bea
\label{econ}
3h (\psi' +h\phi) -(\nabla^2 +3{\kappa})\psi - h\nabla^2 \sigma 
&=& -4\pi Ga^2 \delta\rho \, , \\ 
\label{mcon}
\psi' +h\phi +\kappa\sigma &=& -4\pi Ga^2 (\rho +p)(v+B) \, ,
\eea
derived from the $0-0$ and the $0-i$ components of the Einstein
equations, respectively, where $\sigma=-B+E'$, as given in
Eq.~(\ref{shearscalar}). \\

The $0-i$ component of the Einstein equation for \emph{vector 
perturbations} gives rise to a single constraint equation,
\be
\left(\nabla^2+2\kappa\right)\left(S_i+F^{\prime}_i\right)
=16\pi G a^2 (\rho+p) \left(S_i -v_i\right) \, .
\label{vectorconstraint}
\ee
The $i-j$ component then leads to an evolution equation, 
expressed in terms of the vector-shear $\tau_{ij}$ defined 
in Eq.~(\ref{vecshear}), 
%
%
\be
\tau^{\prime}_{ij} +  h \tau_{ij} =
a^3 2 \pi G \left(\pi_{i|j}+\pi_{j|i} \right) \, .
\label{vectorevol}
\ee
The case where the matter content of the universe is dominated 
by a scalar field is of great interest in cosmology.
In anticipation of the results of Section \ref{scalarfields}
we can now show that in this case the vector perturbations
are zero and stay so. From Eqs.~(\ref{perfT_0i}),(\ref{perfT_ij})
and (\ref{scalT_0i}) and (\ref{scalT_ij}) we see that
the vector matter variables $v_i-S_i$ and $\pi_i$ are zero and 
there are therefore no source terms in the constraint 
equation (\ref{vectorconstraint}) and the evolution equation 
(\ref{vectorevol}). Hence the vector perturbations are zero in 
all of space and stay so as long as the scalar field dominates 
the energy content of the universe and afterwards if no sources 
of vorticity are introduced. 

The only non-zero component of the perturbed Einstein tensor for 
\emph{tensor perturbations} is $\delta G^i_{~j}$. This gives rise to
\be
\label{tensorequofmo}
\left[ h^{\prime\prime}_{ij} + 2h~h^{\prime}_{ij}
+\left(2\kappa -\nabla^2 \right)h_{ij} \right]  =
16 \pi G a^2 {}^{\rm{(tensor)}}\pi_{ij} \,.
\ee
There is no separate conservation equation for the tensor matter
variables. Hence the time evolution of the tensor metric perturbations
$h_{ij}$ is determined solely by the Hubble parameter $h$ and the
scale factor $a$, only sourced by matter perturbations in form of the
tensor-anisotropic stress ${}^{\rm{(tensor)}}\pi_{ij}$.
For scalar and vector matter fields ${}^{\rm{(tensor)}}\pi_{ij}=0$
for linear perturbations and the evolution equation for $h_{ij}$
is source-free or homogeneous.

\subsection{Picking a gauge: three examples}
\label{threegauge}

To illustrate the gauge invariant approach we use the 
gauge dependent equations derived in the last section and
present them in terms of gauge-invariant quantities which
coincide with physical quantities in particular time-slicings.
As examples we work with quantities coinciding with
three specific gauges: the popular longitudinal
gauge, the uniform density gauge and the comoving gauge.

\subsubsection{Governing equations in the longitudinal gauge}

The longitudinal gauge is defined by vanishing shift vector,
$\tilde B=0$, and vanishing anisotropic potential , $\tilde E=0$,
and hence $\tilde\sigma_{\rm{l}}=0$.
Note that the influential report by Mukhanov, Feldman and 
Brandenberger, \cite{MFB}, employs this gauge throughout. 
We now give the {\emph{gauge invariant}} equations of motion in this 
particular gauge. 

We get from the conservation of the 
energy momentum tensor a ``continuity'' and a constraint equation, 
\bea
\delta\tilde\rho_{\rm{l}}' +3h(\delta\tilde\rho_{\rm{l}}
+\delta \tilde p_{\rm{l}}) &=& (\rho+p)
\left[3\tilde\psi_{\rm{l}}'-\nabla^2 \tilde v_{\rm{l}}'\right] \, , \\
\left[(\rho+p)\tilde v_{\rm{l}}\right]'+
\delta \tilde p_{\rm{l}} +{2\over3}(\nabla^2+3{\kappa})\tilde\Pi_{\rm{l}} 
&=& - (\rho+p)\left[\tilde\phi_{\rm{l}} + 4 h\tilde v_{\rm{l}}\right]\, .
\eea
The evolution equations become
\bea
\tilde\psi_{\rm{l}}'' + 2h\tilde\psi_{\rm{l}}' - \kappa\tilde\psi_{\rm{l}} 
+ h\tilde\phi_{\rm{l}}' 
+ (2h'+h^2)\tilde\phi_{\rm{l}} 
&=& 4\pi Ga^2 \left( \delta \tilde p_{\rm{l}} 
+{2\over3}\nabla^2 \tilde\Pi_{\rm{l}} \right) \, ,\\
\label{psiequphi}
\tilde\psi_{\rm{l}} - \tilde\phi_{\rm{l}}   
&=& 8\pi Ga^2 \tilde\Pi_{\rm{l}} \, , 
\eea
and the energy and momentum constraints are
\bea
3h (\tilde\psi_{\rm{l}}' +h\tilde\phi_{\rm{l}}) 
- (\nabla^2 + 3\kappa)\tilde\psi_{\rm{l}} 
&=& -4\pi Ga^2 \delta\tilde\rho_{\rm{l}} \, ,\\ 
\tilde\psi_{\rm{l}}' +h\tilde\phi_{\rm{l}} 
&=& -4\pi G a^2(\rho+p) v_{\rm{l}}  \,.
\eea
From Eq.~(\ref{psiequphi}) we see that in the case of vanishing
anisotropic stresses, as is the case for a perfect fluid or
a scalar field, the curvature perturbation and the lapse
function coincide,
\be
\tilde\psi_{\rm{l}} = \tilde\phi_{\rm{l}} \,.   
\ee
This can simplify calculations considerably.

\subsubsection{Governing equations in the uniform density gauge}

In the uniform density gauge the density perturbation vanishes,
$\delta\tilde\rho=0$. This defines the spatial hypersurfaces. There
is another degree of freedom inside the spatial hypersurfaces and we
can choose either $\tilde B$, $\tilde E$ or $\tilde v$ being zero. We
choose the shift vector $\tilde B=0$, so that
$\tilde\sigma_{\delta\rho}\equiv \tilde E_{\delta\rho}$.  We now give
the {\emph{gauge invariant}} equations of motion in this gauge.

The conservation of the stress energy tensor gives then two
constraint equations
\bea
\label{constener}
\frac{3h}{\rho+p}\delta \tilde p_{\delta\rho} 
&=& 3\tilde\psi_{\delta\rho}^{\prime}
- \nabla^2\left(\tilde v_{\delta\rho}
+ \tilde\sigma_{\delta\rho} \right) \, , \\ 
\left[(\rho+p)\tilde v_{\delta\rho}\right]^{\prime} 
+ \delta \tilde p_{\delta\rho} 
&+& \frac{2}{3}(\nabla^2-3\kappa)\tilde\Pi_{\delta\rho} \nonumber \\
&=& -(\rho+p)\left[\tilde\phi_{\delta\rho}+4h \tilde v_{\delta\rho}
\right] \, .
\eea
It is worth to point out the importance of the perturbed energy
conservation equation in the uniform density gauge,
Eq.~(\ref{constener}) above, for later parts of this work.  In Section
\ref{coquala} we discuss conserved quantities on large scales.
Postponing a detailed discussion to the later section, we can already
see that on large scales, i.e.~when $\nabla^2(\tilde v_{\delta\rho} +
\tilde\sigma_{\delta\rho}) \to 0$, the change in the curvature
perturbation is proportional to the pressure perturbation,
$\tilde\psi_{\delta\rho}^{\prime} \propto
\delta \tilde p_{\delta\rho}$.
Hence the curvature perturbation $\tilde\psi_{\delta\rho}$ is constant
on large scales for a vanishing pressure perturbation.  In Section
\ref{preheatsect} on preheating we will discuss the case in which the
pressure perturbation does not vanish and might give rise to a change
in the curvature perturbation on large scales.

The Einstein evolution equations in the uniform density gauge are 
\bea
\tilde\psi_{\delta\rho}'' + 2h\tilde\psi_{\delta\rho}' 
- \kappa \tilde\psi_{\delta\rho} 
+ h\tilde\phi_{\delta\rho}' &+& (2h'+h^2)\tilde\phi_{\delta\rho} \nonumber \\
 &=& 4\pi Ga^2 \left( \delta \tilde p_{\delta\rho} 
+{2\over3}\nabla^2\tilde\Pi_{\delta\rho} \right) \, ,\\
\tilde\sigma_{\delta\rho}' + 2h\tilde\sigma_{\delta\rho} 
- \tilde\phi_{\delta\rho} + \tilde\psi_{\delta\rho} 
&=& 8\pi Ga^2 \tilde\Pi_{\delta\rho} \, , 
\eea
and the constraint equations are 
\bea
3h (\tilde\psi_{\delta\rho}' +h\tilde\phi_{\delta\rho}) - 
(\nabla^2 + 3{\kappa})\tilde\psi_{\delta\rho} 
&-& h\nabla^2 \tilde\sigma_{\delta\rho} = 0 \, \\ 
\tilde\psi_{\delta\rho}' +h\tilde\phi_{\delta\rho} 
+{\kappa}\tilde\sigma_{\delta\rho} &=& 
-4\pi Ga^2 (\rho +p) \tilde v_{\delta\rho}
\,.
\eea

\subsubsection{Governing equations in the comoving gauge}
\label{uniden}

The comoving gauge is defined by vanishing fluid 3-velocity, 
$\tilde v=0$, and vanishing scalar shift function, 
$\tilde B=0$.

We now give the {\emph{gauge invariant}} equations of motion in this 
particular gauge. We get from the conservation of the 
energy momentum tensor a ``continuity'' and a constraint equation, 
\bea
\delta\tilde\rho_{\rm{m}}' +3h(\delta\tilde\rho_{\rm{m}}
+\delta \tilde p_{\rm{m}}) &=& (\rho+p)
\left[3\tilde\psi_{\rm{m}}'-\nabla^2 \tilde E_{\rm{m}}'\right] \, , \\
\delta \tilde p_{\rm{m}} +{2\over3}(\nabla^2+3{\kappa})\tilde\Pi_{\rm{m}} 
&=& - (\rho+p)\tilde\phi_{\rm{m}} \, .
\eea
The evolution equations become
\bea
\tilde\psi_{\rm{m}}'' + 2h\tilde\psi_{\rm{m}}' - \kappa\tilde\psi_{\rm{m}} 
+ h\tilde\phi_{\rm{m}}' 
&+& (2h'+h^2)\tilde\phi_{\rm{m}} \nonumber \\
&=& 4\pi Ga^2 \left( \delta \tilde p_{\rm{m}} 
+{2\over3}\nabla^2\tilde\Pi_{\rm{m}} \right) \, ,\\
\tilde\sigma_{\rm{m}}' + 2h\tilde\sigma_{\rm{m}} - \tilde\phi_{\rm{m}} 
+ \tilde\psi_{\rm{m}} &=& 8\pi Ga^2 \tilde\Pi_{\rm{m}} \, , 
\eea
and the energy and momentum constraints are
\bea
3h (\tilde\psi_{\rm{m}}' +h\tilde\phi_{\rm{m}}) 
- (\nabla^2 + 3\kappa)\tilde\psi_{\rm{m}} 
- h\nabla^2 \tilde\sigma_{\rm{m}} 
&=& -4\pi Ga^2 \delta\tilde\rho_{\rm{m}} \, \\ 
\tilde\psi_{\rm{m}}' +h\tilde\phi_{\rm{m}} 
+\kappa\tilde\sigma_{\rm{m}} &=& 0 \,.
\eea
We will use an extension of the comoving orthogonal gauge, the total
matter defined in Section \ref{totmat}, in Section \ref{multicomp} on
multi-component fluids and will give the governing equations in the
total matter gauge there.

\subsection{Scalar fields}
\label{scalarfields}

In this section we briefly introduce the stress energy tensor 
for scalar fields, postponing a detailed discussion of 
scalar fields and their dynamics to Section~\ref{asssect}.

\subsubsection{Single scalar field}
\label{singlescalar} 

A minimally coupled scalar field is specified by 
the Lagrangian density
\be
\label{Lscal}
{\cal L}=-{1\over2}\varphi^{;\mu}\varphi_{;\mu} - V(\varphi) \,,
\ee
where $\varphi^{;\mu}=g^{\mu\nu}\varphi_{,\nu}$.
The energy momentum tensor is defined as
\be
T^{\mu}_{~\nu}=-2\frac{\partial {\cal L}}{\partial g_{\mu\nu}}
+\delta^{\mu}_{~\nu}{\cal L} \, ,
\ee
and we therefore get for a scalar field $\varphi$
\be
T^{\mu}_{~\nu}=\varphi^{;\mu}\varphi_{;\nu}
-\delta^{\mu}_{~\nu}\left( V(\varphi)
+{1\over2}\varphi^{;\mu}\varphi_{;\mu} \right)  \,.
\ee
Splitting the scalar field into a homogeneous background field and 
a perturbation, 
\be
\varphi(\eta,x^i)=\varphi_0(\eta)+\delta\varphi(\eta,x^i) \, ,
\ee
and using the definitions above we find for the components of 
the energy momentum tensor of a perturbed scalar 
field without specifying a gauge yet
\begin{eqnarray}
\label{scalT_00}
T^0_{~0} &=& -{1\over2}a^{-2} \varphi'^{~2}_0 - V_0  
+ a^{-2} \varphi_0' \left(\phi~\varphi_0'-\delta\varphi'\right) 
-V_{,\varphi} \delta\varphi \, ,  \\
\label{scalT_0i}
T^0_{~i} &=& -a^{-2} \left(\varphi_0^{\prime} 
\delta\varphi_{,i} \right) \, , \\
%
%
\label{scalT_ij}
T^i_{~j} &=& \left[ {1\over2}a^{-2} \varphi'^{~2}_0 - V_0 
- V_{,\varphi} \delta\varphi +a^{-2} \varphi_0' 
\left(\delta\varphi' -\phi~\varphi_0'\right)\right]~\delta^i_{~j} \,, 
\end{eqnarray}
where $V_{,\varphi}\equiv\frac{dV}{d\varphi}$ and $V_0=V(\varphi_0)$
\footnote{Note that there is a typographical error in Eq.~(6.6) of 
Ref.~\cite{MFB}, in which Mukhanov, Feldman and Brandenberger define
their scalar field energy momentum tensor.}. We also see, by comparing
Eq.~(\ref{scalT_ij}) with Eq.~(\ref{perfT_ij}) above, that scalar
fields neither support anisotropic stresses nor source vector and
tensor perturbations, to first order.

The Klein-Gordon equation or scalar field equation can either be 
derived by using  the conservation of the
stress energy tensor, $T^{\mu\nu}_{~~;\nu}=0$, or by directly varying 
the action for the scalar field, equation~(\ref{Lscal}). 
We find for the background field
\be
\label{KGback}
\varphi_0''+2h\varphi_0'+a^2 V_{,\varphi}=0 \, ,
\ee
and for the  perturbed Klein-Gordon equation for the field fluctuation
%
%
\be
\label{pertkg}
\delta\varphi^{\prime\prime}+2h\delta\varphi^{\prime}-\nabla^2
\delta\varphi+a^2 V_{,\varphi\varphi}\delta\varphi
+2a^2V_{,\varphi}\phi
-\varphi^{\prime}_0\phi^{\prime}-\varphi^{\prime}_0
\left(3\psi^{\prime}-\nabla^2\sigma\right)=0 \, .
\ee
%

\subsubsection{$N$ scalar fields}
\label{multiscalar}

For $N$ minimally coupled scalar fields the Lagrangian density is 
given by 
\be
{\cal L} =-\frac{1}{2} \sum_I 
\left( \varphi^{;\mu}_{I}\varphi_{I;\mu}\right)-V(\varphi_{I}) \, .
\ee
For the general potential $V=V(\varphi_{I})$ the energy momentum
tensor can not be split into a background and a perturbed part. In the
special case of an additive potential of the form $V=\sum_I
V_I(\varphi_{I})$ this is possible, but we postpone this case until we
investigate the assisted inflation model in Section \ref{asssect}.

The total energy-momentum tensor in the background is hence  
\bea
T^0_{~0} &=& -a^{-2} \left( \sum_I  \frac{1}{2}\varphi_{I 0}^{\prime 2} +
a^2 V \right)  \, , \\
T^0_{~j} &=& 0 \, ,  \\
T^k_{~j} &=& a^{-2}
\left(\sum_I \frac{1}{2}\varphi^{\prime~2}_{I 0}-a^2 V
\right)  \delta^k_{~j} \, .
\eea
The perturbed energy-momentum tensor can be be given for each field 
$I$ and is
\bea
T^0_{(I)~0} &=&  - a^{-2}\left( \varphi_{I
0}^{\prime}\delta\varphi_{I}^{\prime} -\varphi_{I 0}^{\prime 2} \phi
+a^2 V_{,\varphi_{I}}
\delta\varphi_{I}\right) \, ,  \\
T^0_{(I)~j} &=& -a^{-2} \varphi_{I 0}^{\prime 2}\delta\varphi_{I,j}
\, ,  \\
T^k_{(I)~j} &=& a^{-2}
\left[ V_{,\varphi_{I}}\delta \varphi_{I} a^2   
+ \varphi^{\prime}_{I 0} \left(\delta\varphi^{\prime}_I
-\phi~\varphi^{\prime}_{I 0}\right)\right]  \delta^k_{~j} \, ,
\eea
where $V_{,\varphi_{I}}\equiv
\frac{\partial V}{\partial\varphi_{I}}$.
As in the single field case there are no vector and tensor parts and
no anisotropic stresses to first order.  The total perturbed energy
momentum tensor for all the fields is then given as the sum of the
perturbed energy momentum tensors of each field,
\be
\delta T^{\mu}_{~\nu}=\sum_I \delta T^{\mu}_{(I) \nu} \, .
\ee
The background Klein-Gordon equation for the $I$-th scalar
field is
\be
\varphi_{I 0}''+2h\varphi_{I 0}'+a^2V_{\varphi_{I}}=0 \,.
\ee
The perturbed Klein-Gordon equation for the $I$-th scalar
field is
\bea
&\delta\varphi_I^{\prime\prime}&+2h~\delta\varphi_I^{\prime}-\nabla^2
\delta\varphi_I + a^2 \sum_{J} V_{,\varphi_I\varphi_J}\delta\varphi_J
\nonumber \\
&\qquad& + 2a^2 V_{,\varphi_I} \phi
-\varphi_{I 0}^{\prime}\phi^{\prime}-\varphi_{I 0}^{\prime}
\left(3\psi^{\prime}-\nabla^2\sigma\right)=0 \, ,
\label{pertkgN}
\eea
where $V_{,\varphi_{I}\varphi_{J}}\equiv
\frac{\partial^2 V}{\partial\varphi_{I}\partial\varphi_{J}}$ and the 
equation is gauge dependent.

\subsubsection{The Klein-Gordon equation in specific gauges}

As in Section \ref{threegauge}, we give here the perturbed
Klein-Gordon equation (\ref{pertkg}) in {\emph{gauge-invariant}} form
in three different gauges, the longitudinal, the comoving and the
constant energy gauge. We give the equations for the case of a single
scalar field, which simplifies the notation, but they can be readily
extended to the multi-field case, as can be seen by comparing
Sections~\ref{singlescalar} and \ref{multiscalar} above.

The Klein-Gordon equation in the \emph{longitudinal gauge} is
\be
\tilde{\delta\varphi}_{\rm{l}}''+2h\tilde{\delta\varphi}_{\rm{l}}'
-\nabla^2\tilde{\delta\varphi}_{\rm{l}}
+a^2 V_{,\varphi\varphi}\tilde{\delta\varphi}_{\rm{l}}
+2a^2V_{,\varphi} \tilde\phi_{\rm{l}}
-\varphi'_0 \tilde\phi_{\rm{l}}'
-3 \varphi'_0 \tilde\psi_{\rm{l}}'=0 \, .
\ee

The Klein-Gordon equation in the \emph{uniform density gauge} 
can be simplified by using the fact that $\tilde{\delta\rho}$ and 
$\tilde{\delta\rho}'$ vanish. We find therefore in addition to the 
Klein-Gordon equation two constraints,
\bea
\label{constrho}
\varphi'_0\tilde{\delta\varphi}'_{\delta\rho}
&+& a^2 V_{,\varphi}\tilde{\delta\varphi}_{\delta\rho}
-\varphi'^2_0\phi_{\delta\rho}=0 \,, \nonumber \\
\tilde{\delta\varphi}''_{\delta\rho} 
&-& 2h\tilde{\delta\varphi}'_{\delta\rho}
+ a^2 V_{,\varphi\varphi}\tilde{\delta\varphi}_{\delta\rho}
-2\varphi''_0\phi_{\delta\rho}-\varphi'_0\phi'_{\delta\rho}=0  \,.
\eea
Hence the perturbed Klein-Gordon equation in the constant energy gauge
reduces, subject to Eqs.~(\ref{constrho}), 
to a constraint equation in the field fluctuation,  
\be
4ha^2 V_{,\varphi}\tilde{\delta\varphi_{\delta\rho}}
+\varphi'_0\nabla^2\tilde{\delta\varphi_{\delta\rho}}
-2\varphi''_0\varphi'_0\phi_{\delta\rho}
-\varphi'^2_0\left(3\psi_{\delta\rho}'
-\nabla^2 \sigma_{\delta\rho} \right)=0 \,,
\ee
which is not very useful, really.
%
%

In the \emph{comoving gauge} we find by comparing Eqs.~(\ref{perfT_0i})
and (\ref{scalT_0i}) that the field fluctuations vanish,
\be
\tilde{\delta\varphi}_{\rm{m}}=0\,.
\ee
The perturbed Klein-Gordon equation then reduces to
a constraint equation for the metric perturbations,
\be
2a^2 V_{,\varphi} \tilde\phi_{\rm{m}} 
-\varphi'_0 \tilde\phi_{\rm{m}}'
-\varphi'_0\left(3\tilde\psi_{\rm{m}}'-\nabla^2\sigma_{\rm{m}}\right)=0 \, ,
\ee
where $\sigma_{\rm{m}}=E'_{\rm{m}}$.

\subsection{Multicomponent fluids}
\label{multicomp}


\subsubsection{Including energy and momentum transfer}

In this section we will consider the evolution of linear scalar
perturbations about a homogeneous and isotropic (FRW) universe, now
containing several interacting fluids.


The metric perturbations are the same as in the single fluid case,
whereas the matter variables are different in the multi-fluid
case for each fluid.
The Einstein equations can be taken from Section 
(\ref{gd_equ_section}) (for each fluid species), but the 
energy and momentum conservation equation acquire new interaction
terms.

The total energy momentum tensor is given as
\be
T_{\mu\nu}=\sum_{\alpha} T_{(\alpha)\mu\nu} \,.
\label{tmunumulti}
\ee
As in the single fluid case the total energy momentum tensor is 
conserved,
\be
T^{\nu\mu}_{~~;\mu}=0 \, ,\nonumber
\ee
but the (non-)conservation of energy-momentum for each fluid
leads to
\be
T^{\mu\nu}_{(\alpha);\mu}=Q^{\mu}_{(\alpha)} \, ,
\ee
where $Q^{\mu}_{(\alpha)}$ is the energy momentum four vector.
We can now give the governing equations in the background. 
The energy momentum four vector in the background is given as
\be
{}^{(0)}Q^{\mu}_{(\alpha)}=(-aQ_{(\alpha)},0,0,0) \, .
\ee
The energy conservation equation for fluid $\alpha$ with
energy density $\rho_\alpha$ and pressure $p_\alpha$
is then
\be
\rho_\alpha' = -3h(1-q_\alpha)(\rho_\alpha+p_\alpha) \, , \\
\ee
where 
\be
q_{(\alpha)}\equiv \frac{a Q_{(\alpha)}}{3h(\rho_\alpha+p_\alpha)} \,,
\ee
parameterises the energy transfer between fluids,
subject to the constraint
\be
\sum_\alpha (\rho_\alpha+p_\alpha) q_\alpha = 0 \,,
\ee
which ensures that the total energy is conserved, as in the single
fluid case. It follows from Eq.~(\ref{tmunumulti}) that the total
density and the total pressure are the sum of the densities and
pressures, respectively, of each fluid component $\alpha$,
\bea
\rho=\sum_{\alpha}\rho_{\alpha} \, , 
\qquad p=\sum_{\alpha}p_{\alpha} \,.
\eea
The equations of motion for the homogeneous background are the same as
in the single component case, see Section (\ref{gd_equ_section}). The
perturbed energy momentum four vector is given as
\bea
\delta Q^0_{(\alpha)} &=& -aQ_{(\alpha)}(\phi+\epsilon_{\alpha}) 
\, , \nonumber \\
\delta Q^j_{(\alpha)} &=& a\left[Q_{(\alpha)}(v+B)
+f_{(\alpha)}\right]_{|}^{~j} \,,
\eea
where we introduced $\epsilon_{\alpha}$ and $f_{(\alpha)}$ to 
describe the energy and momentum transfer, respectively.
Under a scalar gauge transformation $\epsilon_{\alpha}$ transforms
as
\be
\tilde\epsilon_{\alpha}=\epsilon_{\alpha}-\xi^0
\left( \frac{ Q'_{(\alpha)}}{Q_{(\alpha)}}\right) \, ,
\ee
whereas the momentum transfer parameter $f_{(\alpha)}$ is gauge
invariant.

\subsubsection{Scalar perturbations in the multi-fluid case}

We get for the gauge dependent perturbed energy 
(non)-conservation equation
\bea
\delta\rho_{\alpha}' + 3h\left(\delta\rho_{\alpha}+\delta p_{\alpha}\right)
= & \left( \rho_{\alpha} + p_{\alpha}\right) & 
\left[ -\nabla^2 \left(v_\alpha+E' \right) \right. \nonumber \\
&& + 3 \psi' + 3h q_{\alpha}(\phi+\epsilon_{\alpha}) \Big] \, , 
\eea
and for the perturbed momentum equation
\bea
\left[(\rho_\alpha+p_\alpha) (v_\alpha+B) \right]'&+&(\rho_\alpha+
p_\alpha)\left[4h(v_\alpha+B)+\phi\right] \nonumber \\
+\delta p_\alpha+{2\over3}(\nabla^2 + 3\kappa) &\Pi_\alpha&
= h(\rho_\alpha+p_\alpha)\left[3q_\alpha (v+B)+F_\alpha 
\right] \, .
\eea
Since the metric is the same for all the fluids the field equations
do not change compared to the single fluid case, as pointed out in 
the previous section.

\subsubsection{Vector perturbations in the multi-fluid case}

We extend the equation of motion for vector perturbations in the 
single fluid case, Eq.~(\ref{energvec}), to the multi-fluid case:
\be
\left[(\rho_{\alpha}+p_{\alpha})V_{\alpha i}\right]^{\prime} 
+4h(\rho_{\alpha}+p_{\alpha})V_{\alpha i}
-\frac{1}{2}\left(\nabla^2+2\kappa\right) \pi_{\alpha i}  
=h(\rho_{\alpha}+p_{\alpha}) 
\left[\tilde f_{\alpha i}+q_{\alpha}V_{\alpha i}\right] \, ,
\ee
where $V_{\alpha i} \equiv v_{(\alpha)i}-S_i$ is the gauge-invariant
vector velocity perturbation, $\tilde f_{\alpha i}$ is the 
gauge-invariant vector momentum transfer, and 
$\pi_{\alpha i} \equiv \pi_{(\alpha)i}$ is the anisotropic stress
of the species $\alpha$, respectively.

\subsubsection{Tensor perturbations in the multi-fluid case}

The tensor perturbations are coupled directly to the matter only
through the anisotropic tensor stress. In order to extend the equation
of motion for tensor perturbations, Eq.~(\ref{tensorequofmo}), to the
multi-fluid case we therefore simply have to set
\be 
{}^{\rm{(tensor)}}\pi_{ij}=\sum_{\alpha}{}^{\rm{(tensor)}}\pi_{(\alpha)ij} \, ,
\ee
where ${}^{\rm{(tensor)}}\pi_{(\alpha)ij}$ is the anisotropic stress
perturbation for the tensor mode of species $\alpha$.

\subsubsection{Equations of motion for scalar perturbations in
the total matter gauge}
\label{equtotmat}

Although we work in the total matter gauge throughout this section we
omit the ``tilde'' and the subscript ``m'' to denote the chosen gauge,
since there is no confusion possible and the equations appear less
cluttered.

The first-order perturbed Einstein equations in the total matter gauge 
yield two evolution equations
\bea
\label{psievoltotmg}
\psi'' + 2h\psi' - \kappa \psi + h\phi' + (2h'+h^2)\phi
 &=& 4\pi Ga^2 \left( \delta p + {2\over3}\nabla^2\Pi \right) \, ,\\
\label{Eevoltotmg}
\sigma' + 2h\sigma - \phi + \psi &=& 8\pi Ga^2 \Pi \, ,
\eea
plus the energy and momentum constraints
\bea
\label{econtotmg}
3h(\psi'+h\phi) -(\nabla^2 + 3\kappa)\psi+h\nabla^2 \sigma 
&=& -4\pi Ga^2 \delta\rho \, ,\\
\label{mcontotmg}
\psi' +h\phi + \kappa\sigma &=& 0 \, .
\eea
The (non-)conservation of energy-momentum for each fluid
leads to
\bea
T^{~\mu}_{(\alpha)~0;\mu} &=& -3h(\rho_\alpha+p_\alpha)q_\alpha 
 (1+\phi+\epsilon_\alpha) \, ,\\
T^{~\mu}_{(\alpha)~i;\mu} &=& h(\rho_\alpha+p_\alpha) f_{\alpha|i} \, , 
\eea
so that the perturbed energy transfer in the total matter gauge is given
$3h (\rho_\alpha+p_\alpha) q_\alpha \epsilon_\alpha$
and the momentum transfer is given by the gradient of
$h(\rho_\alpha+p_\alpha)f_\alpha$.

This yields the equations of motion for the density and velocity
perturbations 
\bea
\delta\rho_{\alpha}' + 3h\left(\delta\rho_{\alpha}+\delta p_{\alpha}\right)
 &=& \left( \rho_{\alpha} + p_{\alpha}\right) 
\left[ -\nabla^2 \left(v_\alpha+\sigma \right) \right. \nonumber \\
&+& 3 \left(\psi'+h\phi \right) 
+ 3h q_{\alpha}\epsilon_{\alpha} - 3h ( 1-q_{\alpha} )\phi \Big] \, ,
\eea
\be
\left[ ( \rho_\alpha + p_\alpha) v_\alpha \right]'
 = (\rho_\alpha+p_\alpha) \left[ -4hv_\alpha - \phi + h f_\alpha \right]
 - \delta p_\alpha -{2\over3}(\nabla^2+3\kappa)\Pi_\alpha \, .
\ee
The conservation of the total energy-momentum yields one evolution equation
for the density perturbation
\be
\label{rhoevoltotmg}
\delta\rho' +3h(\rho+p)(\delta\rho+\delta p) 
= (\rho+p) (3\psi'-\nabla^2 \sigma) \, ,
\ee
plus a momentum conservation equation, which in this gauge reduces to
an  equation for the hydrostatic equilibrium between the pressure gradient
and the gravitational potential gradient,
\be
\label{peqmtotmg}
\delta p +{2\over3}(\nabla^2+3\kappa)\Pi = - (\rho+p)\phi \, .
\ee

Use of the energy and momentum constraint equations (\ref{econtotmg}),
(\ref{mcontotmg}) and (\ref{peqmtotmg}), considerably simplifies the
evolution equations (\ref{rhoevoltotmg}) and (\ref{Eevoltotmg}) to
give
\bea
\label{drhoprime}
\delta\rho' + 3h\delta\rho &=& -(\nabla^2+3\kappa) 
\left[ (\rho+p)\sigma -2h\Pi \right] \, ,\\
\label{Eprime2}
\sigma' + h \sigma &=& -\Psi - {\delta p \over \rho+p} 
+ \left( 8\pi Ga^2 - {2\over3}{\nabla^2+3{\kappa} \over \rho+p} 
\right) \Pi \, ,
\eea
where the gauge-invariant metric perturbation $\Psi\equiv\psi+h\sigma$ 
(actually the curvature perturbation in the longitudinal gauge, 
Eq.~(\ref{psil}) is related to the density perturbation in 
the total matter gauge by the energy constraint 
equation~(\ref{econ})
\be
\Psi = - {4\pi Ga^2 \over k^2-3{\kappa}} \delta\rho \, .
\ee

\subsubsection{Entropy perturbations}
\label{entropert}

In the case of a single fluid with no entropy perturbations the
pressure perturbation $\delta p$ is simply a function of the density
perturbation, 
\be
\delta p=c^2_s \delta\rho \, , 
\ee
where the adiabatic sound speed is defined as
\be
c_s^2 = {p' \over \rho'} \, .
\ee

In general the pressure perturbation (in any gauge) can be split
into adiabatic and entropic (non-adiabatic) parts,
by writing 
\begin{equation}
\delta p = c_{{\rm s}}^2 \delta\rho + p \Gamma \, ,
\end{equation}
where $\Gamma$ is the dimensionless entropy perturbation. 
The non-adiabatic part can be written as
\be
\label{dpnad}
\delta p_{\rm nad}\equiv  p \Gamma \equiv p' 
\left({\delta p \over p'} - {\delta\rho \over \rho'}\right) \, .
\ee
The non-adiabatic pressure  perturbation $\delta p_{\rm{nad}}$, 
defined in this way is gauge-invariant, and represents the 
displacement between hypersurfaces of uniform pressure and uniform 
density.

We can extend the notion of adiabatic and non-adiabatic
from density and pressure perturbations to perturbations in general:
An adiabatic perturbation is one for which all perturbations $\delta
x$ share a common value for $\delta x/{x'}$, where $x'$ is the
time dependence of the background value of $x$.  If the Universe is
dominated by a single fluid with a definite equation of state, or by a
single scalar field whose perturbations start in the vacuum state,
then only adiabatic perturbations can be supported. If there is more
than one fluid, then the adiabatic condition is a special case, but
for instance is preserved if a single inflaton field subsequently
decays into several components.  However, perturbations in a second
field, for instance the one into which the inflaton decays during
preheating, typically violate the adiabatic condition.

The entropy perturbation between any two quantities (which are
spatially homogeneous in the background) has a naturally
gauge-invariant definition, 
which follows from the obvious extension of Eq.~(\ref{dpnad}),
\begin{equation}
\label{Gamma_xy}
\Gamma_{xy} \equiv {\delta x \over {x'}} - {\delta y \over {y'}}
\,.
\end{equation}

In a multi-component system the entropy perturbation $\Gamma$ can 
be split into two parts
\be
\Gamma = \Gamma_{\rm int} + \Gamma_{\rm rel} \, ,
\ee
where $\Gamma_{\rm int}$ is the intrinsic and $\Gamma_{\rm rel}$
the relative entropy perturbation.
The pressure perturbation in the $\alpha$ fluid is given as
\be
\delta p_\alpha = c_\alpha^2 \delta\rho_\alpha + p_\alpha\Gamma_\alpha \, ,
\ee
where $\Gamma_\alpha$ is intrinsic entropy perturbation of this particular 
fluid and the sound speed of the $\alpha$ fluid is
\be
c_\alpha^2 = {p_\alpha' \over \rho_\alpha'} \, .
\ee
The sum of the intrinsic entropy perturbations in each fluid 
is represented by
\be
p \Gamma_{\rm int} = \sum_{\alpha} p_\alpha \Gamma_\alpha \, .
\ee
There may be some cases in which one can discuss the entropy perturbation 
within a single fluid, however we are more interested in the case where 
the entropy perturbation arises due to relative evolution between two or
more fluids with different sound speeds. This is given by
\be
\label{pGammarel1}
p \Gamma_{\rm rel} \equiv \sum_{\alpha} (c_\alpha^2-c_s^2) 
\delta\rho_\alpha \, .
\ee
The overall adiabatic sound speed is determined by the component
parts\footnote{Note that there is an obvious typo on the
right-hand-side of the first line of Eq.~(5.33) in Kodama and
Sasaki~\cite{KS}.}
\bea
c_s^2 &=& \sum_{\alpha}  \frac{c_\alpha^2 \rho_\alpha'}{\rho'} \\
&=& \sum_{\alpha} c_\alpha^2 (1-q_\alpha) 
{\rho_\alpha+p_\alpha \over \rho+p} \, .
\eea
Substituting this into equation~(\ref{pGammarel1}) we
obtain\footnote{There is a typo in Eq.~(5.36) of Ref.~\cite{KS}. The
second term on the right-hand-side in their final equation for
$p\Gamma_{\rm rel}$ should be multiplied by $\Delta$ (in their
notation). This agrees with the corrected version of the equations
given in Ref.~\cite{HK96}.}
\be
\label{pGammarel2}
p\Gamma_{\rm rel} = {1\over2} \sum_{\alpha,\beta}
{(\rho_\alpha+p_\alpha)(\rho_\beta+p_\beta) \over \rho+p}
(c_\alpha^2-c_\beta^2)S_{\alpha\beta}
+ \sum_\alpha q_\alpha c_\alpha^2 (\rho_\alpha+p_\alpha) \Delta \, ,
\ee
where, following Kodama and Sasaki \cite{KS}, we introduce new
dimensionless variables for the perturbed total and relative energy
densities\footnote{Note that Kodama and Sasaki use $\Delta$ to
describe $\delta\rho/\rho$.}
\bea
\label{Delta}
\Delta &\equiv& {\delta\rho \over \rho+p} \, ,\\
\label{Delta_alpha}
\Delta_\alpha &\equiv& {\delta\rho_\alpha \over \rho_\alpha+p_\alpha} \, , \\
\label{S_ab}
S_{\alpha\beta} &\equiv& \Delta_\alpha - \Delta_\beta \, .
\eea

While this is a useful definition of the entropy perturbation
$S_{\alpha\beta}$ in the absence of energy transfer between the fluids
($q_\alpha=0$) it is not very well suited to (and not gauge invariant
in) the more general case, as can be seen by the presence of adiabatic
perturbation $\Delta$ in the expression for $\Gamma_{\rm
rel}$. Instead it is useful to introduce alternative definitions
\bea
\hat\Delta_\alpha &\equiv&
 {\delta\rho_\alpha \over (1-q_\alpha)(\rho_\alpha+p_\alpha)} \, ,\\
\hat{S}_{\alpha\beta} &\equiv& \hat\Delta_\alpha - \hat\Delta_\beta \, ,
\eea
$\hat{S}_{\alpha\beta}$ is then a gauge invariant definition of the
entropy perturbation, and it allows us to re-write Eq.~(\ref{pGammarel2})
as
\be
\label{pGammarel3}
p\Gamma_{\rm rel} = {1\over2} \sum_{\alpha,\beta}
{(1-q_\alpha)(\rho_\alpha+p_\alpha)(1-q_\beta)(\rho_\beta+p_\beta)
 \over \rho+p}
(c_\alpha^2-c_\beta^2)\hat{S}_{\alpha\beta} \, ,
\ee
which vanishes in the absence of any non-adiabatic perturbation.


\subsection{Refined equations of motion}
\label{refinedequ}

Here we present the equations of motion in terms of Kodama and
Sasaki's redefined density perturbations defined in our
Eqs.~(\ref{Delta}--\ref{S_ab}). After some delicate calculations we
obtain from the equations presented in Section \ref{equtotmat} a
system of coupled first order ordinary differential equations
\bea
\Delta' - 3hc_s^2\Delta
 &=& (k^2-3{\kappa}) \left[ \sigma - {2h\Pi \over \rho+p} \right] \, ,\\
\sigma' + h\sigma
 &=& \left[ {4\pi Ga^2(\rho+p) \over k^2-3{\kappa}}-c_s^2 \right] \Delta
 + {p \over \rho+p}\Gamma \nonumber \\
 &\qquad& 
+ \left[ 8\pi Ga^2 + {2\over3}{k^2-3{\kappa} \over \rho+p} \right] \Pi \, , \\
S_{\alpha\beta}' &=& k^2 v_{\alpha\beta} 
 - 3h \left[ q_\alpha(1+c_\alpha^2)\Delta_\alpha
  - q_\beta(1+c_\beta^2)\Delta_\beta \right] \nonumber \, \\
&& \ \ 
-3h (q_\alpha-q_\beta) 
\left[ {\delta p - 2(k^2-3{\kappa})\Pi/3 \over \rho+p} \right]
 \nonumber \\
&& \ \ 
+3h(E_{\alpha\beta} - \Gamma_{\alpha\beta}) \, , \\
v_{\alpha\beta}' +h v_{\alpha\beta} &=&
 3h\left[ c_\alpha^2v_\alpha - c_\beta^2v_\beta - 
 q_\alpha(1+c_\alpha^2)v_\alpha + q_\beta(1+c_\beta^2)v_\beta \right]
\nonumber \\
&&\ \  + h f_{\alpha\beta} 
 - \left( c_\alpha^2\Delta_\alpha - c_\beta^2\Delta_\beta \right) 
 - \Gamma_{\alpha\beta} - \Pi_{\alpha\beta} \, , 
\eea
where 
\bea
v_{\alpha\beta} &\equiv& v_\alpha - v_\beta \, ,\\
\Gamma_{\alpha\beta} &\equiv&
 {p_\alpha\Gamma_\alpha \over \rho_\alpha+p_\alpha} 
 - {p_\beta\Gamma_\beta \over \rho_\beta+p_\beta} \, ,\\
\Pi_{\alpha\beta} &\equiv& {2(3{\kappa}-k^2) \over 3} 
 \left( {\Pi_\alpha\over\rho_\alpha+p_\alpha}
 - {\Pi_\beta \over \rho_\beta+p_\beta} \right) \, ,\\
E_{\alpha\beta} &\equiv& q_\alpha \epsilon_\alpha 
- q_\beta \epsilon_\beta \, , \\
f_{\alpha\beta} &\equiv& f_\alpha - f_\beta \, .
\eea
The residual dependence on the individual $\Delta_\alpha$ and
$v_\alpha$ on the right-hand-sides of the equations of motion for
$S_{\alpha\beta}$ and $v_{\alpha\beta}$ can be eliminated by
substituting in 
\bea 
\Delta_\alpha &=&
 \Delta + \sum_{\gamma} {\rho_\gamma+p_\gamma \over \rho+p} 
S_{\alpha\gamma} \, , \\ 
v_\alpha &=&
\sum_{\gamma} {\rho_\gamma + p_\gamma \over \rho+p} v_{\alpha\gamma} \, .
\eea
Finally then we obtain
\footnote{These should be compared with Kodama and Sasaki's equations
(5.53) and (5.57). There are some, possibly typographical, errors in
their equations. Again our corrected equations agree with Ref.~\cite{HK96}.}
\bea
S_{\alpha\beta}' && \hspace{-12pt}
 +{3\over2}h \left[ q_\alpha(1+c_\alpha^2) + q_\beta(1+c_\beta^2)
\right] S_{\alpha\beta} \nonumber\\
&& +{3\over2}h \left[ q_\alpha(1+c_\alpha^2) - q_\beta(1+c_\beta^2)
\right] \sum_\gamma {\rho_\gamma+p_\gamma \over \rho+p}
 (S_{\alpha\gamma}+S_{\beta\gamma}) \nonumber \\
&&= k^2 v_{\alpha\beta}
 -3h \left[ c_s^2(q_\alpha-q_\beta) + q_\alpha(1+c_\alpha^2)
-q_\beta(1+c_\beta^2) \right] \Delta  \nonumber \\
&& \quad 
 -3h (q_\alpha-q_\beta) \left[ {p\Gamma 
- 2(k^2-3{\kappa})\Pi/3 \over \rho+p} \right]
 +3h(E_{\alpha\beta} - \Gamma_{\alpha\beta}) \, , \\
v_{\alpha\beta}' && \hspace{-12pt}
 +h\left[ 1-{3\over2}(c_\alpha^2+c_\beta^2) +
{3\over2}q_\alpha(1+c_\alpha^2) + {3\over2}q_\beta(1+c_\beta^2)
\right] v_{\alpha\beta}  \nonumber\\
&& -{3\over2}h
\left[ c_\alpha^2 - c_\beta^2 - q_\alpha(1+c_\alpha^2) +
q_\beta(1+c_\beta^2) \right] \sum_\gamma {\rho_\gamma+p_\gamma \over
\rho+p} (v_{\alpha\gamma}+v_{\beta\gamma}) \nonumber\\
&& = 
-(c_\alpha^2-c_\beta^2)\Delta -{1\over2}(c_\alpha^2+c_\beta^2)
S_{\alpha\beta} \nonumber\\
&& \quad 
- {1\over2} (c_\alpha^2-c_\beta^2) \sum_\gamma {\rho_\gamma+p_\gamma
\over\rho+p} (S_{\alpha\gamma}+S_{\beta\gamma}) 
+h f_{\alpha\beta} -\Gamma_{\alpha\beta} - \Pi_{\alpha\beta} \, .
\eea
These equations provide a closed set of coupled first-order equations
in the absence of intrinsic entropy perturbation ($\Gamma_\alpha=0$)
in the individual fluids, and if one specifies the energy and momentum
transfer between fluids, and their anisotropic stress. 

However they remain rather unsatisfactory due to the presence of the
adiabatic perturbation $\Delta$ as a source term for $q_{\alpha} \neq
0$ on the right-hand-side of the equation of motion for the entropy
perturbation $S_{\alpha\beta}$. It should be possible to conceive of
an adiabatic perturbation on large scales,
$\delta\eta=\delta\rho_\alpha/\rho_\alpha'$ for all fluids $\alpha$
(i.e.~a perturbation along the homogeneous background trajectory),
which remains adiabatic (i.e.~remains along the trajectory) and does
not generate an entropy perturbation even in the presence of energy
transfer between the fluids. This should indeed be possible if we work
in terms of our alternative (gauge-invariant for $q_{\alpha} \neq 0$)
entropy perturbation $\hat{S}_{\alpha\beta}$.

We can then write the equations of motion for $\hat{S}_{\alpha\beta}$
and $v_{\alpha\beta}$ as
\bea
\hat{S}_{\alpha\beta}'
&& \hspace*{-24pt} 
- {1\over2} \left[ {q_\alpha'\over1-q_\alpha} + {q_\beta'\over1-q_\beta}
 - 3h q_\alpha(1+c_\alpha^2) - 3hq_\beta(1+c_\beta^2) \right]
 \hat{S}_{\alpha\beta} \nonumber \\
&& \hspace*{-24pt} 
+ {1\over2} \left[ {q_\alpha'\over1-q_\alpha} - {q_\beta'\over1-q_\beta}
 - 3h q_\alpha(1+c_\alpha^2) + 3hq_\beta(1+c_\beta^2) \right]
\nonumber \\
&&\times \sum_\gamma {(1-q_\gamma)(\rho_\gamma+p_\gamma) \over \rho+p}
(\hat{S}_{\alpha\gamma} + \hat{S}_{\beta\gamma}) 
\nonumber \\
&=& k^2 \hat{v}_{\alpha\beta}
 - {{\kappa} \over h} \left( {q_\alpha \over 1-q_\alpha} - {q_\beta \over
1-q_\beta} \right) \hat{S}_{\Delta{\kappa}}
 + 3h \left( \hat{E}_{\alpha\beta} - \hat{\Gamma}_{\alpha\beta}
\right) \, ,\\
v_{\alpha\beta}' && \hspace{-24pt}
 +h\left[ 1-{3\over2}(c_\alpha^2+c_\beta^2) +
{3\over2}q_\alpha(1+c_\alpha^2) + {3\over2}q_\beta(1+c_\beta^2)
\right] v_{\alpha\beta} \nonumber\\
&& -{3\over2}h
\left[ c_\alpha^2 - c_\beta^2 - q_\alpha(1+c_\alpha^2) +
q_\beta(1+c_\beta^2) \right] \sum_\gamma {\rho_\gamma+p_\gamma \over
\rho+p} (v_{\alpha\gamma}+v_{\beta\gamma}) \nonumber\\
&& = 
-[c_\alpha^2(1-q_\alpha)-c_\beta^2(1-q_\beta)]\Delta
 -{1\over2} [c_\alpha^2(1-q_\alpha)+c_\beta^2(1-q_\beta)]
 \hat{S}_{\alpha\beta} \nonumber\\
&& \quad
- {1\over2} [c_\alpha^2(1-q_\alpha)-c_\beta^2(1-q_\beta)]
 \sum_\gamma {(1-q_\gamma)(\rho_\gamma+p_\gamma)
\over\rho+p} (\hat{S}_{\alpha\gamma}+\hat{S}_{\beta\gamma}) \nonumber\\
&& \quad 
+h f_{\alpha\beta} -\Gamma_{\alpha\beta} - \Pi_{\alpha\beta} \, ,
\eea
where 
\bea
\hat{S}_{\Delta{\kappa}} &\equiv& \Delta - 3\psi \nonumber \\
&=& \left[ 1 + {12\pi G a^2 (\rho+p) \over k^2-3{\kappa}} \right] \Delta
 +3h\sigma \, ,\\
\hat{v}_{\alpha\beta} &\equiv&
 {v_\alpha - (\psi/h)\over 1-q_\alpha} -  {v_\beta - (\psi/h) \over
1-q_\beta} \, ,\\
\hat{E}_{\alpha\beta} &\equiv&
 {q_\alpha\hat\epsilon_\alpha \over 1-q_\alpha} - 
 {q_\beta\hat\epsilon_\beta \over 1-q_\beta} \, ,\\
\hat\Gamma_{\alpha\beta} &\equiv& 
 {p_\alpha\Gamma_\alpha \over (1-q_\alpha)(\rho_\alpha+p_\alpha)} 
 - {p_\beta\Gamma_\beta \over (1-q_\beta)(\rho_\beta+p_\beta)} \, ,
\eea
and we have introduced the non-adiabatic part of the perturbed energy
transfer for each fluid 
\be
\hat{\epsilon}_\alpha \equiv \epsilon_\alpha
 + {Q_\alpha' \over Q_\alpha} {\Delta \over 3h} \, ,
\ee
where $Q_\alpha\equiv3hq_\alpha(\rho_\alpha+p_\alpha)/a$.  This
vanishes if the perturbed energy transfer vanishes on surfaces of
constant total density, and thus is zero if the energy transfer is a
function solely of the total density, or if the perturbations are
purely adiabatic.

We have had to introduce the total density perturbation on uniform
curvature hypersurfaces $\hat{S}_{\Delta{\kappa}}$ and the momentum 
perturbation on uniform curvature hypersurfaces $v_\alpha-(\psi/h)$. 

But there remains a source term on the right hand side of the
$\hat{S}$ equation from adiabatic perturbations,
$\hat{S}_{\Delta{\kappa}}$, for spatially curved background FRW models
($\kappa \neq 0$), which threatens to source entropy perturbations on
large scales by adiabatic ones, in contradiction with our intuition.

\subsection{Conserved quantities on large scales}
\label{coquala}

During the different epochs of the universe 
different forms of matter dominate and 
may later become insignificant or decay altogether.
Instead of following the evolution of the various matter fields
it is more convenient to follow the evolution of a metric 
variable, the curvature perturbation, which is, as we shall 
see below, conserved on large scales for adiabatic perturbations 
on uniform density hypersurfaces.

Bardeen, Steinhardt and Turner constructed a conserved quantity
$\zeta_{\rm{BST}}$ from quantities in the uniform expansion gauge,
where the expansion scalar $\delta\theta=0$, as defined in
Eq.~(\ref{theta}), which can be identified with the curvature
perturbation on uniform density hypersurfaces $\zeta$ in
\cite{BST,MS}. A neat way of showing the constancy of $\zeta$ is by
using the energy conservation equation in the uniform density gauge,
Eq.~(\ref{constener}). This approach, which does not depend on the
Einstein field equations, was introduced first in \cite{separate}.

As shown in Section \ref{entropert} we can split
the pressure perturbation into an adiabatic and a non-adiabatic 
part,
\be
\delta p = c_s^2 \delta\rho + \delta p_{\rm{nad}} \, ,
\ee
where $\delta p_{\rm{nad}}$ is the non-adiabatic pressure 
perturbation defined in Eq.~(\ref{dpnad}). 
Note that the first term in the above equation is zero in 
the uniform density gauge and we can identify $\delta p_{\rm{nad}}$
with the pressure perturbation in the uniform density gauge,
$\tilde{\delta p_{\delta\rho}}$.
Rewriting Eq.~(\ref{constener}) and using the fact that in the 
uniform density gauge $\tilde\psi_{\delta\rho}\equiv -\zeta$ we find
\be
\label{dashzeta}
\zeta^{\prime}=-\frac{h}{\rho+p}\delta p_{\rm{nad}} 
-\nabla^2\left(v+E^{\prime}\right) \, .
\ee
On sufficiently large scales gradient terms can be neglected 
and \cite{David+Tony,GBW2},
\be
\label{dotzeta}
\zeta^{\prime}=-\frac{h}{\rho+p}\delta p_{\rm{nad}} \, .
\ee
It follows that $\zeta$ is conserved on large scales for adiabatic
perturbations, for which $\delta p_{\rm{nad}}=0$.  We emphasize, that
this result has been derived independently from the gravitational 
field equations,
using only the conservation of energy~\cite{separate}! Hence this
result holds in
\emph{any theory of relativistic gravity}
\footnote{Note that related results have been obtained in particular
non-Einstein gravity theories \cite{Hwang2_a,Hwang2_b}.}.

In Ref.~\cite{MFB} the gauge-invariant variable
$\zeta_{\scriptscriptstyle {\rm MFB}}$ is
defined as
\begin{equation}
\zeta_{\scriptscriptstyle {\rm MFB}}
 = \Phi + \frac{2}{3} \frac{ \Phi' + \Phi h}{(1+w)h},
\end{equation}
where $w\equiv p_0/\rho_0$.  On large scales (where we neglect
spatial derivatives) and in flat-space ($\kappa=0$) with vanishing
anisotropic stresses ($\pi^i_{~j}=0$, which requires that
$\Phi=\Psi$~\cite{MFB}) all three quantities $\tilde\psi_{\rm{m}}$,
$\tilde\psi_\rho$ and $\zeta_{\scriptscriptstyle {\rm MFB}}$
coincide. The curvature perturbation, in one or the other of 
these forms,
is often used to predict the amplitude of perturbations re-entering
the horizon scale during the radiation or matter dominated eras in
terms of perturbations that left the horizon during an inflationary
epoch, because they remain constant on super-horizon scales (whose
comoving wavenumber $k\ll h$) for adiabatic
perturbations~\cite{LL93}.

It is illustrative to investigate the coincidence of the three
quantities $\tilde\psi_{\rm{m}}$, $\tilde\psi_\rho$ and 
$\zeta_{\scriptscriptstyle {\rm MFB}}$ more closely.

\subsubsection{Single-field inflation}
\label{sifi}

The specific relation between the inflaton field and curvature
perturbations depends on the choice of gauge.  In practice the
inflaton field perturbation spectrum can be calculated on
uniform-curvature ($\tilde\psi=0$) slices, where the field perturbations
have the gauge-invariant definition~\cite{Mukhanov,MFB}
\be
\tilde\delta\varphi_\psi\equiv\delta\varphi +{\varphi' \over h} \psi \,.
\ee
In the slow-roll limit the amplitude of field fluctuations at horizon
crossing ($\lambda=H^{-1}$) is given by $H/2\pi$ (see 
Section \ref{inhomlinpert}). 
Note that this is the amplitude of the asymptotic solution on large
scales.
This result is independent of the geometry and holds for a massless
scalar field in de Sitter spacetime independently of the gravitational
field equations.

The field fluctuation is then related to the curvature perturbation
on comoving hypersurfaces (on which the scalar field is uniform,
$\delta\varphi_{\rm m}=0$) using Eq.~(\ref{transpsi}), by
\begin{equation}
{\cal R} \equiv \tilde\psi_{\rm m} = 
\frac{h}{{\varphi}'} \delta \varphi_{\psi} \, .
\end{equation}
We will now demonstrate that for adiabatic perturbations we can
identify the curvature perturbation on comoving hypersurfaces, ${\cal
R}$, with the curvature perturbation on uniform-density hypersurfaces,
$-\zeta$.  In an arbitrary gauge the density and pressure perturbations
of a scalar field are given by (see Eqs.~(\ref{scalT_00}) and 
(\ref{scalT_ij}))
\begin{eqnarray}
\delta \rho &=& a^{-2}(\varphi'\,{\delta\varphi'} 
- \phi \varphi'^2) + V_{,\varphi} \, \delta\varphi \,,\\
\delta p &=& a^{-2}(\varphi'\,{\delta\varphi'} 
- \phi \varphi'^2)   - V_{,\varphi} \, \delta\varphi \,,
\end{eqnarray}
where $V_{,\varphi}\equiv dV/d\varphi$.  Thus we find
$\delta\rho-\delta p=2V_{,\varphi}\delta\varphi$. 
The entropy perturbation, Eq.~(\ref{dpnad}), for 
a scalar field is given by
\be
\Gamma_{\rm{scalar}}\propto \left\{\frac{\delta\varphi}{\varphi'}
-\frac{\delta\varphi'-\varphi'^2\phi}{\varphi''-h\varphi'}\right\} \,,
\ee
and from the energy and momentum constraints,
Equations (\ref{econ}) and (\ref{mcon}), for a scalar field we get
\be
\nabla^2\left(\psi-h\,\sigma\right)=4\pi G\left(\varphi'\delta\varphi'
-\varphi'^2\phi-(\varphi''-h\varphi')\delta\varphi\right)\, .
\ee
Hence the entropy perturbation for a scalar field vanishes if gradient
terms can be neglected. Therefore on large scales the scalar field
perturbations become adiabatic and then on uniform-density
hypersurfaces both the density and pressure perturbation must vanish
and thus so does the field perturbation $\delta\varphi_{\delta\rho}=0$
for $V_{\varphi}\neq0$. Hence the uniform-density and comoving
hypersurfaces coincide, and ${\cal R}$ and $-\zeta$ are identical, for
adiabatic perturbations and, from Eq.~(\ref{dotzeta}), are constant on
large scales.

\subsubsection{Multi-component inflaton field}
\label{gencrit}

During a period of inflation it is important to distinguish between
``light'' fields, whose effective mass is less than the Hubble
parameter, and ``heavy'' fields whose mass is greater than the Hubble
parameter. Long-wavelength (super-Hubble scale) perturbations of heavy
fields are under-damped and oscillate with rapidly decaying amplitude
($\langle\phi^2\rangle\propto a^{-3}$) about their vacuum
expectation value as the universe expands. Light fields, on the other
hand, are over-damped and may decay only slowly towards the minimum of
their effective potential. It is the slow-rolling of these light
fields that controls the cosmological dynamics during inflation.

The inflaton, defined as the direction of the classical evolution, is
one of the light fields, while the other light fields (if any) will be
taken to be orthogonal to it in field space.
In a multi-component inflation model there is a family of inflaton
trajectories, and the effect of the orthogonal perturbations is to
shift the inflaton from one trajectory to another.

If all the fields orthogonal to the inflaton are heavy then there is
in effect a unique inflaton trajectory in field space. In this case
even a curved path in field space, after canonically normalizing the
inflaton trajectory, is indistinguishable from the case of a straight
trajectory, and leads to no variation in $\zeta$.

When there are multiple light fields evolving during inflation,
uncorrelated perturbations in more than one field will lead to
different regions that are not simply time translations of each
other. In order to specify the evolution of each locally homogeneous
universe one needs as initial data the value of every cosmologically
significant field. In general, therefore, there will be non-adiabatic
perturbations, $\Gamma_{xy}\neq0$.

If the local integrated expansion, $N$, is sensitive to the
value of more than one of the light fields then $\zeta$ is able to
evolve on super-horizon scales, as has been shown by several authors
\cite{modes_a,modes_b,GBW}.  Note also that the comoving and uniform-density
hypersurfaces need no longer coincide in the presence of non-adiabatic
pressure perturbations.  In practice it is necessary to follow the
evolution of the perturbations on super-horizon scales in order to
calculate the curvature perturbation at later times.
In most models studied so far, the trajectories converge to a unique
one before the end of inflation, but that need not be the case in
general.

The separate universe approach described in Section~\ref{sepsect}
gives a rather straightforward procedure for calculating the evolution
of the curvature perturbation, $\psi$, on large scales based on the
change in the integrated expansion, $N$, in different locally
homogeneous regions of the universe. This approach was developed in
Refs.~\cite{SS,ns,ST} for general relativistic models where scalar
fields dominate the energy density and pressure, though it has not
been applied to many specific models.  In the case of a
single-component inflaton, this means that on each comoving scale,
$\lambda$, the curvature perturbation, $\zeta$, on uniform-density (or
comoving orthogonal) hypersurfaces {\em must} stop changing when
gradient terms can be neglected. More generally,
with a multi-component inflaton, the perturbations generated in the
fields during inflation will still determine the curvature
perturbation, $\zeta$, on large scales, but one needs to follow the
time evolution during the entire period a scale remains outside the
horizon in order to evaluate $\zeta$ at later times. This will
certainly require knowledge of the gravitational field equations and
may also involve the use of approximations such as the slow-roll
approximation to obtain analytic results.

\subsubsection{Preheating}
\label{subsectpreheat}

During inflation, every field is supposed to be in the vacuum state
well before horizon exit, corresponding to the absence of particles.
The vacuum fluctuation cannot play a role in cosmology unless it is
converted into a classical perturbation, defined as a quantity which
can have a well-defined value on a sufficiently long time-scale
\cite{guthpi,Lyth85}.  For every light field this conversion occurs at
horizon exit ($\lambda\sim H^{-1}$).
In contrast, heavy fields become classical, if at all, only when their
quantum fluctuation is amplified by some other mechanism.

There has recently been great interest in models where vacuum
fluctuations become classical (i.e., particle production occurs) due
to the rapid change in the effective mass (and hence the vacuum state)
of one or more fields. This usually (though not always~\cite{Rocky})
occurs at the end of inflation when the inflaton oscillates about its
vacuum expectation value which can lead to parametric amplification of
the perturbations --- a process which has become known as
preheating~\cite{KLS97}.  The rate of amplification tends to be
greatest for long-wavelength modes and this has lead to the claim that
rapid amplification of non-adiabatic perturbations could change the
curvature perturbation, $\zeta$, even on very large
scales~\cite{Betal_a,Betal_b}. Preheating will be discussed in more detail
in Section~\ref{preheatsect}.

Within the separate universes picture, discussed in Section 
\ref{sepsect}, this is certainly possible if
preheating leads to different integrated expansion in different
regions of the universe. In particular by Eq.~(\ref{dotzeta}) the
curvature perturbation $\zeta$ can evolve if a significant
non-adiabatic pressure perturbation is produced on large
scales. However it is also apparent in the separate universes picture
that no non-adiabatic perturbation can subsequently be introduced on
large scales if the original perturbations were purely adiabatic.
This is of course also apparent in the field equations where
preheating can only amplify pre-existing field fluctuations.

Efficient preheating requires strong coupling between the inflaton and
preheating fields which typically leads to the preheating field being
heavy during inflation (when the inflaton field is large).  The strong
suppression of super-horizon scale fluctuations in heavy fields during
inflation means that in this case no significant change in $\zeta$ is
produced on super-horizon scales before back-reaction due to particle
production on much smaller scales damps the oscillation of the
inflaton and brings preheating to an end~\cite{jedam,ivan,LLMW}.

Because the first-order effect is so strongly suppressed in such
models, the dominant effect actually comes from second-order
perturbations in the fields~\cite{jedam,ivan,LLMW}.  The expansion on
large scales is no longer independent of shorter wavelength field
perturbations when we consider higher-order terms in the equations of
motion.  Nonetheless in many cases it is still possible to use linear
perturbation theory for the metric perturbations while including terms
quadratic in the matter field perturbations.
%
%
In Ref.~\cite{LLMW} this was done to show that even allowing for
second-order field perturbations, there is no significant
non-adiabatic pressure perturbation, and hence no change in $\zeta$,
on large scales in the original model of preheating in chaotic
inflation.

More recently a modified version of preheating has been
proposed~\cite{Betal2} (requiring a different model of inflation)
where the preheating field is light during inflation, and the coupling
to the inflaton only becomes strong at the end of inflation. In such a
multi-component inflation model non-adiabatic perturbations are no
longer suppressed on super-horizon scales and it is possible for the
curvature perturbation $\zeta$ to evolve both during inflation and
preheating, as described in Section \ref{gencrit}.

\subsubsection{Non-interacting multi-fluid systems}

In a multi-fluid system we can define
uniform-density hypersurfaces for each fluid and a corresponding
curvature perturbation on these hypersurfaces, 
\be
\zeta_{(\alpha)}\equiv
-\psi-\frac{\delta\rho_{(\alpha)}}{\rho'_{(\alpha)}}h \, .
\ee
Equation~(\ref{dashzeta}) then
shows that $\zeta_{(\alpha)}$ {\em remains constant for 
adiabatic perturbations
in any fluid whose energy--momentum is locally conserved:} 
\be
n^\nu T^{~\mu}_{(\alpha)~\nu;\mu}=0 \, .  
\ee
Using the relation between the curvature perturbation on constant
density hypersurfaces and the density perturbation on constant 
curvature hypersurfaces
\be
\tilde\psi_{\delta\rho}=\frac{h}{\rho'}\tilde{\delta\rho_{\psi}} \, ,
\ee
the total curvature perturbation, i.e.~curvature perturbation
on uniform total density hypersurfaces, can then be defined as
\be
\zeta \equiv -\frac{ h \sum_{\alpha} \delta\rho_{\alpha} }
{\sum_{\alpha} \rho'_{\alpha}} \, ,
\ee
where the sum is over the fluid components.
This quantity is not in general conserved in a multi-fluid 
system in presence of a non-zero relative entropy perturbation
$\Gamma_{\rm{rel}} \neq 0$ in Eq.~(\ref{pGammarel3}).

Thus, for example, in a universe
containing non-interacting cold dark matter plus radiation, which both
have well-defined equations of state ($p_{\rm{m}}=0$ and
$p_\gamma=\rho_\gamma/3$), the curvatures of uniform-matter-density
hypersurfaces, $\zeta_{\rm{m}}$, and of uniform-radiation-density
hypersurfaces, $\zeta_\gamma$, remain constant on super-horizon
scales. The curvature perturbation on the uniform-total-density
hypersurfaces is given by
\begin{equation}
\label{matplusrad}
\zeta = \frac{(4/3)\rho_\gamma\zeta_\gamma + \rho_{\rm{m}}\zeta_{\rm{m}}}
{(4/3)\rho_\gamma + \rho_{\rm{m}}} \,.
\end{equation}
At early times in the radiation dominated era 
($\rho_\gamma\gg\rho_{\rm{m}}$)
we have $\zeta_{\rm{ini}}\simeq\zeta_\gamma$, while at late times
($\rho_{\rm{m}}\gg \rho_\gamma$) we have 
$\zeta_{\rm{fin}}\simeq\zeta_{\rm{m}}$. 
$\zeta$ remains constant throughout only for adiabatic perturbations 
where the uniform-matter-density and uniform-radiation-density 
hypersurfaces coincide, ensuring $\zeta_\gamma=\zeta_{\rm{m}}$. 
The isocurvature (or entropy) perturbation is conventionally 
denoted by the perturbation in
the ratio of the photon and matter number densities
\begin{equation}
\label{Sgammam}
S_{\gamma{\rm{m}}} 
= {\delta n_\gamma \over n_\gamma} 
- {\delta n_{\rm{m}} \over n_{\rm{m}}} =
3 \left( \zeta_\gamma - \zeta_{\rm{m}} \right) \,.
\end{equation}
Hence the entropy perturbation for any two non-interacting fluids
always remains constant on large scales independent of the 
gravitational field equations.
Hence we recover the standard result for the final curvature
perturbation in terms of the initial curvature and entropy
perturbation
\footnote{This result was  derived first  by solving a differential 
equation \cite{ks87}, and then  \cite{David+Tony} by integrating 
Eq.~(\ref{dashzeta}) using Eq.~(\ref{matplusrad}). We have here 
demonstrated that even the integration is unnecessary.}
\begin{equation}
\zeta_{\rm{fin}} = \zeta_{\rm{ini}} - {1\over3} S_{\gamma{\rm{m}}}  \,.
\end{equation}

\subsection{The separate universe picture}
\label{sepsect}

Thus far we have shown how one can use the perturbed field equations,
to follow the evolution of linear perturbations in the metric and
matter fields in whatever gauge one chooses.  This allows one to
calculate the corresponding perturbations in the density and pressure
and the non-adiabatic pressure perturbation if there is one, and see
whether it causes a significant change in $\zeta$.

However, there is a particularly simple alternative approach to
studying the evolution of perturbations on large scales, which has
been employed in some multi-component inflation models
\cite{star,Salo,SS,ns,ST,David+Tony}.  This considers each
super-horizon sized region of the Universe to be evolving like a
separate Robertson--Walker universe where density and pressure may take
different values, but are locally homogeneous.  After patching
together the different regions, this can be used to follow the
evolution of the curvature perturbation with time.  Figure~1 shows the
general idea of the separate universe picture, though really every
point is viewed as having its own Robertson--Walker region surrounding
it.

\begin{figure}
\begin{center}
\includegraphics[width=0.7\textwidth]{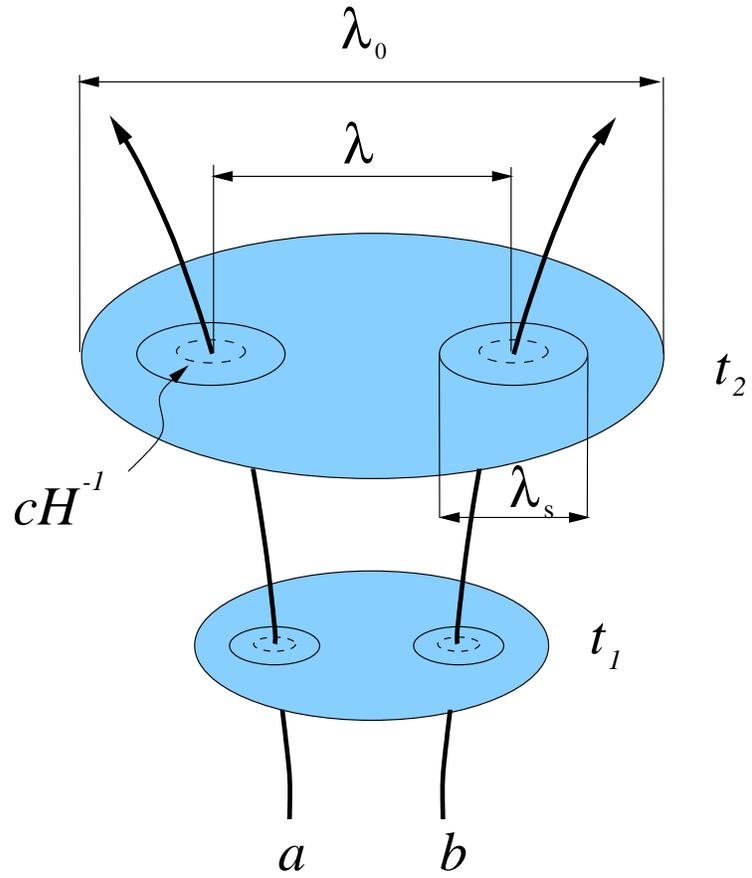}
\caption[sepfig]{\label{sepfig} A schematic illustration of the separate 
universes picture, with the symbols as identified in the text.}
\end{center}
\end{figure}

Consider two such locally homogeneous regions $(a)$ and $(b)$ at fixed
spatial coordinates, separated by a coordinate distance $\lambda$, on
an initial hypersurface (e.g., uniform-density hypersurface) specified
by a fixed coordinate time, $t=t_1$, in the appropriate gauge (e.g.,
uniform-density gauge). The initial large-scale curvature perturbation
on the scale $\lambda$ can then be defined (independently of the
background) as
\begin{equation}
\delta\psi_1\equiv\psi_{a1}-\psi_{b1} \,.
\end{equation}
On a subsequent hypersurface defined by $t=t_2$ the curvature
perturbation at $(a)$ or $(b)$ can be evaluated using
Eq.~(\ref{dotpsi}) [but neglecting $\nabla^2\sigma$] to give~\cite{SS}
\begin{equation}
\psi_{a2} = \psi_{a1} - \delta N_a \,,
\end{equation}
where the integrated expansion between the two hypersurfaces along the
world-line followed by region $(a)$ is  given by
$N_a=N+\delta N_a$, and $N\equiv\ln a$ is the expansion in the
unperturbed background and
\begin{equation}
\delta N_a = \int_1^2 {1\over3} \delta\tilde\theta_a dt \,.
\end{equation}
The curvature perturbation when $t=t_2$ on the comoving scale
$\lambda$ is thus given by
\begin{equation}
\label{dpsi2}
\delta\psi_2 \equiv \psi_{a2}-\psi_{b2}
 = \delta\psi_1 - \left( N_a - N_b \right) \,.
\label{neq}
\end{equation}
In order to calculate the change in the curvature perturbation in any
gauge on very large scales it is thus sufficient to evaluate the
difference in the integrated expansion between the initial and final
hypersurface along different world-lines.

In particular, using \eq{neq}, one can evolve the curvature
perturbation, $\zeta$, on super-horizon scales, knowing only the
evolution of the family of Robertson--Walker universes, which according
to the separate Universe assumption describe the evolution of the
Universe on super-horizon scales:
\begin{equation}
- \Delta\zeta = - \Delta N \,.
\end{equation}
where $-\zeta=\psi_a-\psi_b$ on uniform-density hypersurfaces and
$\Delta N=N_a-N_b$ in Eq.~(\ref{dpsi2}).  As we shall discuss in
Section \ref{preheatsect}, this evolution is in turn specified by the
values of the relevant fields during inflation, and as a result one
can calculate $\zeta$ at horizon re-entry from the vacuum fluctuations
of these fields.

While it is a non-trivial assumption to suppose that every comoving
region {\em well outside the horizon} evolves like an unperturbed
universe, there has to be some scale $\lambda_{\rm s}$ for which that
assumption is true to useful accuracy.  If there were not, the concept
of an unperturbed (Robertson--Walker) background would make no sense.
We use the phrase `background' to describe the evolution on a much
larger scale $\lambda_0$, which should be much bigger even than our
present horizon size, with respect to which the perturbations in
this section are defined. It is important to
distinguish this from regions of size $\lambda_{\rm{s}}$ large enough to be
treated as locally homogeneous, but which when pieced together over a
larger scale, $\lambda$, represent the long-wavelength perturbations
under consideration.  Thus we require a hierarchy of scales:
\begin{equation}
\lambda_0 \gg \lambda \gg \lambda_{{\rm s}} \gtrsim cH^{-1}\, .  
\end{equation} 
Ideally $\lambda_0$ would be taken to be infinite. However it may be
that the Universe becomes highly inhomogeneous on some very much
larger scale, $\lambda_{{\rm e}} \gg\lambda_0$, where effects such as
stochastic or eternal inflation determine the dynamical evolution.
Nevertheless, this will not prevent us from defining an effectively
homogeneous background in our observable Universe, which is governed
by the local Einstein equations and hence impervious to anything
happening on vast scales.  Specifically we will assume that it is
possible to foliate spacetime on this large scale $\lambda_0$ with
spatial hypersurfaces.

When we use homogeneous equations to describe separate regions on
length scales greater than $\lambda_{\rm{s}}$, we are implicitly assuming
that the evolution on these scales is independent of shorter
wavelength perturbations. This is true within linear perturbation
theory in which the evolution of each Fourier mode can be considered
independently, but any non-linear interaction introduces mode-mode
coupling which undermines the separate universes picture.  
The separate universe model may still be used for the evolution of
linear metric perturbation if the perturbations in the total density
and pressure remain small, but a suitable model (possibly a
thermodynamic description) of the effect of the non-linear evolution
of matter fields on smaller scales may be necessary in some cases.  

Adiabatic perturbations in the density and pressure correspond to
shifts forwards or backwards in time along the background
solution, $\delta p/\delta\rho={p'}/\rho'\equiv c_s^2$, and
hence $\Gamma=0$ in Eq.~(\ref{dpnad}). 
For example, in a universe containing only baryonic matter plus
radiation, the density of baryons or photons may vary locally, but the
perturbations are adiabatic if the ratio of photons to
baryons remains unperturbed (cf.~Eq.(\ref{Sgammam})).
Different regions are compelled to undergo the same evolution along a
unique trajectory in field space, separated only by a shift in the
expansion.  
The pressure $p$ thus remains a unique function of the density $\rho$
and the energy conservation equation, $d\rho/dN=-3(\rho+p)$,
determines $\rho$ as a function of the integrated expansion, $N$.
Under these conditions, uniform-density hypersurfaces are separated by
a uniform expansion and hence the curvature perturbation, $\zeta$,
remains constant.

For $\Gamma\neq0$ it is no longer possible to define a simple shift to
describe both the density and pressure perturbation.  The existence of
a non-zero pressure perturbation on uniform-density hypersurfaces
changes the equation of state in different regions of the Universe and
hence leads to perturbations in the expansion along different
worldlines between uniform-density hypersurfaces.
This is consistent with Eq.~(\ref{dotzeta}) which quantifies how the
non-adiabatic pressure perturbation determines the variation of
$\zeta$ on large scales \cite{GBW,David+Tony}.

We defined a generalized adiabatic condition which requires
$\Gamma_{xy}=0$ for any physical scalars $x$ and $y$ in
Eq.~(\ref{Gamma_xy}) in Section~\ref{entropert}. 
In the separate universes picture this condition
ensures that if all field perturbations are adiabatic at any one time
(i.e. on any spatial hypersurface), then they must remain so on large
scales at any subsequent time. Purely adiabatic perturbations can
never give rise to entropy perturbations on large scales as all fields
share the same time shift, $\delta\eta = \delta x/{x'}$, along a
single phase-space trajectory.


                       
\newpage
\section{Quantum fluctuations from inflation}
\label{asssect}
 
In this section we will discuss the generation of quantum 
fluctuations from inflation in the context of a particularly
simple multi-component model: assisted inflation. The 
assisted inflation model is one of the very few models, in 
which it is possible to calculate the 
power spectra of the homogeneous and the inhomogeneous field 
fluctuations analytically, without resorting to the slow 
roll approximation \cite{LL93,David+Tony,LLbook}.

\subsection{The assisted inflation model}

\setcounter{equation}{0}

A single scalar field with an exponential potential is known to drive
power-law inflation, where the cosmological scale factor grows as
$a\propto t^p$ with $p>1$, for sufficiently flat
potentials~\cite{PL,Halliwell,BB,CLW}.  Liddle, Mazumdar and Schunck
\cite{LMS} recently proposed a novel model of inflation driven by
several scalar fields with exponential potentials. Although each
separate potential,
\begin{equation}
\label{Vi}
V_i = V_0 \exp \left( - \sqrt{16\pi\over p_i} {\varphi_i\over m_{\rm Pl}}
\right) \,,
\end{equation}
may be too steep to drive inflation by itself ($p_i<1$), the combined
effect of several such fields, with total potential energy
\begin{equation}
V = \sum_{i=1}^n V_i \,,
\end{equation}
leads to a power-law expansion $a\propto
t^{\bar{p}}$ with~\cite{LMS}
\begin{equation}
\label{barp}
\bar{p} = \sum_{i=1}^n p_i\,,
\end{equation}
provided $\bar{p}>1/3$.
Supergravity theories typically predict steep exponential
potentials, but if many fields can cooperate to drive inflation, this
may open up the possibility of obtaining inflationary solutions in
such models.

Scalar fields with exponential potentials are known to possess
self-similar solutions in Friedmann-Robertson-Walker
models either in vacuum~\cite{Halliwell,BB} or in the presence of
a barotropic fluid~\cite{Wetterich,WCL,CLW,vdH}. In the presence of other
matter, the scalar field is subject to additional friction, due to the
larger expansion rate relative to the vacuum case. This means that a
scalar field, even if it has a steep (non-inflationary) potential may
still have an observable dynamical effect in a radiation or matter
dominated era~\cite{Wetterich95,Joyce_a,Joyce_b,LV}.

The paper of Liddle, Mazumdar and Schunck~\cite{LMS} was the
first to consider the effect of additional scalar fields with
independent exponential potentials. They considered $n$ scalar fields
in a spatially flat Friedmann-Robertson-Walker universe with scale
factor $a(t)$. The Lagrange density for the fields, as an extension
of Eq.~(\ref{Lscal}), is
\begin{equation}
{\cal L} = \sum_{i=1}^n - {1\over2} \left(\nabla\varphi_i\right)^2 - V_i \,,
\end{equation}
with each exponential potential $V_i$ of the form given in Eq.~(\ref{Vi}). 
The cosmological expansion rate is then given by the Friedmann
equation (\ref{fried0}) in coordinate time
\begin{equation}
\label{constraint}
H^2 = {8\pi \over 3m_{\rm Pl}^2} \sum_{i=1}^n \left( V_i + {1\over2}
\dot\varphi_i^2 \right) \,,
\end{equation}
and the individual fields obey the Klein-Gordon 
equation (\ref{KGback}),
\begin{equation}
\label{phieom}
\ddot\varphi_i +3H\dot\varphi_i = - {dV_i \over d\varphi_i} \,.
\end{equation}

\begin{figure}
\begin{center}
\includegraphics[width=0.7\textwidth]{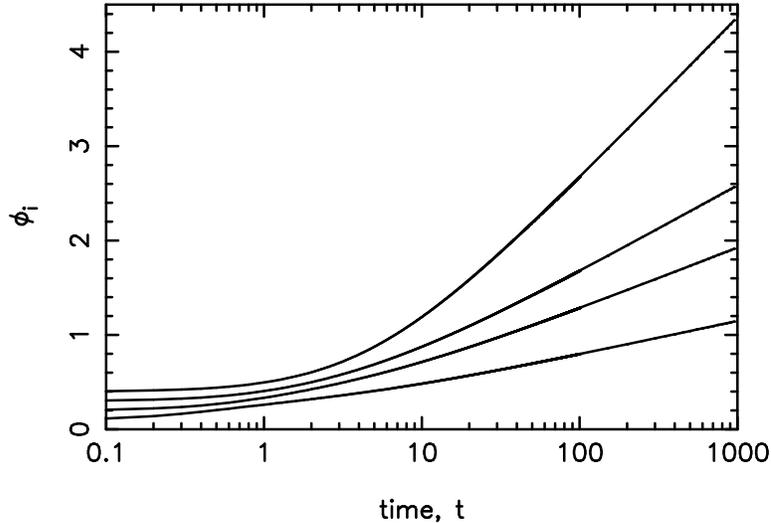}
\caption[Phi fields]{\label{Fone}
Evolution of four fields $\varphi_1$, $\varphi_2$, $\varphi_3$ and
$\varphi_4$ (from bottom to top) during assisted inflation with
$p_1=0.3$, $p_2=1$, $p_3=2$ and $p_4=7$.}
\end{center}
\end{figure}

One can then obtain a scaling solution of the form~\cite{LMS}
\begin{equation}
{\dot\varphi_i^2 \over \dot\varphi_j^2} = {V_i \over V_j} = C_{ij} \,.
\end{equation}
Differentiating this expression with respect to time, and using the
form of the potential given in Eq.~(\ref{Vi}) then implies that
\begin{equation}
{1\over \sqrt{p_i}} \dot\varphi_i - {1\over \sqrt{p_j}} \dot\varphi_j = 0 \,,
\end{equation}
and hence
\begin{equation}
C_{ij} = {p_i \over p_j} \,.
\end{equation}
The scaling solution is thus given by~\cite{LMS}
\begin{equation}
\label{attractor}
{1\over\sqrt{p_i}} \varphi_i - {1\over\sqrt{p_j}} \varphi_j 
 = {m_{\rm Pl} \over \sqrt{16\pi}} \ln {p_j\over p_i} \,.
\end{equation}
A numerical solution with four fields is shown in Fig.~\ref{Fone} as
an example. In Ref.~\cite{LMS} the authors demonstrated the existence
of a scaling solution for $n$ scalar fields written in terms of a
single re-scaled field
$\tilde\varphi=\sqrt{\bar{p}/p_1}\varphi_1$. The choice of $\varphi_1$
rather than any of the other fields is arbitrary as along the scaling
solution all the $\varphi_i$ fields are proportional to one another.

In this part of the thesis we will prove that this scaling solution is the
late-time attractor by choosing a redefinition of fields (a rotation
in field space) which allows us to write down the effective potential
for field variations orthogonal to the scaling solution and show that
this potential has a global minimum along the attractor solution.  In
general the full expression for an arbitrary number of fields is
rather messy so we first give, in Section~\ref{sect2}, the simplest case
where there are just two fields, and then extend this to $n$ fields in
Section~\ref{sectn}.  The resulting inflationary potential is similar to
that used in models of hybrid inflation and we show in
Section~\ref{secthy} that assisted inflation can be interpreted as a
form of ``hybrid power-law inflation''. As in the case of power-law or
hybrid inflation, one can obtain analytic expressions for
inhomogeneous linear perturbations close to the attractor trajectory
without resorting to slow-roll type approximations. Thus we are able
to give exact results for the large-scale perturbation spectra due to
vacuum fluctuations in the fields in Section~\ref{sectpert}. We discuss
our results in Sections~\ref{conc3} and \ref{disc}.

\subsection{Two field model}
\label{sect2}

We will restrict our analysis initially to just two scalar fields, 
$\varphi_1$ and $\varphi_2$, with the Lagrange density
\begin{equation}
\label{V2}
{\cal L} = -{1\over2}(\nabla\varphi_1^2) - {1\over2}(\nabla\varphi_2^2) - 
V_0 \left[ \exp \left( - \sqrt{16\pi\over p_1} {\varphi_1\over m_{\rm
Pl}} \right)
 + \exp \left( - \sqrt{16\pi\over p_2} {\varphi_2\over m_{\rm Pl}}
\right) \right]
\,.
\end{equation}

We define the fields
\begin{eqnarray}
\label{barphi2}
\bar\varphi_2 &=& { \sqrt{p_1} \varphi_1 
+ \sqrt{p_2} \varphi_2 \over \sqrt{p_1 + p_2} } \nonumber \\
&&+ {m_{\rm Pl} \over \sqrt{16\pi(p_1+p_2)}} \left( p_1 \ln
{p_1\over p_1+p_2} + p_2 \ln {p_2 \over p_1+p_2} \right)  \,,\\
\label{barsigma2}
\bar\sigma_2 &=& { \sqrt{p_2} \varphi_1 
- \sqrt{p_1} \varphi_2 \over \sqrt{p_1 + p_2} } 
+ {m_{\rm Pl} \over \sqrt{16\pi}} \sqrt{p_1p_2 \over p_1+p_2}
\ln {p_1 \over p_2} \,,
\end{eqnarray}
to describe the evolution along and orthogonal to the scaling
solution, respectively, by applying a Gram-Schmidt orthogonalisation
procedure.
\begin{figure}
\begin{center}
\includegraphics[width=0.7\textwidth]{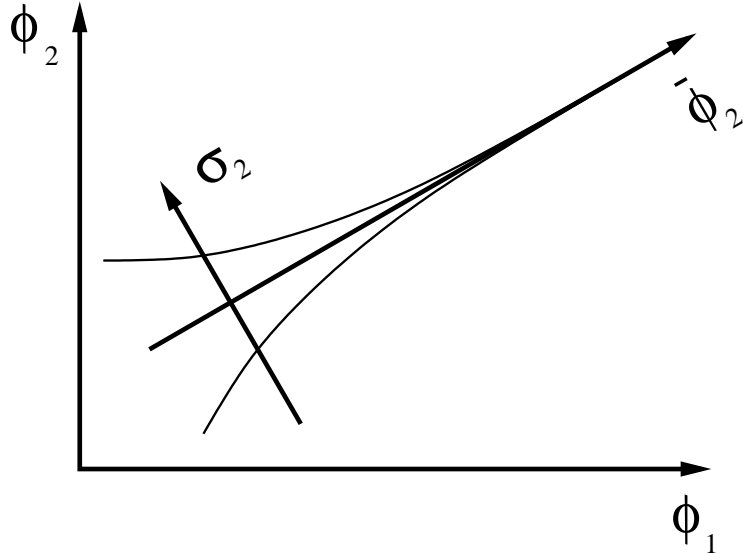}
\caption[Field Rotation]{\label{Frot}
The field rotation in the two field case. The $\bar\varphi_2$ direction
is along the attractor, $\bar\sigma_2$ orthogonal to it.
}
\end{center}
\end{figure}
The re-defined fields $\bar\varphi_2$ and $\bar\sigma_2$ are 
orthonormal linear combinations of the original fields $\varphi_1$ 
and $\varphi_2$.
They represent a rotation, and arbitrary shift of the origin, in
field-space. Fig.~\ref{Frot} illustrates this. 
Thus $\bar\varphi_2$ and $\bar\sigma_2$ have canonical kinetic 
terms, and the Lagrange density given in Eq.~(\ref{V2}) can be written as
\begin{equation}
\label{L2}
{\cal L} = -{1\over2}(\nabla\bar\varphi_2^2) 
- {1\over2}(\nabla\bar\sigma_2^2) 
- \bar{V}(\bar\sigma_2) \exp \left( - \sqrt{16\pi\over p_1+p_2}
{\bar\varphi_2\over m_{\rm Pl}} \right) \,,
\end{equation}
where
\bea
\label{barV2}
\bar{V}(\bar\sigma_2) &=& V_0 \left[
 {p_1\over p_1+p_2}
 \exp \left( -\sqrt{16\pi \over p_1+p_2} \sqrt{p_2\over p_1}
  {\bar\sigma_2\over m_{\rm Pl}} \right) \right. \nonumber \\
&\qquad& + \left. {p_2\over p_1+p_2} 
 \exp \left( \sqrt{16\pi \over p_1+p_2} \sqrt {p_1\over p_2}
 {\bar\sigma_2\over m_{\rm Pl}} \right) \right] \,.
\eea
It is easy to confirm that $\bar{V}(\bar\sigma_2)$ has a global minimum 
value $V_0$ at $\bar\sigma_2=0$, which  implies that $\bar\sigma_2=0$ is 
the late time attractor, which coincides with the scaling solution given in
Eq.~(\ref{attractor}) for two fields.

Close to the scaling solution we can expand about the minimum, to
second-order in $\bar\sigma_2$, and we obtain
\begin{equation}
\label{V2Taylor}
V(\bar\varphi_2,\bar\sigma_2) \approx
 V_0 \left[ 1 + {8\pi \over (p_1+p_2)} {\bar\sigma_2^{~2}\over m_{\rm Pl}^2} 
\right] \exp \left( - \sqrt{16\pi\over p_1+p_2} {\bar\varphi_2\over m_{\rm Pl}}
 \right) \,. 
\end{equation}

Note that the potential for the field $\bar\sigma_2$ has the same 
form as in models of hybrid inflation~\cite{hybrid1_a,hybrid1_b,hybrid2} 
where the inflaton field rolls towards the minimum of a potential 
with non-vanishing potential
energy density ${V}_0$. Here there is in addition a ``dilaton'' field,
$\bar\varphi_2$, which leads to a time-dependent potential energy density
as $\bar\sigma_2\to0$. 
Assisted inflation is related to hybrid
inflation~\cite{hybrid1_a,hybrid1_b,hybrid2} in the
same way that extended inflation~\cite{extinf} was related to Guth's
old inflation model~\cite{Guth}.
As in hybrid or extended inflation, we require a phase transition to bring
inflation to an end. Otherwise the potential given by
Eq.~(\ref{V2Taylor}) leads to inflation into the indefinite future.

\subsection{Many field model}
\label{sectn}

We will now prove that the attractor solution presented in
Ref.~\cite{LMS} is the global attractor for an arbitrary number of
fields with exponential potentials of the form given in
Eq.~(\ref{Vi}), using proof by induction. To do this, we recursively
construct the orthonormal fields and their potential.

Let us assume that we already have $n$ fields $\varphi_i$ with
exponential potentials $V_i$ of the form given in Eq.~(\ref{Vi}) and
that it is possible to pick $n$ orthonormal fields
$\bar\sigma_2,\ldots,\bar\sigma_n$ and $\bar\varphi_n$ such that the sum
of the individual potentials $V_i$ can be written as
\begin{equation}
\label{Vsumn}
\sum_{i=1}^n V_i
= \bar{V}_{n} \exp \left( - \sqrt{16\pi\over \bar p_{n}}
{\bar\varphi_{n}\over m_{\rm Pl}} \right)  \,,
\end{equation}
where we will further assume that $\bar{V}_n= \bar{V}_n
(\bar\sigma_i)$ has a global minimum $\bar V_n(0)=V_0$ when
$\bar\sigma_i=0$ for all $i$ from $2$ to $n$.

It is possible to extend this form of the potential
to $n+1$ fields if we consider an additional field $\varphi_{n+1}$ with
an exponential potential $V_{n+1}$ of the form given in
Eq.~(\ref{Vi}). 
Analogously to the two field case, we define
\begin{eqnarray}
\label{barphin+1}
\bar\varphi_{n+1} &=& { \sqrt{\bar p_n} \bar\varphi_n + \sqrt{p_{n+1}}
\varphi_{n+1} \over \sqrt{\bar p_{n+1} } } \nonumber \\ 
&+& {m_{\rm Pl} \over
\sqrt{16\pi(\bar p_{n+1})}} \left( \bar 
p_{n} \ln {\bar p_n\over \bar p_{n+1}} + p_{n+1} \ln {p_{n+1} \over
\bar p_{n+1}} \right) \,, \\
%
\bar\sigma_{n+1} &=& { \sqrt{p_{n+1}} \bar\varphi_n - \sqrt{\bar p_n}
\varphi_{n+1} \over \sqrt{\bar p_{n+1}} } + {m_{\rm Pl} \over
\sqrt{16\pi}} \sqrt{\bar p_n p_{n+1} \over 
\bar p_{n+1} } \ln {\bar p_n \over p_{n+1} } \,,
\end{eqnarray}
where
\begin{equation}
\label{recurbarp}
\bar p_{n+1} = \bar p_n + p_{n+1} \ .
\end{equation}
Using these definitions we can show that the sum of the $n+1$
individual potentials $V_i$ can be written as
\begin{equation}
\sum_{i=1}^{n+1} V_i
 = \bar{V}_{n+1} \exp \left( - \sqrt{16\pi\over \bar p_{n+1}}
{\bar\varphi_{n+1}\over m_{\rm Pl}} \right)  \,,
\end{equation}
where $\bar{V}_{n+1}=\bar{V}_{n+1}(\bar\sigma_i)$ is given by
\bea
\label{Vbarn+1}
\bar V_{n+1} &=& \bar V_n 
 {\bar p_n \over \bar p_{n+1} }
 \exp \left( -\sqrt{16\pi \over \bar p_{n+1}} \sqrt{p_{n+1}\over \bar p_n}
  {\bar\sigma_{n+1} \over m_{\rm Pl}} \right) \nonumber \\
&+& V_0~ {p_{n+1}\over
 \bar p_{n+1}}  
 \exp \left( \sqrt{16\pi \over \bar p_{n+1}} \sqrt {\bar p_n\over p_{n+1}}
 {\bar\sigma_{n+1} \over m_{\rm Pl}} \right) \,.
\eea
Because we have assumed that $\bar{V}_n$ has a global minimum value 
$\bar{V}_n(0)=V_0$ when $\bar\sigma_i=0$ for all $i$ from $2$ to $n$, 
one can verify that $\bar{V}_{n+1}$ also has a
minimum value $\bar{V}_{n+1}(0)=V_0$ when $\bar\sigma_i=0$, for all
$i$ from $2$ to $n+1$.

However, we have already shown in Section~\ref{sect2} that for two 
fields $\varphi_1$ and $\varphi_2$, we can define two fields $\bar\varphi_2$ 
and $\bar\sigma_2$,
given in Eqs.~(\ref{barphi2}) and~(\ref{barsigma2}) whose combined
potential given in Eq.~(\ref{L2}) is of the form required in
Eq.~(\ref{Vsumn}), with $\bar{p}_2=p_1+p_2$. Hence we can write the
potential in the form given in Eq.~(\ref{Vsumn}) for $n$ fields, for all
$n\geq2$, with 
\begin{equation}
\label{newbarp}
\bar p \equiv \bar p_n = \sum_{i=1}^n p_i \ .
\end{equation}
Equations (\ref{barphi2}) and (\ref{barphin+1}) then lead us to the
non-recursive expression for the ``weighted mean field''
\begin{equation}
\bar\varphi \equiv \bar\varphi_n = \sum_{i=1}^n \left( \sqrt{ { p_i \over
\bar{p} } }  
\varphi_i + {m_{\rm Pl} \over \sqrt{16\pi\bar{p}}} ~p_i \ln {p_i\over \bar{p}}
 \right) \,,
\end{equation}
which describes the evolution along the scaling solution.
This is simply a rotation in field space plus an arbitrary shift, chosen
to preserve the form of the potential given in Eq.~(\ref{Vsumn}).
The $n-1$ fields $\bar\sigma_i$ describe the evolution orthogonal to
the attractor trajectory.

The potential $\bar{V}_n$ has a global minimum at $\bar\sigma_i=0$, which
demonstrates that this is the stable late-time attractor.
{}From Eqs.~(\ref{barV2}) and~(\ref{Vbarn+1}) we get a closed
expression for $\bar{V}_n$, 
\begin{eqnarray}
\label{Vbarn}
\bar{V}_n &=& V_0 \left\{ \frac{p_1}{\bar p} \exp \left[
-\frac{\sqrt{16\pi}}{m_{\rm Pl}} ~\sum^{n}_{i=2}
\sqrt{\frac{p_{i}}{\bar p_{i} \bar p_{i-1}} } ~\bar \sigma_i \right] 
\right. \nonumber \\ 
&+&\left.~\sum^{n-1}_{i=2} \frac{p_{i}}{\bar p} \exp \left[
\frac{\sqrt{16\pi}}{m_{\rm Pl}} \left( 
\sqrt{\frac{\bar p_{i-1}}{\bar p_{i} p_{i}} } ~\bar \sigma_i
- \sum^{n}_{j=i+1} \sqrt{\frac{ p_{j}}{\bar p_{j} \bar p_{j-1}} }
~\bar \sigma_j \right) \right] \right. \nonumber \\ 
&& \qquad \left. + ~ \frac{p_n}{\bar p} \exp \left[
\frac{\sqrt{16\pi}}{m_{\rm Pl}} \sqrt{\frac{\bar p_{n-1}}{\bar p
p_{n}} } ~\bar \sigma_n  
\right]  \right\} .
\end{eqnarray}
Close to the attractor trajectory (to second order in $\bar\sigma_i$)
we can write a Taylor expansion for the potential
\begin{equation}
\label{VnTaylor}
\sum_{i=1}^n V_i
\approx V_0 \left( 1 +  {8\pi \over {\bar{p} m_{\rm Pl}^2}} \sum_{j=2}^{n}
 \bar\sigma_j^2  
\right) \exp \left( - \sqrt{16\pi\over \bar p}
{\bar\varphi_{n}\over m_{\rm Pl}} \right)  \,.
\end{equation}
Note that this expression is dependent only upon $\bar p$ and not on the 
individual $p_i$.

\subsection{Stringy hybrid inflation}
\label{secthy}

The form of the potentials in Eqs.~(\ref{V2Taylor})
and~(\ref{VnTaylor}) is reminiscent of the effective potential
obtained in the Einstein conformal frame from Brans-Dicke type
gravity theories \cite{thesis}.  
The appearance of the weighted mean field, $\bar\varphi$, as a
``dilaton'' field in the potential suggests that the matter Lagrangian
might have a simpler form in a conformally related frame. 
If we work in terms of a conformally re-scaled metric
\begin{equation}
\label{tildeg}
\tilde{g}_{\mu\nu} = \exp \left( - \sqrt{ 16\pi \over \bar{p}
} {\bar\varphi \over m_{\rm Pl}} \right) g_{\mu\nu} \,,
\end{equation}
then the Lagrange density given in Eq.~(\ref{L2}) becomes
\begin{equation}
\label{confL}
\tilde{{\cal L}} = \exp \left( \sqrt{ 16\pi \over \bar{p} }
 {\bar\varphi \over m_{\rm Pl}} \right) 
  \times \left\{ - {1\over2}
 \left(\tilde\nabla\bar\varphi\right)^2 - \sum_{i=2}^{n} {1\over2}
 \left(\tilde\nabla\bar\sigma_i\right) - \bar{V} \right\} \,,
\end{equation}

In this conformal related frame the field $\bar\varphi$ is non-minimally
coupled to the gravitational part of the Lagrangian.  The original
field equations were derived from the full action, including the
Einstein-Hilbert Lagrangian of general relativity,
\begin{equation}
S = \int d^4x \sqrt{-g} \left[ {m_{\rm Pl}^2 \over 16\pi} R + {\cal L}
\right] \,,
\end{equation}
where $R$ is the Ricci scalar curvature of the metric $g_{\mu\nu}$. In
terms of the conformally related metric given in Eq.~(\ref{tildeg})
this action becomes (up to boundary terms~\cite{Wands94})
\begin{equation}
S = \int d^4x \sqrt{-\tilde{g}} e^{-\varphi_{\rm{dil}}} \left[ 
{m_{\rm Pl}^2 \over 16\pi} \tilde{R}  
- \omega (\tilde\nabla\varphi_{\rm{dil}})
-{1\over2} \sum_{i=1}^{n-1} (\tilde\nabla\bar\sigma_i)^2 - \bar{V}
\right] \,,
\end{equation}
where we have introduced the dimensionless dilaton field
\begin{equation}
\varphi_{\rm{dil}} = - \sqrt{16\pi \over \bar{p}} 
{\bar\varphi \over m_{\rm Pl}} \,,
\end{equation}
and the dimensionless Brans-Dicke parameter
\begin{equation}
\omega = {\bar{p} - 3 \over 2} \,.
\end{equation}

Thus the assisted inflation model is identical to $n-1$ scalar fields
$\bar\sigma_i$ with a hybrid inflation type potential
$\bar{V}(\bar\sigma_i)$ in a string-type gravity theory with dilaton,
$\varphi_{\rm{dil}}\propto\bar\varphi$. However, we note that 
in order to obtain
power-law inflation with $\bar{p}\gg1$ the dimensionless constant
$\omega$ must be much larger than that found in the low-energy limit
of string theory where $\omega=-1$.

\subsection{Perturbations about the attractor}
\label{sectpert}

The redefined orthonormal fields and the potential allow us to give
the equations of motion for the independent degrees of freedom. If we
consider only linear perturbations about the attractor then the
energy density is independent of all the fields except $\bar\varphi$, and
we can solve the equation for the $\bar \varphi$ field 
analytically.

The field equation for the weighted mean field is 
\begin{equation}
\ddot{\bar\varphi} + 3H\dot{\bar\varphi} = \sqrt{{16\pi \over \bar p}} {V
\over m_{\rm Pl}} \,.
\end{equation}
Along the line $\bar\sigma_i=0$ for all $i$ in field space,
we have
\begin{equation}
V = V_0 \exp \left( - \sqrt{16\pi\over \bar{p}}
 {\bar\varphi\over m_{\rm Pl}} \right) \,,
\end{equation}
and the well-known power-law solution~\cite{PL} 
with $a\propto t^{\bar{p}}$ is the late-time attractor~\cite{Halliwell,BB} 
for this potential, where
\begin{equation}
\bar\varphi(t)=\bar\varphi_{0} \ln\left(\frac{t}{t_0}\right),
\end{equation}
and $\bar\varphi_{0}=m_{\rm{Pl}}\sqrt{\bar p/4\pi}$ and
$t_0=m_{\rm{Pl}}\sqrt{\bar p/8\pi V_0(3\bar p-1)}$. 

\subsubsection{Homogeneous linear perturbations}

The field equations for the $\bar \sigma_i$ fields are
\begin{equation}
\ddot{\bar\sigma_i} + 3H\dot{\bar\sigma_i} + \frac{\partial
V}{\partial \bar\sigma_i}=0. 
\end{equation}
where the potential $V$ is given by Eqs.~(\ref{Vsumn}) and~(\ref{Vbarn}),
and the attractor solution corresponds to $\bar\sigma_i=0$.
Equation~(\ref{VnTaylor}) shows that we can neglect the back-reaction
of $\bar\sigma_i$ upon the energy density, and hence the cosmological
expansion, to first-order and the field equations have the solutions 
\begin{equation}
\label{sigmat}
\bar\sigma_i(t)=\Sigma_{i+}t^{s_{+}} + \Sigma_{i-}t^{s_{-}},
\end{equation}
where
\begin{equation}
s_{\pm}=\frac{3\bar p -1}{2}
 \left[-1 \pm \sqrt {\frac{3(\bar{p}-3)}{3\bar p -1}} ~ \right], 
\end{equation}
for $\bar{p}>3$, confirming that $\sigma_i=0$ is indeed a local
attractor. In the limit $\bar p \to\infty$ we 
obtain $s_i=-2$. For $1<\bar{p}<3$ the perturbations are under-damped 
and execute decaying oscillations about $\bar\sigma_i=0$.

\begin{figure}
\begin{center}
\includegraphics[width=0.7\textwidth]{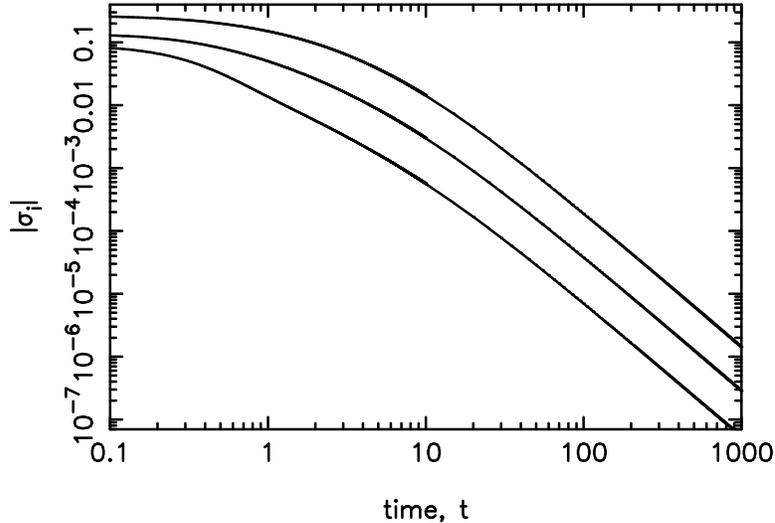}
\caption[Sigma fields]{\label{Ftwo}
Evolution of the fields $\bar\sigma_1$, $\bar\sigma_2$, and
$\bar\sigma_3$, orthogonal to the scaling solution, in the assisted
inflation model shown in Fig.~\ref{Fone}.}
\end{center}
\end{figure}

The form of the solutions given in Eq.~(\ref{sigmat}) for
$\bar\sigma_i(t)$ close to the attractor is the same for all the
orthonormal fields $\bar\sigma_i$, as demonstrated in Fig.~\ref{Ftwo}.
Their evolution is independent of the individual $p_i$ and determined
only by the sum, $\bar{p}$, as expected from the form of the
potential given in Eq.~(\ref{VnTaylor}).

\subsubsection{Inhomogeneous Linear Perturbations}
\label{inhomlinpert}

Conventional hybrid inflation and power-law inflation are two of the
very few models~\cite{LL93} in which one can obtain exact analytic
expressions for the spectra of vacuum fluctuations on all scales
without resorting to a slow-roll type approximation. In the case of
hybrid inflation, this is only possible in the limit that the inflaton
field $\sigma$ approaches the minimum of its potential and we can
neglect its back-reaction on the metric~\cite{GBW}. As the present
model is so closely related to both power-law and hybrid inflation
models close to the attractor, it is maybe not surprising then that we
can obtain exact expressions for the evolution of inhomogeneous linear
perturbations close to the scaling solution.

We will work in terms of the redefined fields $\bar\varphi$ and
$\bar\sigma_i$, and their perturbations on spatially flat
hypersurfaces~\cite{MW}.  In the limit that $\bar\sigma_i\to0$
we can neglect the back-reaction of the $\bar\sigma_i$ field upon the
metric and the field $\bar\varphi$.  Perturbations in the field
$\bar\varphi$ then obey the usual equation for a single field driving
inflation~\cite{Mukhanov}, and perturbations in the field
$\bar\sigma_i$ evolve in a fixed background.  Defining
\begin{eqnarray}
u & = & a \delta\bar\varphi \,,\\
\bar v_i & = & a \delta\bar\sigma_i \,,
\end{eqnarray}
we obtain the decoupled equations of motion for perturbations with
comoving wavenumber $k$,
\begin{eqnarray}
u_k'' + \left( k^2 - {z'' \over z} \right) u_k &=& 0 \,,\\
\bar v_{ik}'' + \left( k^2 + a^2 {d^2 V\over d\bar\sigma_i^2} - {a'' \over a}
\right) \bar v_{ik} &=& 0 \,,
\end{eqnarray}
where~\cite{Mukhanov} $z\equiv a^2\bar\varphi'/a'$ and a prime denotes
differentiation with respect to conformal time $\eta\equiv \int dt/a$.
For power-law expansion we have $z\propto a\propto
(-\eta)^{-\bar{p}/(\bar{p}-1)}$ and thus
\begin{equation}
{a'' \over a} = {\bar{p}(2\bar{p}-1) \over (\bar{p}-1)^2} ~ \eta^{-2}.
\end{equation}
We also have $aH\propto -\bar{p}/((\bar{p}-1)\eta)$ which gives
\begin{equation}
a^2 {d^2 V\over d\sigma^2}
 =\frac{ 2 (3 \bar{p}-1)}{(\bar p -1)^2} ~ \eta^{-2},
\end{equation}
where we have used the fact that $d^2V/d\bar\sigma_i^2=16\pi V/m_{\rm
Pl}^2$ along the attractor. 
The equations of motion therefore become
\begin{eqnarray}
u_k'' + \left( k^2 - {\nu^2 - (1/4) \over \eta^2} \right) u_k &=& 0 \,,\\
\bar v_{ik}'' + \left( k^2 - {\lambda^2 - (1/4) \over \eta^2} \right)
\bar v_{ik} &=& 0 \,,
\end{eqnarray}
where
\begin{eqnarray}
\nu &=& {3\over2} + {1\over \bar{p}-1} \,,\\
\lambda &=& {3\over2} {\sqrt{(\bar{p}-3)(\bar{p}-1/3)} \over \bar{p}-1}
\,,
\end{eqnarray}
and the general solutions in terms of Hankel functions are
\begin{eqnarray}
u_k = U_1 (-k\eta)^{1/2} H^{(1)}_\nu (-k\eta)
 + U_2 (-k\eta)^{1/2} H^{(2)}_\nu (-k\eta) \,, \\
\bar v_{ik} = V_{1i} (-k\eta)^{1/2} H^{(1)}_\lambda (-k\eta)
 + V_{2i} (-k\eta)^{1/2} H^{(2)}_\lambda (-k\eta) \,.
\end{eqnarray}
Taking only positive frequency modes in the initial vacuum state for
$|k\eta|\gg1$ and normalising requires $u_k$ and $\bar v_{ik}\to
e^{-ik\eta}/\sqrt{2k}$, which gives the vacuum solutions
\begin{eqnarray}
u_k &=& \frac{1}{2}(-\pi\eta)^{1/2} e^{\frac{\pi}{2}(\nu +1)i} 
H^{(1)}_\nu (-k\eta),  \\
\bar v_{ik} &=& \frac{1}{2}(-\pi\eta)^{1/2} e^{\frac{\pi}{2}(\lambda +1)i} 
H^{(1)}_\lambda (-k\eta), 
\end{eqnarray}
In the opposite limit, i.e., $-k\eta \to 0$, we use the limiting form
of the Hankel functions, $H_{\nu}^{(1)}(z) \sim -(i/\pi)\Gamma (\nu)
z^{-\nu}$, and therefore on large scales, and at late times, we obtain
\begin{eqnarray}
u_k&\to& \frac{2^{\nu-1}}{\sqrt{\pi k}}e^{i\frac{\pi}{2}\nu} 
\left(-k\eta\right)^{\frac{1}{2}-\nu}\Gamma(\nu), \\
\bar v_{ik}&\to& \frac{2^{\lambda-1}}{\sqrt{\pi k}}e^{i\frac{\pi}{2}\lambda} 
\left(-k\eta\right)^{\frac{1}{2}-\lambda}\Gamma(\lambda) \,.
\end{eqnarray}

The power spectrum of a Gaussian random field $X$ is conventionally
given by \cite{LLbook}
\be
\label{powerspectrum}
{\cal P}_{X}(k) \equiv \left(\frac{k^3}{2\pi^2}\right)
|X_k|^2  \, .  
\ee
The power spectra on large scales for the field perturbations
$\delta\bar\varphi$ and $\delta\bar\sigma_i$ are thus
\begin{eqnarray}
{\cal P}_{\delta\varphi}^{~~1/2}&=&\frac{C(\nu)}{\left(\nu-\frac{1}{2}
\right)}~\frac{H}{2\pi}~\left(-k\eta\right)^{\frac{3}{2}-\nu}, \\
{\cal P}_{\delta\sigma_i}^{~~1/2}&=&\frac{C(\lambda)}
{\left(\nu-\frac{1}{2}\right)}~\frac{H}{2\pi}~
\left(-k\eta\right)^{\frac{3}{2}-\lambda}, 
\end{eqnarray}
where we have used $\eta=-(\nu-1/2)/(aH)$ and we define
\begin{equation}
C(\alpha) \equiv
\frac{2^{\alpha}\Gamma(\alpha)}{2^{\frac{3}{2}}\Gamma(\frac{3}{2})}.
\end{equation}
Both the weighted mean field $\bar\varphi$ and the orthonormal fields
$\bar\sigma_i$ are ``light'' fields ($m^2<3H^2/2$) during assisted
inflation (for $\bar{p}>3$) and thus we obtain a spectrum of
fluctuations in all the fields on large scales.
Note that in the de Sitter limit, $\bar{p}\to\infty$ and thus $\nu \to
3/2$ and $\lambda\to3/2$, we have ${\cal P}_{\delta\varphi}^{~1/2} \to
H/2\pi$, and ${\cal P}_{\delta\sigma_i}^{~1/2} \to H/2\pi$.

Denoting the scale dependence of the perturbation spectra by
\be
\Delta n_x \equiv \frac{d \ln {\cal P}_x} {d \ln k} \, ,
\ee
we obtain
\begin{eqnarray}
\label{nphi}
\Delta n_{\delta\varphi}&=&3-2\nu=-\frac{2}{\bar p - 1}, \\
\Delta n_{\delta\bar\sigma_i}&=&3-2\lambda=3 \left( 
1-{\sqrt{(\bar{p}-3)(\bar{p}-1/3)} \over \bar{p}-1} \right). 
\end{eqnarray}

\subsection{Conclusions}
\label{conc3}

We have shown that the assisted inflation model, driven by many scalar
fields with steep exponential potentials, can be better understood by
performing a rotation in field space, which allows us to re-write the
potential as a product of a single exponential potential for a
weighted mean field, $\bar\varphi$, and a potential $\bar{V}_n$ for
the orthogonal degrees of freedom, $\bar\sigma_i$, which has a global
minimum when $\bar\sigma_i=0$. This proves that the scaling solution
found in Ref.~\cite{LMS} is indeed the late-time attractor.

The particular form of the potential which we present for scalar
fields minimally-coupled to the spacetime metric, can also be obtained
via a conformal transformation of a hybrid inflation type inflationary
potential~\cite{hybrid1_a,hybrid1_b,hybrid2} with a non-minimally 
coupled, but otherwise massless, dilaton field, 
$\varphi_{\rm{dil}}\propto\bar\varphi$. 
Thus we see that assisted inflation can be understood as a form of 
power-law hybrid inflation, where the false-vacuum energy density is 
diluted by the evolution of the dilaton field.

We have also been able to give exact solutions for inhomogeneous
linear perturbations about the attractor trajectory in terms of our
rotated fields.
Perturbations in the weighted mean field $\bar\varphi$ corresponds to the 
perturbations in the density on the uniform curvature hypersurfaces,
or equivalently, perturbations in the curvature of uniform
density hypersurfaces:
\begin{equation}
\zeta = -{H  \over \dot{\bar\varphi}} \delta\bar\varphi  \, .
\end{equation}
These perturbations are along the attractor trajectory, and hence
describe adiabatic curvature perturbations.
The perturbation spectrum for $\zeta$ is
\begin{equation}
{\cal P}_{\zeta} = \left(\frac{H}{\dot{\bar\varphi}}\right)^2
 {\cal P}_{\delta\varphi}
\end{equation}
which will be ${\cal P}_{\zeta}^{1/2}= H^2/(2\pi \dot{\bar\varphi})$
in the de Sitter limit (where $\bar{p}\to\infty$). 
The spectral index of the curvature perturbations on large scales is
thus given from Eq.~(\ref{nphi}) as
\begin{equation}
n_s \equiv 1 + {d \ln {\cal P}_\zeta \over d\ln k} = 1 - {2\over
\bar{p}-1} \,,
\end{equation}
and is always negatively tilted with respect to the
Harrison-Zel'dovich spectrum where $n_s=1$. Note that in the de Sitter
limit ($\bar{p}\to\infty$) we recover the result of Ref.~\cite{LMS}.

First-order perturbations in the fields orthogonal to the weighted
mean field are isocurvature perturbations during inflation. 
Vacuum fluctuations lead to a positively tilted spectrum.
The presence of non-adiabatic perturbations can lead to a
more complicated evolution of the large-scale curvature perturbation than
may be assumed in single-field inflation models~\cite{Salo,SS,GBW2}
as discussed in Section \ref{cosmopertsect} and in particular
Section \ref{coquala}.
However, we have shown that these perturbations decay relative to the
adiabatic perturbations and hence we recover the single field limit at
late times. In particular we find that the curvature perturbation
$\zeta$ becomes constant on super-horizon scales during inflation. Note, 
however, that assisted inflation must be ended by a phase
transition whose properties are not specified in the model. If this 
phase transition is sensitive to the isocurvature (non-adiabatic)
fluctuations orthogonal to the attractor trajectory, then the
curvature perturbation, $\zeta$, during the subsequent radiation
dominated era may not be simply related to the curvature perturbation
during inflation.

Since the publication of the first two papers on assisted inflation,
\cite{LMS,MW}, the model has enjoyed considerable further interest 
by several authors. In Ref.~\cite{edetal}, Copeland, Mazumdar and
Nunes extended the model presented above to include cross-couplings
between the scalar fields.  Kanti and Olive proposed a realisation of
assisted inflation based on the compactification of a five-dimensional
Kaluza-Klein model, extending the model to include standard
chaotic-type potentials
\cite{kanti1,kanti2}.  Green and Lidsey showed that an arbitrary
number of scalar fields with exponential potentials can be obtained
from compactification of a higher dimensional theory
\cite{anne+jim}. They used a rotation in field space to show that the
system reduces to a single scalar field with a single exponential
potential.  Coley and van den Hoogen investigated the dynamics of
assisted inflation, including a barotropic fluid and curvature as well
as scalar fields with exponential potentials, applying methods of
dynamical systems
\cite{coley}.  Lately Kaloper and Liddle used the assisted inflation
mechanism in connection with chaotic inflation and showed that the
dynamics does not become independent of the initial conditions at late
times \cite{kaloper}.


\newpage
\section{Preheating}
\label{preheatsect}


\setcounter{equation}{0}

The standard inflationary paradigm is an extremely successful model in
explaining observed structures in the Universe (see
Refs.~\cite{LL93,LLbook,David+Tony} for reviews). The inhomogeneities
originate from the quantum fluctuations of the inflaton field, which
on being stretched to large scales become classical perturbations as
shown in the preceding section. The field inhomogeneities generate a
perturbation in the curvature of uniform density hypersurfaces,
$\zeta$, as shown in the preceding section and later on these
inhomogeneities are inherited by matter and radiation when the
inflaton field decays. In the simplest scenario, the curvature
perturbation on scales much larger than the Hubble length is constant,
and in particular is unchanged during the inflaton decay.  This
enables a prediction of the present-day perturbations which does not
depend on the specific cosmological evolution between the late stages
of inflation and the recent past (say, before nucleosynthesis).

It has recently been claimed~\cite{Betal_a,Betal_b} that this 
simple picture may
be violated if inflation ends with a period of preheating, a violent
decay of the inflaton particles into another field (or even into
quanta of the inflaton field itself).  Such a phenomenon would
completely undermine the usual inflationary picture, and indeed the
original claim was that large-scale perturbations would be amplified
into the non-linear regime, placing them in conflict with observations
such as measurements of microwave background anisotropies. Given the
observational successes of the standard picture, these claims demand
attention.

In Section~\ref{gencrit} and Ref.~\cite{separate}, we discuss the
general criteria under which large-scale curvature perturbations can
vary. As has been known for some time, this is possible provided there
exist large-scale non-adiabatic pressure perturbations, as can happen
for example in multi-field inflation
models~\cite{modes_a,modes_b,GBW,SS,David+Tony}.  Under those
circumstances a significant effect is possible during preheating,
though there is nothing special about the preheating era in this
respect and this effect always needs to be considered in any
multi-component inflation model.

In this section we perform an analysis of the simplest preheating model,
as discussed in Ref.~\cite{Betal_a,Betal_b}. We identify two 
possible sources of
variation of the curvature perturbation. One comes from large-scale
isocurvature perturbations in the preheating field into which the
inflaton decays; we concur with the recent analyses of Jedamzik and
Sigl \cite{jedam} and Ivanov \cite{ivan} that this effect is
negligible due to the rapid decay of the background value of the
preheating field during inflation. However, we also show that in fact
a different mechanism gives the dominant contribution, which is
second-order in the field perturbations coming from short-wavelength
fluctuations in the fields.  Nevertheless, we show too that this
effect is completely negligible, and hence that preheating in this
model has no significant effect on large-scale curvature
perturbations.

\subsection{Perturbation evolution}

We describe the perturbations via the curvature perturbation on
uniform-density hypersurfaces, denoted $\zeta$
\footnote{This is the
notation of Bardeen, Steinhardt and Turner~\cite{BST}. General issues
of perturbation description and evolution are discussed Section
\ref{coquala} and Ref.~\cite{separate}. The curvature perturbation of
comoving spatial hypersurfaces, ${\cal R}$ \cite{LL93,David+Tony}, is
the same as $-\zeta$ well outside the horizon for adiabatic
perturbations, since the two coincide in the large-scale limit, see
Section \ref{sifi}.}.
In linear theory the evolution of $\zeta$ is well known, and arises from
the non-adiabatic part of the pressure perturbations,
Eq.~(\ref{dpnad}).

On large scales anisotropic stress can be ignored when the matter
content is entirely in the form of scalar fields, and in its absence
the non-adiabatic pressure perturbation determines the variation of
$\zeta$, and rewriting Eq.~(\ref{dotzeta}) we get 
\cite{GBW,David+Tony}
\begin{equation}
\label{dzetadN}
{d\zeta \over dN} =  -\frac{1}{\rho+p}\, \delta p_{\rm{nad}} \,,
\end{equation}
where $N \equiv \ln a$ measures the integrated expansion. 
The uniform-density hypersurfaces become
ill-defined [cf.~the definition of $\zeta$, Eq.~(\ref{defzeta})] 
if the density is not a strictly
decreasing function along worldlines between hypersurfaces of uniform
density, and one might worry that this undermines the above
analysis. However we can equally well derive this evolution equation
in terms of the density perturbation on spatially-flat hypersurfaces,
$\delta\rho_{\psi}\equiv-(d\rho/dN)\zeta$, which remains
well-defined. Spatially-flat hypersurfaces are automatically separated
by a uniform integrated expansion on large scales, so the perturbed
continuity equation in this gauge takes the particularly simple form
\begin{equation}
\label{ddeltarhor}
{d\delta\rho_{\psi} \over dN} = -3(\delta\rho_{\psi}+\delta 
p_{{\psi}}) \,.
\end{equation}
{}From this one finds that $\delta\rho_\psi\propto d\rho/dN$ for adiabatic 
perturbations and hence again we recover a constant value for $\zeta$. 
However it is clearly possible for entropy perturbations to cause a
change in $\zeta$ on arbitrarily large scales when the non-adiabatic
pressure perturbation is non-negligible.

\subsection{Preheating}
\label{preheatingmainsection}

During inflation, the reheat field into which the inflaton field
decays possesses quantum fluctuations on large scales just like the
inflaton field itself. As these perturbations are uncorrelated with
those in the inflaton field, the adiabatic condition will not be
satisfied, and hence there is a possibility that $\zeta$ might vary on
large scales.  Only direct calculation can demonstrate whether the
effect might be significant, and we now compute this effect
in the simplest preheating model, as analyzed in
Ref.~\cite{Betal_a,Betal_b}. This is a chaotic inflation 
model with scalar field potential
\begin{equation}
\label{massivemodel}
V(\varphi,\chi) = \frac{1}{2} \, m^2\varphi^2 +  \frac{1}{2} \, 
g^2\varphi^2\chi^2 \,,
\end{equation}
where $\varphi$ is the inflaton and $\chi$ the reheat field, and $m$
and $g$ are the inflaton mass and the coupling of the reheat field to
the inflaton, respectively.  Slow-roll inflation proceeds with
$\varphi \gtrsim m_{{\rm Pl}}$ and $g\chi\ll m$. The effective mass of
the $\chi$ field is $g \varphi$ and thus will be much larger than the
Hubble rate, $H \simeq \sqrt{4\pi/3} \, m \varphi/m_{{\rm Pl}}$, for
$g \gg m/m_{{\rm Pl}} \sim 10^{-6}$. Throughout this section, we use
the symbol `$\simeq$' to indicate equality within the slow-roll
approximation.

This model gives efficient preheating, since the effective mass of 
$\chi$ oscillates about zero with large amplitude.
In most other models of inflation, preheating is less 
efficient or absent, because the mass oscillates about a nonzero
value and/or has a small amplitude.

Any variation of $\zeta$ during preheating will be driven by the
(non-adiabatic part of) the $\chi$ field perturbation. Our calculation
takes place in three steps. The first is to compute the perturbations
in the $\chi$ field at the end of inflation. The second is to compute
how these perturbations are amplified during the preheating epoch by
the strong resonance. Finally, the main part of the calculation is to
compute the change in $\zeta$ driven by these $\chi$ perturbations.

\subsubsection{The initial quantum fluctuation of the $\chi$-field}

Perturbations in the $\chi$ field obey the wave equation
\begin{equation}
\label{wavechi} 
\ddot{\delta\chi} + 3H\dot{\delta\chi} + 
\left(\frac{k^2}{a^2}+g^2\varphi^2\right)\delta\chi =0 \,.
\end{equation}
The slow-roll conditions ensure that the $\chi$ field remains in the 
adiabatic vacuum state for a massive field \cite{BiDa}
\begin{equation}
\delta\chi_k\simeq {a^{-\frac{3}{2}}~e^{-i\omega t}\over\sqrt{2\omega}} \,,
\end{equation}
where $\omega^2=k^2/a^2+g^2\varphi^2$. This is an approximate solution 
to Eq.~(\ref{wavechi}) provided 
\begin{equation}
\nu \equiv \frac{m_{\chi}}{H} \simeq \sqrt{\frac{3}{4\pi}}
        \frac{g\,m_{{\rm Pl}}}{m} \gg 1 \,,
\end{equation}
where $m_\chi \equiv g\varphi$ is the effective mass of 
the $\chi$ field.

Hence the power spectrum (defined in Eq.~(\ref{powerspectrum}))
for long-wavelength fluctuations ($k\ll m_\chi$) in the $\chi$ field
simply reduces to the result for massive field in flat space
\begin{equation}
\label{powspec1}
{\cal P}_{\delta\chi} \simeq \frac{1}{4\pi^2m_\chi}
        \left(\frac{k}{a}\right)^3 \,.
\end{equation}
Physically, this says that at all times the expansion of the Universe
has a negligible effect on the modes as compared to the mass.  In
particular, at the end of inflation we can write
\begin{equation}
\label{powspec2}
\left. {\cal P}_{\delta\chi} \right|_{{\rm end}} \simeq \frac{1}{\nu}
        \left(\frac{H_{{\rm end}}}{2\pi}\right)^2
        \left(\frac{k}{k_{{\rm end}}}\right)^3 \,.
\end{equation}
The power spectrum has a spectral index
$n_{_{\delta\chi}} = 3$. This is the extreme limit of the mechanism
used to give a blue tilt in isocurvature inflation scenarios 
\cite{LM_a,LM_b}.

\subsubsection{Parametric resonance}

After inflation, the inflaton field $\varphi$ oscillates. 
Strong parametric resonance may now occur, amplifying the initial 
quantum fluctuation in $\chi$ to become a perturbation of the classical 
field $\chi$. The condition for this is
\begin{equation}
q \equiv \frac{g^2 \varphi_{\rm{ini}}^2}{4m^2} \gg 1 \,,
\end{equation}
where $\varphi_{\rm{ini}}$ is the initial amplitude of the 
$\varphi$-field oscillations.  

We model the effect of preheating on the amplitude
of the $\chi$ field following Ref.~\cite{KLS97} as
\begin{equation}
\label{chievol}
{\cal P}_{\delta\chi} =
\left. {\cal P}_{\delta\chi} \right|_{{\rm end}} \; e^{2\mu_{k} m\Delta 
t} \,,
\end{equation}
and the Floquet index $\mu_k$ is taken as 
\begin{equation}
\label{defmuk}
\mu_k\simeq\frac{1}{2\pi} 
        \ln \left( 1+ 2e^{-\pi\kappa^2} \right) 
\,,
\end{equation}
with
\begin{equation}
\label{defkappa}
\kappa^2
\equiv \left(\frac{k}{k_{{\rm max}}}\right)^2
\equiv \frac{1}{18\sqrt{q}}\left(\frac{k}{k_{{\rm end}}}\right)^2 
 \,.
\end{equation}
For strong coupling ($q \gg 1$), we have $\kappa^2\ll1$ for
all modes outside the Hubble scale after inflation
ends $(k \le k_{{\rm end}})$.  Therefore $\mu_k\approx
\ln3/2\pi\approx 0.17$ is only very weakly dependent on the wavenumber
$k$. Combining Eqs.~(\ref{powspec2}) and (\ref{chievol}) gives
\begin{equation}
\label{powspec3}
{\cal P}_{\delta\chi} \simeq {1\over \nu} \left(\frac{H_{{\rm
        end}}}{2\pi}\right)^2\left(\frac{k}{k_{{\rm end}}}\right)^3
        \; e^{2\mu_km\Delta t} \,.
\end{equation}

\subsubsection{Change in the curvature perturbation on large scales}

In order to quantify the effect that parametric growth of the $\chi$ field
fluctuations during preheating might have upon the standard
predictions for the spectrum of density perturbations after inflation,
we need to estimate the change in the curvature perturbation $\zeta$
on super-horizon scales due to entropy perturbations on large-scales.

The density and pressure perturbations due to first-order
perturbations in the inflaton field on large scales (i.e.~neglecting
spatial gradient terms) are of order $g^2\varphi^2\chi\delta\chi$. Not
only are the field perturbations $\delta\chi$ strongly suppressed on
large scales at the end of inflation [as shown in our
Eq.~(\ref{powspec2})] but so is the background field $\chi$.
We can place an upper bound on the size of the background field by
noting that in order to have slow-roll chaotic inflation (dominated by
the $m^2\varphi^2/2$ potential) when any given mode $k$ which we are
interested in crossed outside the horizon, we require $\chi\ll
m/g$. The large effective mass causes this background field to decay,
just like the super-horizon perturbations, and at the end of inflation
we require $\chi\ll m/g(k/k_{\rm end})^{3/2}$ when considering
preheating in single-field chaotic inflation. Combining this with
Eq.~(\ref{powspec2}) we find that the spectrum of density or pressure
perturbations due to linear perturbations in the $\chi$ field has an
enormous suppression for $k\ll k_{\rm end}$:
\begin{equation}
\left. {\cal P}_{\chi\delta\chi} \right|_{\rm end} \ll 
\sqrt{{4\pi\over3}} \left( {m\over gm_{\rm Pl}} \right)^3 
\left( {m_{\rm Pl}H_{\rm end} \over 2\pi} \right)^2 
\left( {k\over k_{\rm end}} \right)^6 
\end{equation}
Effectively the density and pressure perturbations have no term linear
in $\delta \chi$, because that term is multiplied by the background
field value which is vanishingly small. 

By contrast the second-order pressure perturbation is of order
$g^2\varphi^2\delta\chi^2$ where the power spectrum of $\delta\chi^2$ is
given by~\cite{axion}
\begin{equation}
\label{sqrspec}
{\cal P}_{\delta\chi^2} \simeq {k^3\over2\pi} \int_0^{k_{\rm cut}}
\frac{{\cal P}_{\delta\chi}(\left|{\bf k}'\right|)
{\cal P}_{\delta\chi}(\left|{\bf k}-{\bf k}'\right|)}
{\left|{\bf k}\right|^{\prime3}\left|{\bf k}-{\bf k}' \right|^3} 
d^3{\bf k}' \,.
\end{equation}
We impose the upper limit $k_{\rm cut}\sim k_{\rm max}$ to 
eliminate the ultraviolet divergence associated with the vacuum state.
Substituting in for ${\cal P}_{\delta\chi}$ from
Eq.~(\ref{powspec2}), we can write
\begin{equation}
\left. {\cal P}_{\delta\chi^2} \right|_{\rm end}
 = {8\pi\over9} \left( {m\over gm_{\rm Pl}} \right)^2 
\left( {H_{\rm end} \over
2\pi} \right)^4 \left( {k_{\rm cut} \over k_{\rm end}} \right)^3 
\left( {k \over k_{\rm end}} \right)^3 \,.
\end{equation}
Noting that $H_{\rm end}\sim m$ and $k_{\rm cut}\sim k_{\rm max}\sim
q^{1/4}k_{\rm end}$, it is evident that the second-order effect will
dominate over the linear term for $k<g^{1/2}q^{1/3}k_{\rm
end}$.

The leading-order contributions to the pressure and density
perturbations on large scales are thus
\begin{eqnarray}
\label{deltarho}
\delta\rho &=& m^2\varphi\,\delta\varphi + \dot\varphi\,\dot{\delta\varphi}
-\dot\varphi^2\phi+{1\over2} g^2\varphi^2\delta\chi^2 
+ {1\over2}\dot{\delta\chi}^2 \,,\\
\label{deltap}
\delta p &=& - m^2\varphi\,\delta\varphi + \dot\varphi\,\dot{\delta\varphi}
-\dot\varphi^2\phi - {1\over2} g^2\varphi^2\delta\chi^2 + {1\over2}\dot{\delta\chi}^2 \,.
\end{eqnarray}
We stress that we will still only consider first-order perturbations
in the metric and total density and pressure, but these include
terms to second-order in $\delta\chi$.
{}From Eqs.~(\ref{dpnad}), (\ref{deltarho}) and~(\ref{deltap}) we
obtain
\begin{equation}
\label{deltapnad2}
\delta p_{\rm nad} =
\frac {-m^2\varphi\,\dot{\delta\chi}^2 + \ddot\varphi
g^2\varphi^2\delta\chi^2}{3H\dot\varphi}
\,,
\end{equation}
where the long-wavelength solutions for vacuum fluctuations in the
$\varphi$ field obey the adiabatic condition \cite{GWBM}
\be
\frac{\delta\varphi}{\dot\varphi}
=\frac{\dot{\delta\varphi}-\dot\varphi\phi}{\ddot\varphi} \, .
\ee
Inserted into \eq{dzetadN}, this gives the rate of change of $\zeta$.

Note that the non-adiabatic pressure will diverge periodically when
$\dot\varphi=0$ as the comoving or uniform density hypersurfaces become
ill-defined.  Such a phenomenon was noted in the single-field context
by Finelli and Brandenberger \cite{FB}, who evaded it by instead using
Mukhanov's variable $u=a\delta\varphi_{\psi}$ which renders well-behaved
equations. Linear perturbation
theory remains valid as there are choices of hypersurface, such as the
spatially-flat hypersurfaces, on which the total pressure perturbation
remains finite and small.  In particular, we can calculate the change
in the density perturbation due to the non-adiabatic part of the
pressure perturbation on spatially-flat hypersurfaces from
Eq.~(\ref{ddeltarhor}), which yields
\begin{equation}
\Delta\rho_{\rm nad} = -3\int \delta p_{\rm nad} H dt \, .
\end{equation}
Even though $\delta p_{\rm nad}$ contains poles whenever $\dot\varphi=0$,
the integrated effect remains finite whenever the upper and lower limits
of the integral are at $\dot\varphi\neq0$.
{}From this density perturbation calculated in the spatially-flat gauge
one can reconstruct the change in the curvature perturbation on uniform
density hypersurfaces
\begin{equation}
\Delta\zeta =  H {\Delta\rho_{\rm nad} \over \dot\rho} \,.
\end{equation}

Substituting in our expression for $\delta p_{\rm nad}$ we obtain
\begin{equation}
\label{Deltazeta}
\Delta\zeta = -{1\over \dot\varphi^2} 
\int \left( 1+ {2m^2\varphi\over3H\dot\varphi}
\right) g^2\varphi^2 \left|\delta\chi^2\right| H\, dt \,,
\end{equation}
where we have averaged over short timescale oscillations of the
$\chi$-field fluctuations to write $\left|\dot{\delta\chi}^2\right| =
g^2\varphi^2 \left|\delta\chi^2\right|$.
To evaluate this we take the usual adiabatic evolution for the
background $\varphi$ field after the end of inflation
\begin{equation}
\varphi = \varphi_{\rm{ini}} \, {\sin(m\Delta t) \over m\Delta t} \,,
\end{equation}
and time-averaged Hubble expansion
\begin{equation}
H = {2m \over 3(m\Delta t + \Theta)} \,,
\end{equation}
where $\Theta$ is an integration constant of order unity.
The amplitude of the $\chi$-field fluctuations also decays proportional
to $1/\Delta t$ over a half-oscillation from $m\Delta t=n\pi$ to
$m\Delta t=(n+1)\pi$, with the stochastic growth in particle
number occurring only when $\varphi=0$.
Thus evaluating $\Delta\zeta$ over a half-oscillation $\Delta t=\pi/m$
we can write
\begin{equation}
\label{ourintegral}
\Delta\zeta = -{2g^2 |\delta\chi^2| x_n^4 \over 3m^2} 
 \int_{x_n}^{x_{n+1}} \left( {1\over x+\Theta} + {s\over s'} \right)
{s^2\over x^2} dx \,,
\end{equation}
where $x=m\Delta t$, $s(x)=\sin x/x$, $x_n=n\pi$ and a dash indicates
differentiation with respect to $x$. The integral is dominated by the
second term in the bracket which has a pole of order 3 when $s'=0$.
Although $s/s'$ diverges, it yields a finite contribution to the
integral which can be evaluated numerically. For $x_n\gg1$ the integral
is very well approximated by $24/x_n^4$, independent of the integration
constant $\Theta$.

This expression gives us the rate of change of the curvature
perturbation $\zeta$ due to the pressure of the field fluctuations
$\delta\chi^2$ over each half-oscillation of the inflaton field $\varphi$. 
Approximating the sum over several oscillations as a smooth integral and
using Eq.~(\ref{chievol}) for the growth of the $\chi$-field fluctuations
during preheating (neglecting the weak $k$-dependence of the Floquet
index, $\mu_k$, on super-horizon scales) we obtain
\begin{equation}
\label{zetanad}
\zeta_{\rm nad} = -{16 g^2 \over 2\pi\mu} {\left|\delta\chi^2\right|
_{\rm end} \over m^2} e^{2\mu m\Delta t} \,.
\end{equation}

The statistics of these second-order fluctuations are non-Gaussian,
being a $\chi^2$-distribution. Both the mean and the variance of
$\zeta_{{\rm nad}}$ are non-vanishing. The mean value will not
contribute to density fluctuations, but rather indicates that the
background we are expanding around is unstable as energy is
systematically drained from the inflaton field.  We are interested in
the variance of the curvature perturbation, and in particular the
change of the curvature perturbation power spectrum on super-horizon
scales which is negligible if the power spectrum of $\zeta_{{\rm
nad}}$ on those scales is much less than that of $\zeta$ generated
during inflation, the latter being required to be of order $10^{-10}$
to explain the COBE observations.

To evaluate the power spectrum for $\zeta_{\rm nad}$ we must evaluate
the power spectrum of $\delta\chi^2$ which is given 
by substituting ${\cal P}_{\delta\chi}$, from
Eq.~(\ref{powspec3}), 
into Eq.~(\ref{sqrspec}). This gives
\begin{equation}
{\cal P}_{\delta\chi^2} = {2\over3\nu^2} \left( {H_{\rm end} \over
2\pi} \right)^4 \left( {k_{\rm max} \over k_{\rm end}} \right)^3 
\left( {k \over k_{\rm end}} \right)^3 I(\kappa,m\Delta t)\,,
\end{equation}
where 
\begin{equation}
\label{integral}
I(\kappa,m\Delta t)\equiv \frac{3}{2} \int_0^{\kappa_{\rm cut}} \!\! d\kappa'
\int_0^\pi \! d\theta \, e^{2(\mu_{\kappa'}+\mu_{\kappa-\kappa'})m\Delta t} 
\kappa'^2 \sin\theta \,,
\end{equation}
$\kappa= k/k_{\rm max}$ as defined in Eq.~(\ref{defkappa}), 
and $\theta$ is the angle between ${\bf k}$ and ${\bf k}'$. 
Note that at the end of inflation we have $I(\kappa,0)=\kappa_{\rm
cut}^3\sim 1$, and ${\cal P}_{\delta\chi^2}\propto k^3$. 
This yields
\begin{equation}
\label{Pzetanad}
{\cal P}_{\zeta_{\rm nad}} 
\simeq {2^{9/2}3 \over \pi^5\mu^2} \left({\varphi_{\rm{ini}} 
\over m_{\rm Pl}}\right)^2 
\left({H_{\rm end} \over m}\right)^4 g^4 q^{-1/4} \left({k\over k_{\rm
end}}\right)^3 I(\kappa,m\Delta t) \,.
\end{equation}

One might have thought that the dominant contribution to $\zeta_{{\rm
nad}}$ on large scales would come from $\delta\chi$ fluctuations on
those scales, and that is indeed the presumption of the calculation of
Bassett et al.~\cite{Betal_a,Betal_b}.  However, in fact the integral is
initially dominated by $k' \sim k_{{\rm cut}}$, namely the shortest
scales. The reason for this is the steep slope of ${\cal
P}_{\delta\chi}$; were it much shallower (spectral index less than
3/2), then the dominant contribution would come from large scales.

To study the scale dependence of $I(\kappa,m\Delta t)$ and hence
${\cal P}_{\zeta_{\rm nad}}$ at later times, we can
expand $\mu_{\kappa-\kappa'}$ for $\kappa \kappa'\ll 1$ as 
\begin{equation}
\mu_{\kappa-\kappa'} = \mu_{\kappa'} 
+\frac{2\kappa'\cos\theta}{2+e^{\pi \kappa'^2}}\, 
\kappa + {\cal O}(\kappa^2)\,.
\end{equation}
We can then write the integral in Eq.~(\ref{integral}) as
\begin{equation}
I(\kappa,m\Delta t)
 = I_0(m\Delta t)+ {\cal O}(\kappa^2)\,,
\end{equation}
where first-order terms, ${\cal O}(\kappa)$, vanish by symmetry and
\begin{equation}
I_0(m\Delta t)=\frac{3}{2}
\int_0^{\kappa_{\rm cut}} e^{4 \mu_{\kappa'} m \Delta t}
\kappa'^2 d\kappa' \,.
\end{equation}
Thus the scale dependence of ${\cal P}_{\zeta_{\rm nad}}$ remains $k^3$
on large-scales for which $\kappa\ll 1$.

\begin{figure}
\begin{center}
\includegraphics[width=0.7\textwidth]{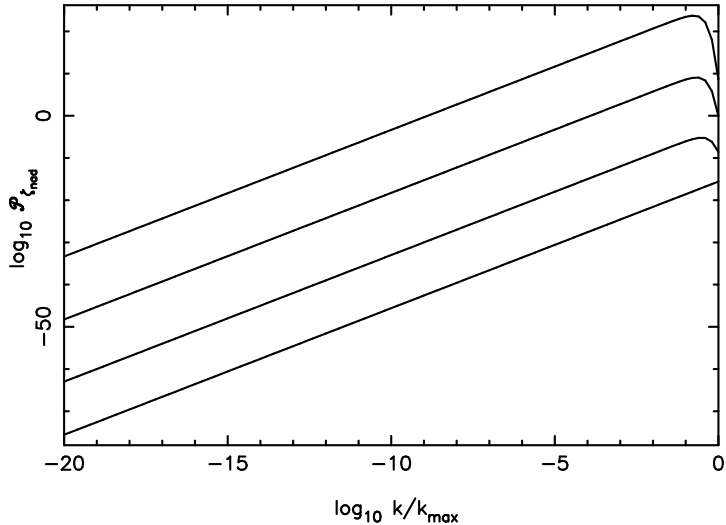}\\
\caption[specint]{\label{specint} The power spectrum of the
non-adiabatic curvature perturbation ${\cal P}_{\zeta_{{\rm nad}}}$,
shown at four different times: from bottom to top $m\Delta t = 0$,
$50$, $100$ and $150$. The parameters used were $g=10^{-3}$,
$m=10^{-6} m_{{\rm Pl}}$ and $k_{\rm cut}=k_{\rm max}$.}
\end{center}
\end{figure}

At late times these integrals become dominated by the 
modes with $\kappa'^2\ll (m\Delta t)^{-1}$ which are preferentially
amplified during preheating. These are longer wavelength 
than $k_{{\rm cut}}$, but still very short compared to the scales 
which give rise to large scale structure in the present Universe.
{}From Eq.~(\ref{defmuk}) we have
$\mu_{\kappa'} \approx \mu_0 - \kappa'^2/3$, for $\kappa'^2\ll1$, where
$\mu_0=(\ln 3)/2\pi$, which gives the asymptotic behaviour at late
times
\begin{equation}
I_0 \simeq 0.86 (m\Delta t)^{-3/2}e^{4\mu_0 m\Delta t} \,.
\end{equation}
Thus although the rate of growth of ${\cal P}_{\zeta_{\rm nad}}$
becomes determined by the exponential growth of the long-wavelength
modes, the scale dependence on super-horizon scales remains
proportional to $k^3$ for $\kappa \lesssim (m\Delta t)^{-1/2}$. This
ensures that there can be no significant change in the curvature
perturbation, $\zeta$, on very large scales before back-reaction on
smaller scales becomes important and this phase of preheating
ends when $m\Delta t\sim100$~\cite{KLS97}.

Numerical evaluation of Eq.~(\ref{Pzetanad}) confirms our analytical
results, as shown in Fig.~\ref{specint}. For $k \ll k_{{\rm
max}}$, the spectral index remains $k^3$ during preheating. Observable
scales have $\log_{10} k/k_{{\rm max}} \simeq -20$. \\

\subsection{Conclusions}

Our result shows that because of the $k^3$ spectrum of $\delta \chi$,
which leads to a similarly steep spectrum for $\zeta_{{\rm nad}}$,
there is a negligible effect on the large-scale perturbations before
the resonance ceases. The fluctuations in $\chi$ grow largest on
small scales, so that backreaction can turn off the 
resonant amplification on these scales before the curvature
perturbation can become non-linear on very large scales.

This contrasts with the original findings of Ref.~\cite{Betal_b}, in
which Bassett et al.~report the results of their investigation of the
model given by Eq.~(\ref{massivemodel}).  They solved the field
equations for particular modes to linear order numerically, assuming
that the initial $\chi$ spectrum is flat.  For strong parametric
resonance this would lead to an exponential amplification of the
curvature perturbation $\zeta$ on super-horizon scales into the
non-linear regime before backreaction can shut down the
resonance. This would undermine the standard cosmological model, in
which the curvature perturbation is constant on large scales.

Our result was derived analytically, with exemption of the integral in
Eq.~(\ref{ourintegral}), which we had to evaluate numerically.  The
use of an analytic approximation, as described in Section 
\ref{preheatingmainsection},
allowed us to calculate the power spectrum of the $\chi$ fluctuations
and hence enables us to calculate their effect over the whole range of
scales.
%
The suppression of the large-scale perturbations in $\delta\chi$,
discussed in Refs.~\cite{jedam,ivan}, means that large-scale
perturbations in $\delta\chi$ are completely unimportant. However, it
turns out that they don't give the largest effect, which comes from
the short-scale modes which dominate the integral for $\zeta_{{\rm
nad}}$. Nevertheless, even they give a negligible effect, again with a
$k^3$ spectrum. Indeed, that result with hindsight can be seen as
inevitable; it has long been known
\cite{causal_a,causal_b,causal_c} that local processes conserving 
energy and momentum cannot generate a tail shallower than $k^3$ (with
our spectral index convention) to large scales, which is the Fourier
equivalent of realizing that in real space there is an upper limit to
how far energy can be transported. Any mechanism that relies on
short-scale phenomena, rather than acting on pre-existing large-scale
perturbations, is doomed to be negligible on large scales.

%
%

\newpage
\section{Concluding remarks and discussion}
\label{disc}

\setcounter{equation}{0}

\subsection{Summary}

In Section \ref{cosmopertsect} we presented the framework of
cosmological perturbation theory for single and multiple fluids.
This allowed us to introduce the notion of gauge-invariant
perturbations and to provide their governing equations, which we then
used throughout the rest of this work.

In Sections~\ref{sepsect} and \ref{coquala}, we have identified the
general condition under which the super-horizon curvature perturbation
on spatial hypersurfaces can vary as being due to differences in the
integrated expansion along different worldlines between hypersurfaces.
As long as linear perturbation theory is valid, then, when spatial
gradients of the perturbations are negligible, such a situation can be
described using the separate universes picture, where regions are
evolved according to the homogeneous equations of motion.

Equivalently, we can study the model under consideration using the
perturbed energy conservation equation.  As discussed in detail in
Section~\ref{coquala} and in Ref.~\cite{separate}, large-scale
curvature perturbations can vary provided there is a significant
non-adiabatic pressure perturbation. This is always possible in
principle if there is more than one field or fluid, and since for
example preheating usually involves at least one additional field into
which the inflaton resonantly decays, such variation is in principle
possible during preheating.
In particular, the curvature perturbation on uniform-density
hypersurfaces, $\zeta$, can vary only in the presence of a significant
non-adiabatic pressure perturbation.  The result follows directly from
the local conservation of energy--momentum and is independent of the
gravitational field equations. Thus $\zeta$ is conserved on
sufficiently large scales in any metric theory of gravity, including
scalar--tensor theories of gravity or induced four-dimensional gravity
in the brane-world scenario \cite{MWBH99,separate}.

Multi-component inflaton models are an example where non-adiabatic
perturbations may cause the curvature perturbation to evolve on
super-horizon scales. \\


In Section \ref{asssect} we have shown that the recently proposed
model of assisted inflation, driven by many scalar fields with steep
exponential potentials, can be better understood by performing a
global rotation in field space, which allowed us to re-write the
potential as a product of a single exponential potential for a
weighted mean field, $\bar\varphi$, and a potential $\bar{V}_n$ for
the orthogonal degrees of freedom, $\bar\sigma_i$, which has a global
minimum when $\bar\sigma_i=0$.

We have also been able to give exact solutions for inhomogeneous
linear perturbations about the attractor trajectory in terms of our
rotated fields.
Perturbations in the weighted mean field $\bar\varphi$ corresponds to the 
perturbations in the density on the uniform curvature hypersurfaces,
or equivalently, perturbations in the curvature of constant
density hypersurfaces and hence constitute
perturbations  along the attractor trajectory. They therefore 
describe adiabatic curvature perturbations.

First-order perturbations in the fields orthogonal to the weighted
mean field are isocurvature perturbations during inflation. 
Vacuum fluctuations lead to a positively tilted spectrum.
%
%
We have shown that the non-adiabatic perturbations decay relative to the
adiabatic perturbations and hence we recover the single field limit at
late times. In particular we find that the curvature perturbation
$\zeta$ becomes constant on super-horizon scales during inflation. 

Recently field rotations have been used in a more general context than
assisted inflation \cite{GWBM}. It has been shown that for multiple
scalar fields a local rotation in field space can always be used to
separate out adiabatic and entropic modes. 
It was shown that the non-adiabatic part of the pressure perturbation
on large scales is proportional to the curvature of the trajectory
in field space, and that $\zeta$ remains constant on large scales
for straight line trajectories such as the assisted inflation 
attractor with $\bar\sigma_i=0$. \\


In Section~\ref{preheatsect} we have focussed on the simplest
preheating model, as discussed in Ref.~\cite{Betal_a,Betal_b}.  We
have identified the non-adiabatic pressure, and shown that the
dominant effect comes from second-order perturbations in the
preheating field.  Further, the effect is dominated by perturbations
on short scales, rather than from the resonant amplification of
non-adiabatic perturbations on the large astrophysical
scales. Nevertheless, we have shown that the contribution has a $k^3$
spectrum to large scales, rendering it totally negligible on scales
relevant for structure formation in our present Universe by the time
backreaction ends the resonance. Amongst models of inflation involving
a single-component inflaton field, this model gives the most
preheating, and so this negative conclusion will apply to all such
models.

In Ref.~\cite{Betal2} Bassett et al.~have suggested large effects
might be possible in more complicated models. They consider two types
of model. In one kind, inflation takes place along a steep-sided
valley, which lies mainly along the direction of a field $\varphi$ but
with a small component along another direction $\chi$. In this case,
one can simply define the inflaton to be the field evolving along the
valley floor, and the second heavy field lies orthogonal to it.
Taking that view, there is no reason to expect the preheating of the
heavy field to give rise to a bigger effect than in the simpler model
considered in this paper.

In the second kind of model, the reheat field is light during
inflation, and this corresponds to a two-component inflaton field. As
has long been known, there can indeed be a large variation of $\zeta$
in this case, which can continue until a thermalized
radiation-dominated universe has been established. Indeed, in models
where one of the fields survives to the present Universe (for example
becoming the cold dark matter), variation in $\zeta$ can continue
right to the present. This variation is due to the presence on large
scales of classical perturbations in both fields (properly thought of
as a multi-component inflaton field) generated during inflation, and
the effect of these must always be considered in a multi-component
inflation model, with or without preheating. \\

Recently, using the results presented in Sections~\ref{cosmopertsect}
and~\ref{preheatsect}, we were able to present the first calculations
predicting the over-production of primordial black holes (PBHs) due to
preheating \cite{GM}.
The scales which we are interested in pass
outside the Hubble radius towards the end of inflation. 
The fluctuations in the preheat field $\chi$ are amplified during
preheating and large PBHs will be formed, when these scales re-enter
the Hubble radius after reheating during the radiation dominated era,
if the fluctuations are sufficiently large.
We found that for a wide range of parameter values in the massive
inflaton model given by Eq.~(\ref{massivemodel}) PBHs are
over-produced \emph{before} back-reaction can shut down the resonant 
amplification.
This poses a serious problem for the standard theory of preheating 
as described in \cite{KLS97}.

\subsection{Extensions}
\label{extensions}

A natural extension of the cosmological perturbation theory 
of 4-dimensional spacetimes, perturbed around a FRW 
background, is to higher dimensional theories.
Non-Einstein gravity (in our four spacetime dimensions) may for example
emerge \cite{SMS99} from theories involving a large extra dimension
\cite{hw_a,hw_b,RS99_a,RS99_b}.
Recently there has been a lot of interest in this so called brane-world
scenario~\cite{BDEL}. In this model matter is confined to the world
volume of a 4-dimensional three-brane, whereas gravity lives as well
in the higher dimensional bulk spacetime. The simplest case of such a
higher dimensional spacetime is a 5-dimensional spacetime, in which
4-dimensional slices correspond to ``our'' standard FRW universe.

In order to study inhomogeneous perturbations we will pick a specific
form for the unperturbed 5-d spacetime metric that accommodates
spatially flat FRW cosmological solutions on the brane, 
\begin{equation}
\label{backmetric}
ds_5^2 = - n^2(t,y) d\eta^2 + a^2(t,y) \delta_{ij}dx^i dx^j
 + dy^2 \, ,
\end{equation}
where $a=a(t,y)$ and $n=n(t,y)$ are scale factors and $y$ is the
``new'' fifth dimension. The line element given above includes anti-de
Sitter spacetime as a special case and has been extensively studied in
the literature~\cite{BDEL}.

We consider arbitrary linear perturbations about the background
metric. In keeping with the standard approach of cosmological
perturbation theory~\cite{Bardeen}, as presented in Section
\ref{cosmopertsect} of this thesis, we will introduce scalar, vector
and tensor perturbations defined in terms of their properties on the
3-spaces at fixed $t$ and $y$ coordinates.

We can write the most general metric perturbation to first-order as
\begin{equation}
\label{pertmetric}
g_{AB}= \left(
\begin{array}{ccc}
-n^2(1+2\phi) & na(B_{|i} - S_i) & n\phi_y \nonumber\\
na(B_{|j} - S_j) &
a^2\left[ (1-2\psi)\gamma_{ij} + 2E_{|ij} + 2F_{(i|j)} + h_{ij}
\right] & a(B_{y|i} - S_{yi})
 \nonumber\\
n\phi_y & a(B_{y|i} - S_{yi}) & 1+2\phi_{yy} \nonumber
\end{array}
\right) \,,
\end{equation}
where $\phi$, $B$, $\psi$, $E$, $\phi_y$ and $\phi_{yy}$ are scalars,
$S$, $F$, and $S_y$ are (divergence-free) 3-vectors, and $h_{ij}$ is a
(transverse and traceless) 3-tensor.
The reason for splitting the metric perturbation into these three
types is that they are decoupled in the linear perturbation equations,
as in the 4-dimensional case (for a separable metric Ansatz).

In the perturbed spacetime there is a gauge-freedom in the definition
of the scalar and vector perturbations, as in the standard case 
described in Section~\ref{cosmopertsect}. Under a first-order coordinate
transformation, $x^A\to x^A+\xi^A$, which we will write here as
\begin{eqnarray}
\label{shift}
t &\to& t + \delta t \nonumber \\
x^i &\to& x^i + \delta x_|^{~i} + \delta x^i \\
y &\to& y + \delta y \nonumber
\end{eqnarray}
where $\delta t$, $\delta x$ and $\delta y$ are scalars and $\delta
x^i$ is a (divergence-free) 3-vector, the perturbations transform as
\begin{eqnarray}
\phi &\to& \phi - \dot{\delta t} - \frac{\dot n}{n}\delta t
- \frac{n'}{n} \delta y \, , \qquad
B \to B + \frac{n}{a} \delta  t
- \frac{a}{n} \dot{\delta x} \, , \nonumber \\
S_i &\to& S_i + \frac{a}{n} \dot{\delta x_i} \, , \qquad
B_y \to B_y - a \delta x' - \frac{1}{a} \delta y \, , \nonumber \\
S_{yi} &\to& S_{yi} + a \delta {x_i}' \, ,  \qquad
\psi \to \psi +  \frac{\dot a}{a} \delta t
+ \frac{a'}{a} \delta y \,, \\
E &\to& E - \delta x \, , \qquad
F_i \to F_i -\delta x_i \, , \nonumber \\
\phi_y &\to& \phi_y  +n \delta t'
- \frac{1}{n} \dot{\delta y}\, , \qquad
\phi_{yy} \to \phi_{yy} - \delta y' \, , \nonumber
\end{eqnarray}
where dot and prime denote differentiation with respect to coordinate
time $t$ and bulk coordinate $y$, respectively. Comparing these
equations with the transformation properties of scalar and vector
perturbations in the 4-dimensional case, Eq.~(\ref{transphi})
-(\ref{vectran}), we see that there is now even more scope for
``gauge-ambiguities''. But we can easily circumvent any such
difficulties by, again, simply fixing the gauge and constructing
gauge-invariant quantities.
For example, we can construct two gauge-invariant vector 
perturbations from the vector perturbations given above,
\bea
\tilde S_i=S_i+\frac{a}{n}\dot F_i \, , \\
\tilde S_{yi} = S_{yi} +a F_i' \,.
\eea
Although we have ``only'' introduced one more dimension to the
problem the number of degrees of freedom increases quite 
dramatically. In the 4-dimensional case we had four scalar
perturbations, $\phi$, $B$, $\psi$ and $E$, and two vector
perturbations, $F_i$ and $S_i$. This allowed us to construct 
two gauge-invariant scalar perturbations and one vector 
gauge-invariant perturbation (see Section \ref{gicomb}).

In the 5-dimensional case we have seven scalar perturbations, $\phi$,
$B$, $\psi$, $E$, $B_y$, $\phi_y$ and $\phi_{yy}$, and three vector
perturbations, $F_i$, $S_i$ and $S_{yi}$.  This allows us to construct
four gauge-invariant scalar perturbations and two vector
gauge-invariant perturbations.  The theory of cosmological
perturbations in more than four dimensions will therefore be much
richer than in four dimensions.

First steps towards solving these problems have already been
undertaken \cite{mukhoyama,kodamabrane,langloisbrane,vandenbruck}.
For example, our proof of the constancy of the curvature perturbation
$\zeta$ on large scales, Eq.~(\ref{dashzeta}), which follows from the
local conservation of energy-momentum independent of the form of the
gravitational field equations, validates a recent discussion
\cite{MWBH99} of chaotic inflation in the brane-world scenario, which
relied on that equation.

\subsection{Outlook}

The theory of cosmological perturbations has made dramatic progress
since the seminal works of Lifshitz \cite{lifshitz} and Bardeen 
\cite{Bardeen}. Nevertheless, the theory is far from complete with 
many open questions remaining.

What should be done next? Although the governing equations for 
multi-component systems with energy exchange have been developed
\cite{KS} and applied to simple systems of fluids \cite{ks87}
quite some time ago, the system of equations used is not entirely
satisfactory. Using the formalism outlined in this work in Section 
\ref{cosmopertsect} and using insight into multi-component systems 
gained in Section \ref{asssect} and \cite{GWBM} it should be possible
to cast the governing equations for multi-component systems with
energy exchange into a new gauge-invariant form. Splitting the system
again into adiabatic and entropic perturbations it should be possible
to show that the adiabatic perturbations are sourced by the entropic
perturbations, whereas the entropic perturbations can't be sourced by
the adiabatic ones on large scales. Interesting systems would, for
example, include quintessence or a form of collisional dark matter as
well as ``standard'' forms of matter.

One obvious extension of the cosmological perturbation formalism is to
higher dimensional theories, as outlined in the previous section
\ref{extensions}. Even if it should turn out that we do not live on a
4-dimensional 3-brane embedded in a higher dimensional space, we can
learn a lot about standard 4-dimensional cosmological perturbations
and improve our technical expertise from investigating these theories.

Recently we were able to show in Ref.~\cite{GM} that the preheating 
model studied in Section \ref{preheatsect} leads to an overproduction
of primordial black holes (PBH) during the radiation dominated era 
before backreaction can shut down the amplification of the field
fluctuations. This constitutes the first consistent investigation
of this kind. But is this a generic feature of preheating, or is there
something wrong with the backreaction mechanism?  
It would therefore be interesting to study other models of preheating
with respect to their PBH generation behaviour. \\




\newpage


%
%
\newpage
\appendix				
\section{Appendix}

\setcounter{equation}{0}
\setcounter{page}{1}
\renewcommand{\thepage}{A.\arabic{page}}
\def\theequation{A.\arabic{equation}}

\subsection{Notation}

The sign convention is (+++) in the classification of Ref.~\cite{MTW}. \\
The number of spatial dimensions is throughout the main part of the thesis 
$n=3$, but we do not fix  $n$ in the appendix. \\
{Tensor indices:} \\
%
Greek indices, such as $\alpha,\beta,\dots,\mu,\nu,\dots$, run from 
0 to $n$, that is over all dimensions. 
Latin indices, such as $a,b,\dots,i,j,\dots$, run from 1 to $n$, that
is only over spatial dimensions. \\
%
%
The connection coefficient is defined as
\be 
\Gamma^{\gamma}_{\beta\mu}= {1\over2}g^{\alpha\gamma} 
\left( g_{\alpha\beta, \mu} + g_{\alpha\mu,\beta} -
 g_{\beta\mu,\alpha} \right) \,.
\ee
The Riemann tensor is defined as 
\be
R^{\alpha}_{~\beta\mu\nu}=
\Gamma^{\alpha}_{\beta\nu,\mu}-\Gamma^{\alpha}_{\beta\mu,\nu}+
\Gamma^{\alpha}_{\lambda\mu}\Gamma^{\lambda}_{\beta\nu}-
\Gamma^{\alpha}_{\lambda\nu}\Gamma^{\lambda}_{\beta\mu} \,.
\ee
The Ricci tensor is a contraction of the Riemann tensor and given by
\be
R_{\mu\nu}=R^{\alpha}_{~\mu\alpha\nu} \,,
\ee
and the Ricci scalar is given by contracting the Ricci tensor
\be
R=R^{\mu}_{~\mu} \,.
\ee
The Einstein tensor is defined as
\be
G_{\mu\nu}=R_{\mu\nu}-\frac{1}{2}g_{\mu\nu}R \,.
\ee
The covariant derivatives are denoted by
\bea
&{}_{;\mu}&\equiv\nabla_{\mu}~~~ {\rm{Covariant~differentiation~with~ 
respect~to}}~g_{\mu\nu} \,, \nonumber \\ 
&{}_{|i}&\equiv\nabla_i~~~ {\rm{Covariant~differentiation~with~ 
respect~to}}~\gamma_{ij} \,.  \nonumber 
\eea
Throughout this work we use the units $c=\hbar=1$.

\subsubsection{Time variables}

Throughout this work we use as time variable the conformal time 
$\eta$, if not stated otherwise. Conformal time is related to
 coordinate time $t$ by
\be
dt=a d\eta \, .
\ee
The time derivatives are denoted by
\bea
&\dot{}&\equiv\frac{d}{dt}~~~{\rm{Time~derivative~with~respect~to
~coordinate~time}} \,, \nonumber \\
&{}^{\prime}&\equiv\frac{d}{d\eta}~~~{\rm{Time~derivative~with~respect~to
~conformal~time}} \,. \nonumber
\eea
Another useful time variable is proper time $\tau$. The definition
follows from the line element $ds^2$ in Eq.~(\ref{ds}) as
\be
d\tau=a\sqrt{1+2\phi}~d\eta \, ,
\ee
and can be rewritten, to first-order in  $\phi$, as
\be
d\tau=a\left(1+\phi\right)d\eta \, .
\ee
%
\subsubsection{Metric perturbation variables}

Before relating our notation to other works, we give the linearly
perturbed metric tensor around a FRW background already presented in
Section \ref{metrictensorsection}.

The covariant metric tensor is given by
\be
g_{\mu\nu}=a^2(\eta) \left( 
\begin{array}{cc} 
-(1+2\phi) & B_{|i} - S_i\nonumber\\ 
B_{|j} - S_j& (1-2\psi)\gamma_{ij} + 2E_{|ij} + F_{i|j} + F_{j|i} + h_{ij}
 \nonumber
\end{array} 
\right)\,,
\ee
and the contravariant metric tensor is
\be
g^{\mu\nu}=a^{-2}(\eta) \left( 
\begin{array}{cc} 
-(1-2\phi) & B_{|}^{~i}-S^i \nonumber\\ 
B_{|}^{~j}-S^j &
 (1+2\psi)\gamma^{ij} - 2E_{|}^{~ij} -F^{i~j}_{~|} - F^{j~i}_{~|} - h^{ij}
 \nonumber
\end{array} 
\right)\,.
\ee

The notation of Kodama and Sasaki~\cite{KS} is related to our notation 
as follows 
\footnote{The notation of Bardeen \cite{Bardeen} is similar to that 
in \cite{KS}, but the harmonic functions are denoted $Q^{(0)}$.}. 
The scalar perturbations are 
\begin{eqnarray}
\phi \equiv AY \, ,
&\qquad& 
B_{,i} \equiv \frac{1}{k}(BY)_{,i} \, , \nonumber \\
E \equiv \frac{H_T Y}{k^2} \, ,
&\qquad& 
\psi \equiv -\left( H_L+{1\over3}H_T \right) Y \, .
\end{eqnarray}
Another useful relation is $\psi-\frac{1}{n}\nabla^2 E \equiv-H_L Y$.
The alternative notation ${\cal R} = - \frac{\psi}{Y}$ is also used. 
The vector perturbations are
\begin{eqnarray}
S_i \equiv B^{(1)}Y^{(1)}_i \, ,
&\qquad& 
F_{i} \equiv -\frac{1}{k}H_T^{(1)} Y^{(1)}_{i} \, .
\end{eqnarray}
The tensor perturbations are related by 
\be
h_{ij} \equiv 2 H_T^{(2)} Y^{(2)}_{ij} \, .
\ee
The scalar harmonic functions are defined as
\bea
\nabla^2 Y&=&-k^2 Y \, ,~~~~~~
Y_i \equiv -\frac{1}{k}Y_{|i} \, , \nonumber \\
Y_{ij} &\equiv& \frac{1}{k^2}Y_{|ij}+\frac{1}{n}\gamma_{ij}Y \, ,
\eea
where $\nabla^2\equiv\nabla^i\nabla_i$.
The vector valued harmonic functions are defined by 
\begin{eqnarray}
(\nabla^2 + k^2) Y^{(1)}_i&=&0 \, ,
\qquad 
Y^{(1)i}_{~~~i}=0 \, , \nonumber \\
Y^{(1)}_{ij} &\equiv& -\frac{1}{2k} \left( Y^{(1)}_{i|j} 
+ Y^{(1)}_{j|i} \right) \, .
\end{eqnarray}
Note as well that $ Y^{(1)~j}_{ij|} = \frac{1}{2k} 
\left( k^2 - (n-1) \kappa \right) Y^{(1)}_i $.
The tensor valued harmonic functions are defined by 
\begin{eqnarray}
(\nabla^2 + k^2) Y^{(2)}_{ij}=0 \, ,
&\qquad& 
Y^{(2)i}_{~~~~i}=0 \, , \nonumber \\
Y^{(2)ij}_{~~~~|j}=0 \, .
\end{eqnarray}
%

\subsubsection{Matter variables}

The scalar matter variables in the notation used by us compare
to the variables used by Kodama and Sasaki in \cite{KS} as
\begin{eqnarray}
\delta\rho \equiv \delta Y \, ,
&\qquad& 
v_{,i} \equiv - \frac{(v Y)_{,i}}{k} \, , \nonumber \\
\delta p \equiv p \pi_L Y \, ,
&\qquad& 
\pi \equiv \frac{p}{k^2} \pi_T Y \, ,
\end{eqnarray}
and $\pi^i_{~j}=p \pi_T Y^i_{~j}$. 
The vector matter variables are related by
\bea
v_i \equiv v^{(i)} Y_i \, ,
\qquad \pi_{(i|j)} \equiv  \pi^{(1)}_T Y^{(1)}_{ij} \, ,
\eea
and the tensor matter variables by
\be
{}^{(tensor)}\pi_{ij} \equiv \pi^{(2)}_T Y^{(2)}_{ij} \, .
\ee

\subsection{The connection coefficients}

The connection coefficients in the FRW background are
\bea
\Gamma^0_{00}& = h \, ,  &\Gamma^0_{0i} =0 \, ,  \nonumber \\
\Gamma^0 _{ij}& = h\gamma_{ij} \, , 
&\Gamma^i_{00} = 0 \, ,  \nonumber \\
\Gamma^i_{j0}& = h \delta^i_j \, , 
&\Gamma^i_{jk} = {}^{(n)}\Gamma^i_{jk} \, ,
\eea
where ${}^{(n)}\Gamma^i_{jk}$ is the connection of the n-sphere.
The perturbed connection coefficients for the scalar perturbations
corresponding to the metric given in Eq.~(\ref{gmunu}) are
\begin{eqnarray}
&\delta\Gamma^0_{00}& = \phi'   \, ,
~~~\Gamma^0_{0i} = \phi_{,i} + h B_{,i} \, ,  \nonumber \\
&\delta\Gamma^0 _{ij}& = -\left( B-E' -2hE \right)_{|ij} 
+ \left( -2h \left( \phi+\psi \right) 
+\psi' \right) \gamma_{ij}   \, , \nonumber  \\
&\delta\Gamma^i_{j0}& = - \psi'  \delta^i_j 
+ E'^{~i}_{~|~j} \nonumber \, ,  \\
&\delta\Gamma^i_{00}& = \left( hB +B' + \phi \right)_{|}^{~i}  
\nonumber \, , \\  
&\delta\Gamma^i_{jk}& =  - h\gamma_{jk} B_{|}^{~i}
-\psi_{,k}\delta^i_j - \psi_{,j}\delta^i_k + \psi_{|}^{~i}\gamma_{jk} 
\nonumber \\
&\qquad& \qquad + E_{|j~k}^{~~i} +  E_{|k~j}^{~~i} + E_{|jk}^{~~~i} \, .  
\end{eqnarray}
The perturbed connection coefficients for the vector perturbations
are
\begin{eqnarray}
&\delta\Gamma^0_{00}& =0  \, ,
~~~\Gamma^0_{0i} = -h S_i \, ,  \nonumber \\
&\delta\Gamma^0 _{ij}& = \frac{1}{2}\left(S_{i,j}+S_{j,i}\right)
+\frac{1}{2}\left(F_{i|j}+F_{j|i}\right)'
+h\left(F_{i|j}+F_{j|i}\right) \, , \nonumber  \\
&\delta\Gamma^i_{j0}& = \frac{1}{2} \left(F^{i}_{~|j}+F_{j|}^{~i}\right)'
+\frac{1}{2} \left(-S^i_{~,j}+S_{j,}^{~i}\right) \, , \nonumber \\
&\delta\Gamma^i_{00}& = -\left( {S^i}' + h S^i \right)
\nonumber \, , \\  
&\delta\Gamma^i_{jk}& = h S^i \gamma_{jk}
+\frac{1}{2}\left\{F^i_{~|jk}+F^{~~i}_{j|~k}+F^i_{~|kj}+F_{k|~j}^{~~i}
\right. \nonumber \\
&\qquad& \qquad \left. -F_{j|k}^{~~~i}-F_{k|j}^{~~~i} \right\} \, . 
\end{eqnarray}
For the tensor perturbations we find
\bea
&\delta\Gamma^0_{00}& =\Gamma^0_{0i} =\Gamma^i_{00}= 0 \, , \nonumber \\
&\delta\Gamma^i_{0j}&=\frac{1}{2} h^{\prime i}_{~j} \, , \nonumber \\
&\delta\Gamma^0_{ij}&=\frac{1}{2} h^{\prime}_{ij}+ h~h_{ij} \, , \nonumber \\
&\delta\Gamma^i_{jk}&=\frac{1}{2}
\left(h^i_{~j|k}+h^i_{~k|j}-h_{jk|}^{~~~i}\right) \, .
\eea
Note that $h$ is the Hubble parameter in conformal time and
not related to the tensor perturbation $h_{ij}$.

\subsection{The Einstein tensor components}

In the background the Einstein tensor takes the form
\bea
G^0_{~0} &=& -\frac{n(n-1)}{2a^2} \left(h^2+\kappa\right) \, ,\nonumber \\
G^i_{~j} &=& -\frac{n-1}{a^2} \left( \frac{a^{\prime\prime}}{a}
+\frac{n-4}{2}+h^2+\frac{n-2}{2} \kappa \right)~\delta^i_{~j} \, ,\nonumber \\
G^0_i &=& G^i_0 = 0 \, .
\eea
Note that $G^0_0-nG^i_i=\frac{n(n-1)}{a^2}
\left(h^2-h^{\prime}+\kappa\right)$. \\
For the scalar perturbations we find
\bea
\delta G^0_{~0} &=& \frac{n-1}{a^2}\left[ nh^2\phi +h\nabla^2 B
+nh\left(\psi^{\prime}-\frac{1}{3}\nabla^2E^{\prime}\right)
-\left(\nabla^2+n\kappa\right)\psi\right] \, , \nonumber \\
\delta G^0_{~i} &=& \frac{n-1}{a^2}\left[h\phi+ \psi^{\prime}
+\kappa\left(B-E^{\prime}\right)\right]_{|i} \, , \nonumber \\
\label{deltaGij}
\delta G^i_{~j} &=& \frac{n-1}{a^2}\left[ 
\left\{2\frac{a^{\prime\prime}}{a}+(n-4)h^2\right\} \phi + 
h\phi^{\prime} +\psi^{\prime\prime}+h\psi+(n-2)h\psi
\right. \nonumber \\
&\qquad& \qquad  -(n-2)\kappa\psi \Bigg] \delta^i_{~j}
+\frac{1}{a^2} \left[\nabla^2 D ~\delta^i_{~j} -D^{~i}_{|~j} \right] \, ,
\eea
where $D=\phi-(n-2)\psi-h\sigma-\sigma^{\prime}-(n-2)h\sigma$.
The trace of the spatial part of the perturbed Einstein tensor is 
obtained by contracting Eq.~(\ref{deltaGij}) to get 
\bea
\label{trGij}
\delta G^i_{~i} &=&
\frac{n(n-1)}{a^2}\left[ 
\left\{2\frac{a^{\prime\prime}}{a}+(n-4)h^2\right\} \phi + 
h\phi^{\prime} +\psi^{\prime\prime}+h\psi+(n-2)h\psi \right.\nonumber \\
&\qquad& \qquad  -(n-2)\kappa\psi\Bigg] 
+ \frac{(n-1)}{a^2} \nabla^2 D \, .
\eea
For the vector perturbations we find
\bea
\delta G^0_{~0} &=& 0 \, , \nonumber \\
\delta G^0_{~i} &=& - \frac{(n-1)\kappa+\nabla^2}{2a^2}
\left(  F_{i}^{\prime} +S_i \right) \, , \nonumber \\
\delta G^i_{~j} &=& \left( \frac{1}{2a^2} \right)
\left[ \left( \frac{\partial}{\partial\eta}+ (n-1)h \right) 
\left( S^i_{|j}+S_{j|}^{~i} + F^{i\prime}_{~|j}+F_{j|}^{\prime~i} 
\right) \right] \, .
\label{vectorGij}
\eea
Note that the last expression in Eq.~(\ref{vectorGij}) above  can be 
rewritten in terms of the vector shear $\tau_{ij}$, given in 
Eq.~(\ref{vecshear}), as
\be
\delta  G_{ij} =\left( \frac{1}{a^3} \right)
\left[ \tau^{\prime}_{ij}+(n-2) h \tau_{ij} \right] \,.
\ee

For the tensor perturbations the Einstein tensor components
are
\bea
\delta G^0_{~0} &=& 0 \, , 
~~~~~~~\delta G^0_{~i} = 0  \, , \nonumber \\
\delta G^i_{~j} &=& \left( \frac{1}{2a^2} \right)
\left[ h^{i~\prime\prime}_{~j} + h(n-1)h^{i~\prime}_{~j}
+\left(2\kappa-\nabla^2\right)h^{i}_{~j} \right] \, .
\eea


\newpage
\begin{center}
\subsection{List of Symbols}


\begin{tabular}{l l}
$a$ & Scale factor \\

$a_{\mu}$	&	Acceleration \\

$c_{\rm{s}}$ 	& 	Adiabatic sound speed \\

$c_J$ 	& 	Adiabatic sound speed of Jth component\\

$f_{(J)}$	&	Momentum transfer perturbation of Jth component\\

$g$	&	Coupling of preheat field $\chi$ to the inflaton \\


$g_{\mu\nu}$	&	Metric tensor \\

$\tilde g_{\mu\nu}$	&	Conformally rescaled metric tensor \\

$h$	&	$a'/a$ \\

$h_{ij}$	&	Tensor metric perturbation \\

$k$	&	Comoving wavenumber \\

$m$	&	Mass of the inflaton field \\

$m_{\rm{Pl}}$	&	Planck mass \\

$m_{\chi}$	& 	Mass of preheat field $\chi$ \\

$n$	&	Scale factor \\

$n_{\rm{s}}$	&	Spectral index of curvature perturbations \\

$\Delta n_X$	&	Scale dependence of perturbation spectrum 
of a quantity $X$\\

$p$	&	Pressure \\

$p_J$	&	Pressure of Jth component\\

$p_i$	&	Exponent of ith exponential potential \\

$\bar p$	&	``Mean'' exponent \\

$q$	&	Dimensionless coupling parameter \\

$q_{(J)}$	&	Reduced energy tranfer parameter of Jth component\\

$ds$	&	Infinitesimal line element \\

$t$	&	Coordinate time \\

$\Delta t$	&	Coordinate time elapsed since end of inflation \\

$u$	&	Rescaled perturbation in  mean scalar field, 
		$a\delta\bar\varphi$ \\

$u^{\mu}$	&	4-velocity \\

$v$	&	Scalar velocity perturbation \\

$v_J$	&	Scalar velocity perturbation of Jth component\\

$\bar v_J$	&	Rescaled perturbation in Jth orthogonal field, 
		$a\delta\bar\sigma_J$ \\

$v^i$	&	Vector velocity perturbation \\

$x^i$	&	Spatial coordinate \\

$y$	&	``Fifth'' dimension \\

$z$	&	Mukhanov potential, $a^2\bar\varphi' /a'$ \\

\end{tabular}

\newpage
\begin{tabular}{l l}

$B$	&	Shift vector (scalar metric perturbation) \\

$B_y$	&	Shift vector (scalar metric perturbation, braneworld) \\

$C_{ij}$	&	Scaling solution \\

$E$	&	Anisotropic stress perturbation (scalar metric perturbation) \\

$F_i$	&	Vector metric perturbation \\

$G$	&	Newton's constant \\

$G_{\mu\nu}$	&	Einstein tensor \\

$H$	&	Hubble parameter with respect to coordinate time \\

$H^{(1)}_{\nu}$	& 	Hankel function of the first kind of degree $\nu$ \\

$K_{\mu\nu}$	&	Extrinsic curvature \\

$L$	&	Lagrangian \\

${\cal{L}}$	&	Lagrangian density \\	

$N$	&	Number of e-folds (integrated expansion)\\

$N^{\mu}$	&	Unit time-like vector field \\

$P_{\mu\nu}$	&	Projection tensor, $g_{\mu\nu}+N_{\mu}N_{\nu}$ \\

$Q_{(J)}$	&	Energy transfer parameter of the Jth component \\

$Q^{\mu}_{(J)}$	& 	Energy momentum four vector of the Jth component\\

${\cal{P}}_X$	&	Power spectrum of a quantity $X$ \\

$R$	&	Ricci scalar \\

$R_{\mu\nu}$	&	Ricci tensor \\

${}^{(3)}R$	&	Intrinsic spatial curvature \\

${\cal{R}}$	&	Curvature perturbation in comoving gauge \\

$S$	&	Action \\

$S_i$	& 	Vector metric perturbation \\

$S_{yi}$	& 	Vector metric perturbation (braneworld) \\

$S_{IJ}$	&	Entropy perturbation \\

$\hat S_{IJ}$	&	Reduced entropy perturbation \\

$T_{\mu\nu}$	&	Energy momentum tensor \\

$V$	&	Velocity perturbation in comoving total matter gauge \\

$V(\varphi)$	&	Potential of scalar field \\

$\bar V_n$	& 	``Mean'' potential of n scalar fields \\

\end{tabular}


\newpage
\begin{tabular}{l l}

$\chi$	&	Preheat field \\

$\delta^{\mu}_{~\nu}$	&	Kronecker delta \\

$\epsilon_{(J)}$	&	Energy transfer perturbation \\

$\eta$	&	Conformal time \\

$-\zeta$	&	Curvature perturbation in constant density gauge \\

$\gamma_{ik}$	&	Metric tensor on 3-D space with 
constant curvature $\kappa$ \\


$\kappa$	&	Curvature of background spacetime \\

$\mu_k$		& 	Floquet index \\ 

$\omega$	&	Dimensionless Brans-Dicke parameter \\

$\omega_{\mu\nu}$	&	Vorticity \\

$\pi^i$	&	Anisotropic stress vector \\

$\pi^{\mu}_{~\nu}$	&	Anisotropic stress tensor \\

${}^{\rm{tensor}}\pi^{i}_{j}$	&	Tensorial anisotropic 
stress tensor \\

$\varphi$	&	Scalar field \\

$\bar\varphi$	&	Weighted mean field \\

$\varphi_{\rm{dil}}$	&	Dimensionless dilaton field \\

$\varphi_{\rm{ini}}$	& 	Initial amplitude of inflaton 
oscillation \\ 

$\phi$	&	Lapse function (scalar metric perturbation) \\

$\phi_y$	&	Extra scalar potential in braneworld \\

$\phi_{yy}$	&	Extra scalar potential in braneworld \\

$\psi$	&	Curvature perturbation (scalar metric perturbation) \\

$\rho$	&	Energy density \\

$\rho_J$	&	Energy density of Jth component \\

$\sigma$	&	Shear scalar \\

$\sigma_J$	&	Jth field orthogonal to mean field \\

$\sigma_{\mu\nu}$	&	Shear tensor \\

$\tau_{ij}$	&	Gauge invariant vector perturbation \\

$\theta$	&	Expansion \\

$\xi$	&	Arbitrary scalar function \\

$\xi^0$	&	Arbitrary scalar function \\

$\bar\xi^i$	&	Arbitrary vector valued function \\

$\Delta$	&	Dimensionless density perturbation \\

$\Delta_J$	&	Dimensionless density perturbation of Jth component\\

$\hat\Delta_J$	&	Reduced dimensionless density perturbation 
of Jth component\\

$\Gamma$	&	Entropy perturbation \\ 	

$\Gamma(x)$	&	Gamma function \\

$\Gamma_{\rm{int}}$	&	Intrinsic entropy perturbation \\ 	

$\Gamma_{\rm{rel}}$	&	Relative entropy perturbation \\ 	

$\Gamma_{J}$	&	Intrinsic entropy perturbation of 
Jth component\\ 	

$\Pi$	&	Scalar anisotropic stress tensor \\

$\Phi$	&	Bardeen potential (lapse function in longitudinal gauge) \\

$\Psi$	&	Bardeen potential (curvature perturbation in 
longitudinal gauge) \\

%
%
%
\end{tabular}


\newpage

\end{center}

%
%



\end{document}